\documentclass[aps,prd,nofootinbib,onecolumn,a4paper]{revtex4}
\usepackage{bm,latexsym,amsmath,amssymb,amsfonts,mathrsfs}
\usepackage[dvipdfmx]{graphicx}
\usepackage[hang,small,bf]{caption}
\usepackage[subrefformat=parens]{subcaption}
\usepackage{ascmac}
\usepackage{physics}
\usepackage{color,float}

\begin{document}

\title{Effects of photon field on entanglement generation in charged particles
}%

\author{Yuuki Sugiyama}
 \email{sugiyama.yuki@phys.kyushu-u.ac.jp}
\affiliation{Department of Physics, Kyushu University, 744 Motooka, Nishi-Ku, Fukuoka 819-0395, Japan}

\author{Akira Matsumura}
 \email{matsumura.akira@phys.kyushu-u.ac.jp}
\affiliation{Department of Physics, Kyushu University, 744 Motooka, Nishi-Ku, Fukuoka 819-0395, Japan}
 
\author{Kazuhiro Yamamoto}
 \email{yamamoto@phys.kyushu-u.ac.jp}
\affiliation{Department of Physics, Kyushu University, 744 Motooka, Nishi-Ku, Fukuoka 819-0395, Japan}
\affiliation{
Research Center for Advanced Particle Physics, Kyushu University, 744 Motooka, Nishi-ku, Fukuoka 819-0395, Japan}

%\date{\today}
\begin{abstract}
The Bose–Marletto–Vedral (BMV) experiment~\cite{Bose2017,Marlleto2017} is a proposal for testing the quantum nature of gravity with entanglement 
due to Newtonian gravity.
This proposal has stimulated controversy on how the entanglement due to Newtonian gravity is related to the essence of quantum gravity 
and the existence of gravitons.
Motivated by this, we analyze the entanglement generation between two charged particles coupled to a photon field. 
We assume that each particle is in a superposition of two trajectories and that the photon field is initially in a coherent state. 
Based on covariant quantum electrodynamics, the formula for the entanglement negativity of the charged particles is derived for the first time. 
Adopting simple analytic trajectories of the particles, we demonstrate the entanglement between them.
It is observed that the entanglement is suppressed by the decoherence due to the vacuum fluctuations of the photon field. 
We also find that the effect of quantum superposition of bremsstrahlung appears in the entanglement negativity formula.
The similar structures between the gravity theory and electromagnetic theory suggests that a similar feature may be observed in the entanglement generation by quantum gravitational radiation.
\end{abstract}

\maketitle

\section{Introduction\label{intro}}
The quantum field theory (QFT) is one of the most successful theories to explain the motion of particles and the interactions among them.
However, the QFT of gravity has not been completed. 
It is unclear whether gravity is described by quantum mechanics or not \cite{Feynmann,Carney2019}, and many efforts have been made to test
%evaluate 
the quantum nature of gravity.
In recent years, the proposal of the BMV experiment~\cite{Bose2017,Marlleto2017} for testing the quantum nature of gravity has attracted considerable attention. 
In this proposal, it was proposed that quantum entanglement due to the Newtonian potential between two masses may be evidence of quantum gravity. 
Triggered by previous interesting works, the Newtonian entanglement has been evaluated in several experimental proposals: matter-wave interferometers \cite{Nguyen2020, Miki2021}, mechanical oscillators \cite{Krisnanda2020, Qvafort2020},
optomechanical systems 
\cite{Balushi2018,Miao2020,Matsumura2020,Miki2022}, hybrid systems \cite{Carney2021a,Pedernales2021, LG}, and others.

Entanglement due to gravity will be an important milestone for quantum gravity; 
however, the implication of the BMV experiment is still under debate \cite{Christodoulou2019,Marshman2020,Carney2021b,Belenchia2018, Matsumura2021, Danielson2021,Bose2022}. 
For example, the role of dynamical gravitons in Newtonian entanglement is not obvious. 
This is because the Newtonian potential comes from the constraint equation in the Einstein gravity and does not describe the dynamical degrees of the freedom of gravity. 
To clarify this kind of question, it is necessary to analyze entanglement generation in the context of QFT. 
A crucial step in this direction is to understand the features of quantized fields that appear in entanglement.

The primary purpose of this study is to proceed with the step based on quantum electrodynamics (QED). 
Particularly, we evaluate the effect of a photon field on the entanglement generation between two charged particles.
We assume that each of the charged particles is in a superposition of two trajectories and that the photon field coupled with them is initially in a coherent state. 
This setting is an extension of that considered in \cite{Ford1993,Breuer2001}, where quantum decoherence and phase shift due to a photon field were discussed.
In \cite{Bassi2017, Riedel2013, Blencone2013, Suzuki2015,Kanno2021a,Kanno2021b}, quantum decoherence due to gravitons was also evaluated for a massive object in a superposition state.
In the present paper, using the extended model, we derive the formula of the entanglement negativity of two charged particles for the first time. 
We use the formula to exemplify the entanglement behavior of the charged particles. 
Through the analysis, we find that two quantum phenomena, the vacuum fluctuations of photon field and the quantum superposition of bremsstrahlung, appear in the entanglement negativity formula. 
Particularly, the decoherence due to the fluctuating photon field suppresses the entanglement generation in the charged particles.
We also demonstrate that this decoherence becomes significant when the decoherence due to the photon emission occurs, 
which could be significantly related to each other. 
We infer that the above observed features are universal in the entanglement behavior of two masses coupled to a quantized gravitational field.

The present paper is organized as follows.
The entanglement generation by the Coulomb potential is studied in Sec. \ref{secII}.
In Sec. \ref{secIII}, we consider the dynamics of the charged particles in a spatial superposition.
We first introduce a single charged particle model that interacts with a photon field.
We then extend the above model to that with two charged particles.
We derive the reduced density matrix of the charged particles to discuss the entanglement generation.
In Sec. \ref{secIV}, we evaluate the entanglement generation for two specific configurations.
We discuss the reason for the effect of the difference of the two configurations on the entanglement generation between the two charged particles in Sec. \ref{secV}.
Sec. \ref{secVI} presents the summary and conclusions.
In Appendix \ref{BRST}, we explain the BRST formalism for the gauge fixing in the present paper.
In Appendix \ref{sec:Tr}, we compute the inner product introduced in Eq. \eqref{rhop} and derive Eqs. \eqref{GammaPP'} and \eqref{PhiPP'}. 
In Appendix \ref{sec:LW}, we derive the field strength of the photon field caused by a charged particle in motion.
In Appendix \ref{expansion}, we explain the $1/c$ expansion of the phase shift in the non-relativistic regime, where $c$ is the speed of light.
In Appendix \ref{detailc}, we present some details of the calculation in Sec. \ref{secIII}.
Throughout the present paper, we use the convention $(-, +, +, +)$.
We note that the charge $e=\sqrt{4\pi\alpha}$ is a dimensionless parameter with the fine-structure constant $\alpha=1/137$, and we use the natural units $c=\hbar=\epsilon_{0}=1$ while we recover $c$ and $\hbar$ as necessary.

\section{Entanglement due to Coulomb interaction of two charged particles \label{secII}}
\subsection{Time evolution of two charged particles with Coulomb interaction}
\begin{figure}[H]
  \centering
  \includegraphics[width=0.6\linewidth]{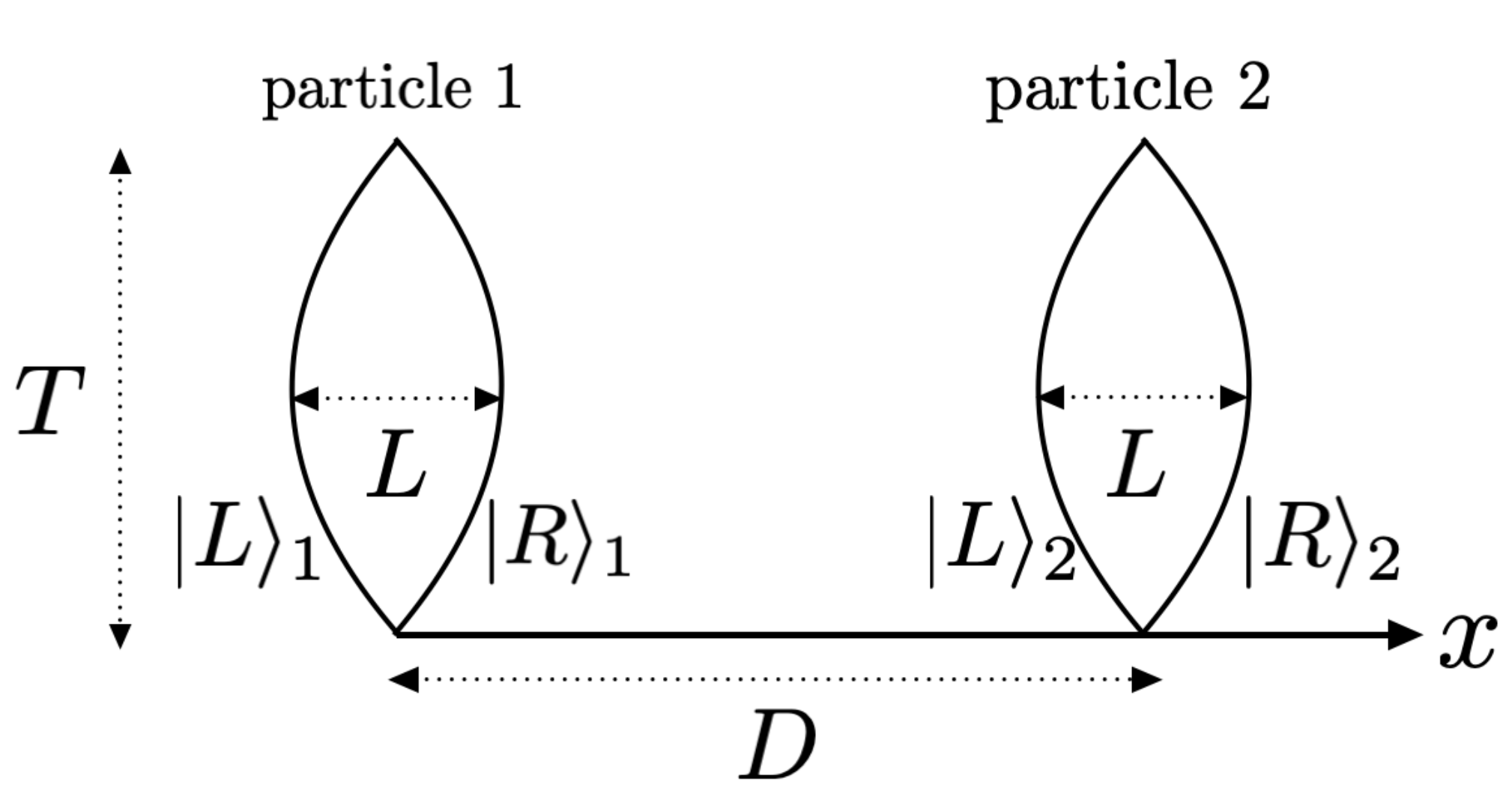}
  \caption{Configuration of trajectories of two charged particles. 
The length scale of each superposition is $L$, the coordinate time during which each particle is superposed is $T$, and the particles are initially separated by the distance $D$. }
  \label{fig:NTDL}
\end{figure}

In this section, we present the entanglement generation for two charged particles 1 and 2 each in a superposition of two trajectories (see Fig. \ref{fig:NTDL}). 
These particles are coupled with each other by the Coulomb potential. 
The total Hamiltonian is
\begin{equation}
\hat{H}=\hat{H}_{1}+\hat{H}_{2}+\hat{V}_{\text{12}}, \quad
\hat{V}_{12}=\frac{e^2}{4\pi}\frac{1}{|\hat{\mathbf{x}}_{1}-\hat{\mathbf{x}}_{2}|},
\label{coul}
\end{equation}
where 
$\hat{H}_{1}$ and 
$\hat{H}_{2}$ are the Hamiltonians of the charged particles 1 and 2, 
$\hat{V}_{12}$ is the interaction Hamiltonian between them with the coupling constant 
${e}$, and 
$\hat{\mathbf{x}}_{1}$ and 
$\hat{\mathbf{x}}_{2}$ denote each position operator of the two charged particles.
In the following computation, we do not need the explicit forms of
$\hat{H}_1$ and $\hat{H}_2$. 
As we will mention after Eq. \eqref{approxX}, they are implicitly given by specifying the trajectories of each particle.
\noindent
Each of the two charged particles at $t=0$ is in the spatially superposed state
\begin{align}
|\Psi(0)\rangle&=\frac{1}{{2}}\sum_{\text{P}, \text{Q}=\text{R}, \text{L}}|\text{P}\rangle_{1}|\text{Q}\rangle_{2},
\end{align}
where $|\text{R}\rangle_{1}$ ($|\text{R}\rangle_{2}$) and $|\text{L}\rangle_{1}$ ($|\text{L}\rangle_{2}$) are the states with the wave packets localized around positions $\bm{x}=\bm{X}_{1{\text{R}}}(t=0)$ ($\bm{x}=\bm{X}_{2{\text{R}}}(t=0)$) and $\bm{x}=\bm{X}_{1{\text{L}}}(t=0)$ ($\bm{x}=\bm{X}_{2{\text{L}}}(t=0)$), respectively. 
We assume the following approximation,
\begin{align}
\hat{\mathbf{x}}^{\text{I}}_{1}(t)|\text{P}\rangle_{1} \approx \mathbf{X}_{{1{\text{P}}}}(t)|\text{P}\rangle_{1},
\quad
\hat{\mathbf{x}}^{\text{I}}_{2}(t)|\text{Q}\rangle_{2} \approx \mathbf{X}_{{2{\text{Q}}}}(t)|\text{Q}\rangle_{2},
\label{approxX}
\end{align}
where $\hat{\mathbf{x}}^{\text{I}}_{1}(t)=e^{it(\hat{H}_1+\hat{H}_2)}\hat{\mathbf{x}}_{1}e^{-it(\hat{H}_1+\hat{H}_2)}$ and $\hat{\mathbf{x}}^{\text{I}}_{2}(t)=e^{it(\hat{H}_1+\hat{H}_2)}\hat{\mathbf{x}}_{2}e^{-it(\hat{H}_1+\hat{H}_2)}$ are the position operators in the interaction picture.
These assumptions are valid \cite{Breuer2001} when the de Brogile wavelength $\lambda_{\text{dB}}$ of the charged particle is much smaller than the width $\Delta x$ of its wave packet 
($\lambda_{\text{dB}} \ll \Delta x$).
The trajectories of each particle 
$\bm{X}_{1\text{P}}(t)$ and 
$\bm{X}_{2\text{Q}}(t)$ are determined by the Hamiltonians 
$\hat{H}_1$ 
and 
$\hat{H}_2$. 
In our computation, we specify the trajectories by hand. 

The evolved state $|\Psi(T)\rangle$ is
\begin{align}
|\Psi(T)\rangle
&=e^{-i\hat{H}T} |\Psi(0)\rangle\nonumber \\
&=e^{-iT(\hat{H}_1+\hat{H}_2)} \text{T}\exp\left[-i\int_{0}^{T} dt \frac{e^2}{4\pi}\frac{1}{|\hat{\mathbf{x}}^{\text{I}}_{1}(t)-\hat{\mathbf{x}}^{\text{I}}_{2}(t)|} \right] |\Psi(0) \rangle \nonumber \\
&\approx \frac{1}{2}e^{-iT(\hat{H}_1+\hat{H}_2)}\sum_{\text{P,Q=R, L}}e^{-i\Phi_{\text{PQ}}}|\text{P}\rangle_{1} \otimes |\text{Q} \rangle_{2},
\end{align}
where $\text{T}$ is the time-ordered product, and the approximation \eqref{approxX} was used in the third line.
The phase shift
\begin{align}
\Phi_{\text{PQ}}=\int_{0}^{T} dt \frac{e^2}{4\pi}\frac{1}{|{\mathbf{X}}_{{1{\text{P}}}}(t)-{\mathbf{X}}_{{2{\text{Q}}}}(t)|}
\end{align}
%%%%
is induced by the Coulomb potential between particles 1 and 2.
The density matrix of those particles is
%%%%
\begin{equation}
\rho_{\text{c}}=|\Psi(T)\rangle \langle\Psi(T)|
=\frac{1}{4}\sum_{\text{P}, \text{Q}=\text{R}, \text{L}}
\sum_{\text{P}', \text{Q}'=\text{R}, \text{L}}
e^{-i\Phi_{\text{PQ}}+i\Phi_{\text{P}'\text{Q}'}}
\,|\text{P}_\text{f}\rangle_1\langle \text{P}'_\text{f}| \otimes |\text{Q}_\text{f}\rangle_2\langle \text{Q}'_\text{f}|,
\label{rhoc}
\end{equation}
%%%%
where 
$|\text{P}_\text{f}\rangle_{1}=e^{-i\hat{H}_1T}|\text{P}\rangle_{1}$ and $|\text{Q}_\text{f}\rangle_{2}=e^{-i\hat{H}_2T}|\text{Q}\rangle_{2}$ are the states of the charged particles 1 and 2 moving along trajectories P and Q, respectively.

\subsection{Entanglement behavior of two charged particles}

Here, we adopt the negativity
$\mathscr{N}$ \cite{Vidal2002} to determine whether the state of two charged particles is entangled or not.
We consider a density matrix 
$\rho$ of a bipartite system AB. 
The negativity is introduced as follows: 
%%%%
\begin{equation}
\mathscr{N}=\sum_{\lambda_\text{i}<0} |\lambda_i|,
\label{negativity}
\end{equation}
%%%%
where $\lambda_i$ are the negative eigenvalues of the partial transposition 
$\rho^{\text{T}_{\text{A}}}$ with the elements 
$\langle a| \langle b| \rho^{\text{T}_{\text{A}}}|a' \rangle |b'\rangle=\langle a'| \langle b| \rho |a \rangle |b'\rangle$ in a basis 
$\{|a \rangle |b \rangle \}_{a,b}$ of the system AB. 
If the negativity does not vanish, then the system is entangled, which follows by the positive partial transpose criterion \cite{Horodecki1996,Peres1996}. 
Additionally, the nonzero negativity is the necessary and sufficient condition for the entanglement of a two-qubit or a qubit-qutrit system 
\cite{Horodecki1996}.
Particularly, there is only one negative eigenvalue $\lambda_{\text{min}}$ 
of the partial transposed density matrix of a two-qubit system 
\cite{Sanpera1998, Verstraete2001}.
We rewrite the negativity as
%%%%
\begin{equation}
\mathscr{N}=\text{max}\big[-\lambda_{\text{min}},0\big].
\label{N}
\end{equation} 
%%%%

The minimum eigenvalue of the partial transpose of the density matrix \eqref{rhoc} is 
%%%%
\begin{equation}
\lambda_\text{min}=-\frac{1}{2}\Big|\sin\Big[\frac{\Phi_\text{c}}{2}\Big]
\Big|,
\label{eigenc}
\end{equation}
where
$\Phi_{\text{c}}$ is given as
\begin{align}
\Phi_{\text{c}}
=
-\frac{e^2}{4\pi}\int_{0}^{T}dt
\Big(
\frac{1}{|{\mathbf{X}}_{{1{\text{R}}}}(t)-{\mathbf{X}}_{{2{\text{R}}}}(t)|}
-\frac{1}{|{\mathbf{X}}_{{1{\text{R}}}}(t)-{\mathbf{X}}_{{2{\text{L}}}}(t)|}
-\frac{1}{|{\mathbf{X}}_{{1{\text{L}}}}(t)-{\mathbf{X}}_{{2{\text{R}}}}(t)|}
+\frac{1}{|{\mathbf{X}}_{{1{\text{L}}}}(t)-{\mathbf{X}}_{{2{\text{L}}}}(t)|}\Big).
\label{phic}
\end{align}
To evaluate 
$\Phi_{\text{c}}$ and the negativity \eqref{N}, we consider the trajectories
\begin{align}
\bm{X}_{1\text{P}}(t)=\Big[\epsilon_\text{P} X(t), 0,0 \Big]^\text{T}, 
\quad 
\bm{X}_{2\text{Q}}
=\Big[\epsilon_{\text{Q}}X(t)+D,0,0 \Big]^\text{T}, 
\quad 
X(t)=8L\Big(1-\frac{t}{T} \Big)^2 \Big(\frac{t}{T}\Big)^2
\label{traj}
\end{align}
where 
$\epsilon_{\text{R}}=-\epsilon_{\text{L}}=1$, 
$L$ is the length scale of each superposition, 
$T$ is the time scale during which each particle is superposed 
and 
$D$ is the initial distance between those particles (see Fig. \ref{fig:NTDL}).
The quantity 
$\Phi_\text{c}$ is given by
\begin{align}
\Phi_{\text{c}}
&=
-\frac{e^2}{4 \pi} 
\int^T_0 dt 
\Big[
\frac{2}{D}
-
\Big(\frac{1}{D-2X(t)}+\frac{1}{D+2X(t)}\Big)
\Big].
\label{nphicexp2}
\end{align}
Now, we recover the light velocity 
$c$ and the reduced Planck constant 
$\hbar$.
We focus on the two regimes $cT \gg D \sim L$ and $cT \gg D \gg L$, in which the charged particles move with non-relativistic velocities 
($cT \gg L$).
In the regime 
$cT \gg D\sim L$,
the above formula of $\Phi_\text{c}$ and the minimum eigenvalue \eqref{eigenc} are computed numerically.
In the regime $cT \gg D \gg L$, the quantity $\Phi_\text{c}$ \eqref{nphicexp2} and the minimum eigenvalue \eqref{eigenc} are approximated as
\begin{align}
\Phi_{\text{c}} \approx \frac{64 e^{2}}{315 \pi \hbar c} \frac{c T L^{2}}{D^{3}},
\quad
\lambda_{\text{min}} \approx -\frac{16 e^{2}}{315 \pi \hbar c} \frac{c T L^{2}}{D^{3}},
\end{align}
where 
$\mathcal{O}(L^3/D^3)$ was ignored, and the Taylor expansion $\sin{\Phi_{\text{c}}/2} \approx \Phi_{\text{c}}/2$ was used.

\begin{figure}[H]
\centering
\begin{minipage}[b]{0.45\linewidth}
  \includegraphics[width=1\linewidth]{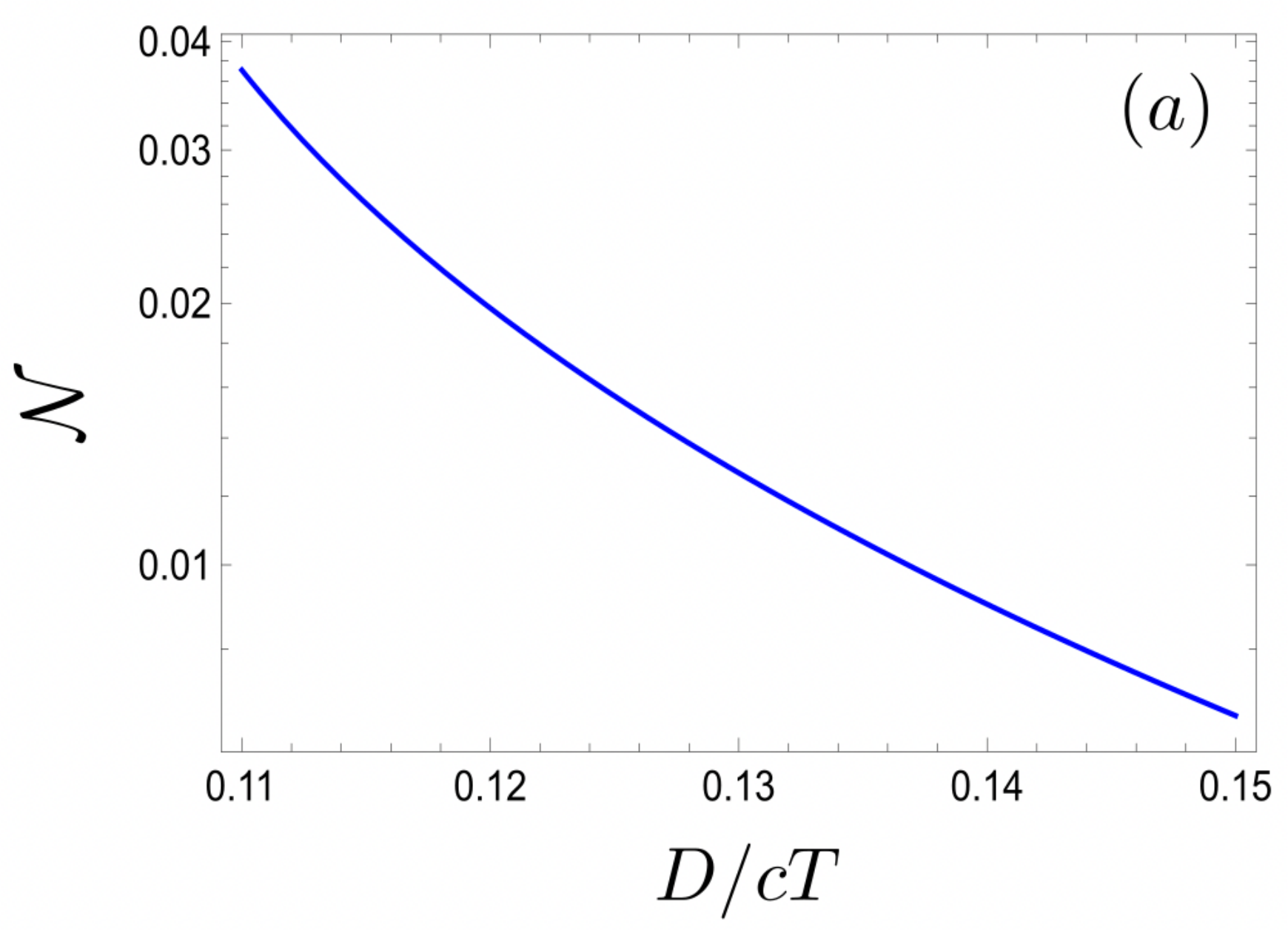}
  \end{minipage}
\begin{minipage}[b]{0.45\linewidth}
  \includegraphics[width=1\linewidth]{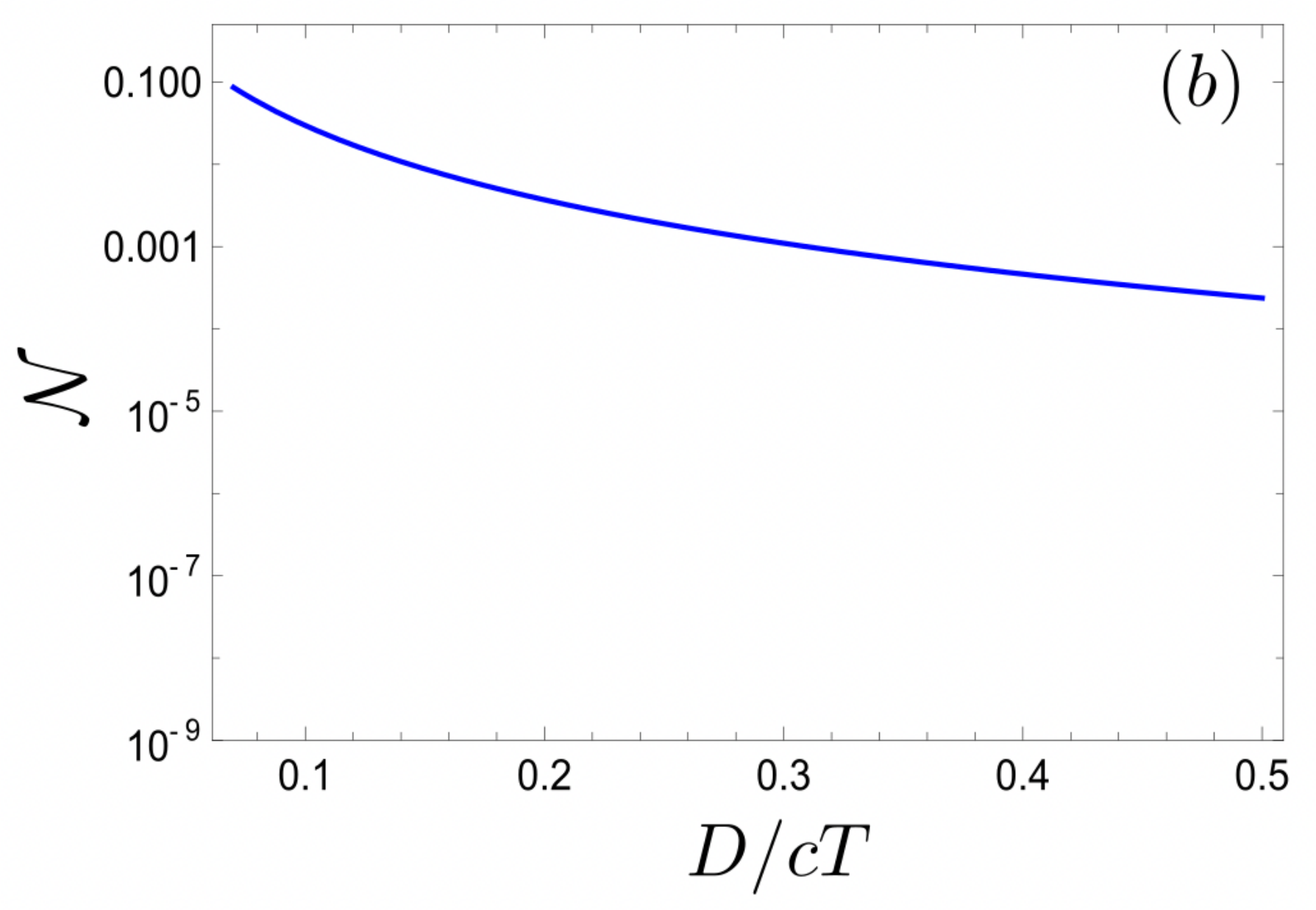}
   \end{minipage}
  \caption{Negativity $\mathscr{N}$ induced by the Coulomb potential between the charged particles. 
  We adopted $L/cT=0.1$.}
  \label{fig:NlTL-D}
\end{figure}

Fig. \ref{fig:NlTL-D} (a) and (b) show the negativity in the regime
$cT \gg D\sim L$ and 
$cT \gg D\gg L$.
These results show that the negativity decreases as the ratio $D/cT$ increases.
Because the negativity is always positive, the two charged particles 1 and 2 interacting with the Coulomb potential are entangled in the regimes
$cT \gg D \sim L$ and 
$cT \gg D \gg L$.
%%%%
%%%%
In the quantum information theory, it is well known that the entanglement between two systems cannot be created by local operations and classical communications (LOCC) \cite{Horodecki2009}. 
This means that it is impossible to create entanglement by classical interaction.
It immediately follows that if the Coulomb interaction entangles two charged particles, then the interaction is quantum.
In fact, the Coulomb interaction has the origin of the momentum transfer by the exchange of virtual (off-shell) photons~\cite{Peskin}.
However, the effect of a real (on-shell) photon field on entanglement generation is not obvious. 
In the next two sections, based on QED, we evaluate the entanglement generation between two charged particles.
We first introduce the model of a single charged particle interacting with a photon field, and then extend it to the model of two charged particles.

\section{Dynamics of charged particles coupled with a photon field\label{secIII}}

We consider the dynamics of charged particles coupled with a photon field, where the charged particles are each in a superposition of trajectories.
After a brief review of the model of a single charged particle, we extend it to the model of two charged particles. 
For the covariant quantization of the electromagnetic field, we use the BRST formalism \cite{WeinbergQFT2} in the Feynman gauge.
The details of the BRST formalism are presented in Appendix \ref{BRST}.

\subsection{Model of a single charged particle
}

We consider a single charged particle and a photon field coupled to it.
The total Hamiltonian in the Schr\"odinger picture is 
%%%%
\begin{equation}
\hat{H}=\hat{H}_\text{p}+\hat{H}_{\text{ph}}+\hat{V}, \quad
\hat{V}=\int d^3x \hat{J}_{\mu}(\mathbf{x})\hat{A}^{\mu}(\mathbf{x}),
\end{equation}
%%%%
where 
$\hat{H}_\text{p}$ is the Hamiltonian of the charged particle,
$\hat{H}_{\text{ph}}$ is the free Hamiltonian of the photon field, 
and 
$\hat{V}$ is their interaction Hamiltonian.
$\hat{J}_\mu $ is the current operator of the charged particle, and 
$\hat{A}^\mu $ is the photon field operator (the U(1) gauge field).

We assume that the charged particle is superposed in two different trajectories R and L. 
The charged particle is initially in the superposed state of
$|\text{R}\rangle$ and 
$|\text{L}\rangle$, where 
$|\text{R} \rangle (|\text{L} \rangle)$ is the state that the particle will go through a trajectory R (L).
%%%%
\begin{figure}[H]
  \centering
  \includegraphics[width=0.3\linewidth]{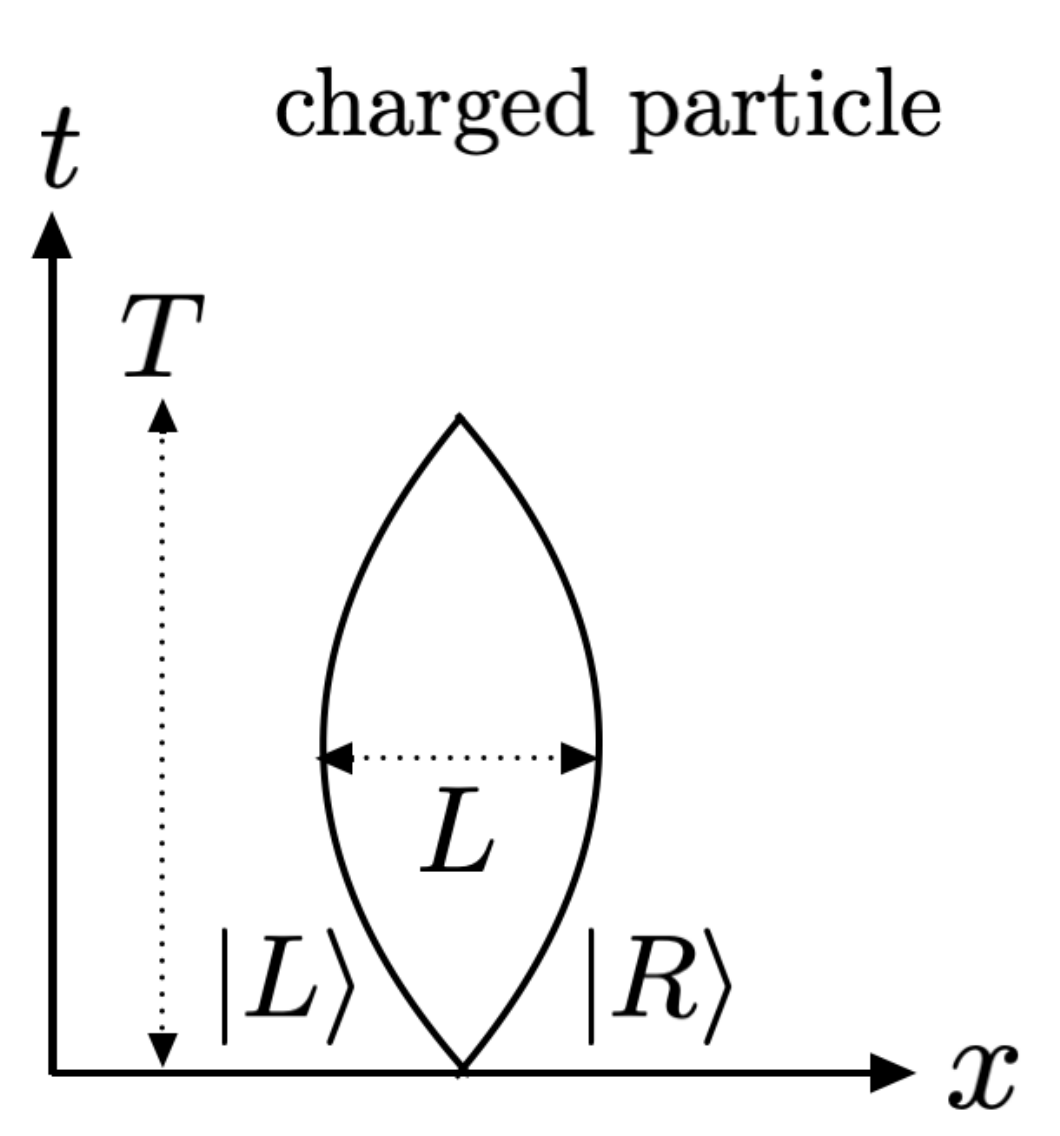}
  \caption{Configuration of a single charged particle trajectory.}
  \label{fig:stra}
\end{figure}
%%%%
\noindent
The photon field is assumed to be initially in a coherent state. 
Then the total initial state at the time
$t=0$ is
%%%%
\begin{align}
|\Psi(0)\rangle=\frac{1}{\sqrt{2}}\big(|\text{R}\rangle + |\text{L} \rangle\big)\otimes |\alpha \rangle_{\text{ph}},
\label{inistate1}
\end{align}
%%%%
where 
$|\alpha \rangle_\text{ph}=\hat{D}(\alpha)|0\rangle_\text{ph}$ is the coherent state of the photon field. 
Here, 
$|0\rangle_\text{ph}$ is the vacuum state satisfying
$\hat{a}_\mu (\bm{k})|0 \rangle_\text{ph}=0$, and 
$\hat{D}(\alpha)$ is the unitary operator referred to as a displacement operator defined as
%%%%
\begin{equation}
\hat{D}(\alpha)=\exp \left[
\int d^3k (\alpha^\mu (\bm{k})\hat{a}^\dagger_\mu (\bm{k})-h.c.)
\right],
\label{D}
\end{equation}
%%%%
where the complex function 
$\alpha^\mu (\bm{k})$ characterizes the amplitude and phase of the initial photon field. 
The form of the complex function 
$\alpha^\mu (\bm{k})$ is restricted by the auxiliary condition in the BRST formalism.
Because we will find that the entanglement between two charged particles does not depend on 
$\alpha^\mu (\bm{k})$ in Sec. III A, the details on 
$\alpha^\mu(\bm{k})$ are omitted here. 
The details are presented in Appendix A.
The coherent state $|\alpha\rangle_{\text{ph}}$ is interpreted as a state in which there is a mode of the electromagnetic field following Gauss's law due to the presence of charged particles.

We assume that the current operator 
$\hat{J}^\mu_\text{I} (x)=e^{i\hat{H}_0 t} \hat{J}^\mu (0, \bm{x}) e^{-i\hat{H}_0 t} $ in the interaction picture defined with 
$\hat{H}_{0}=\hat{H}_\text{p}+\hat{H}_{\text{ph}}$
is approximated by a classical current as
%%%%
\begin{equation}
\hat{J}_\text{I}^\mu (x)|\text{P} \rangle \approx 
J^\mu_\text{P} (x)|\text{P} \rangle, \quad
J^\mu_\text{P} (x)=e \int d\tau \frac{dX^\mu_\text{P}}{d\tau} \delta^{(4)}(x-X_\text{P} (\tau)),
\label{approx1}
\end{equation}
%%%%
where 
$\text{P}=\text{R}, \text{L}$,
$e$ is an electric charge, and 
$X^\mu_\text{P} (\tau)$ represents each trajectory of the charged particle. 
This approximation is valid for the following two assumptions \cite{Breuer2001}: the first assumption is that the de Brogile wavelength is smaller than the wave packet width of the particle.
The second assumption is that the Compton wavelength $\lambda_{\text{C}}$ of the charged particle is much shorter than the wavelength of the photon field
$\lambda_\text{ph}$ (for example, the wavelength of the photon field emitted from the charged particle) ($\lambda_{\text{C}} \ll \lambda_\text{ph}$).
Under this condition,\ the process of a pair creation and annihilation is neglected.

The evolution of the initial state $|\Psi(0)\rangle$ is 
%%%%
\begin{align}
|\Psi(T)\rangle
&=e^{-i\hat{H}T} |\Psi(0)\rangle\nonumber \\
&=e^{-i\hat{H}_0T} \text{T}\exp\Big[-i\int_{0}^{T} dt \hat{V}_\text{I}(t) \Big] |\Psi(0) \rangle \nonumber \\
&=e^{-i\hat{H}_0T} \text{T}\exp\Big[-i\int_{0}^{T} dt \int d^3x \hat{J}^{\mu}_\text{I} (x) \hat{A}^\text{I}_\mu (x) \Big]
\frac{1}{\sqrt{2}}\sum_\text{P=R,L}|\text{P} \rangle \otimes |\alpha \rangle_\text{ph} \nonumber\\
&\approx \frac{e^{-i\hat{H}_0T}}{\sqrt{2}}\sum_\text{P=R,L}|\text{P}\rangle \otimes \hat{U}_{\text{P}} |\alpha \rangle_{\text{ph}},
\end{align}
%%%%
where the approximation in \eqref{approx1} was used in the fourth line, 
$\hat{V}_{\text{I}}(t) = e^{i\hat{H}_{0}t}\hat{V}e^{-i\hat{H}_{0}t}$ and 
$\hat{A}^{\text{I}}_\mu (x) = e^{i\hat{H}_{0}t}\hat{A}(0,\bm{x})e^{-i\hat{H}_{0}t}$.
$``\, \text{T}\,"$ in the second and third lines denotes the time ordered product.
The operator $\hat{U}_\text{P}$ is given by
%%%%
\begin{align}
\hat{U}_\text{P}
&=\text{T}\exp\Big[-i\int_{0}^{T} dt \int d^3x J^\mu_\text{P} (x) \hat{A}^\text{I}_\mu (x) \Big]\nonumber \\
&=\exp\Big[-i\int d^4x J^\mu_\text{P} (x) \hat{A}^\text{I}_\mu (x) -\frac{i}{2} \int d^4x \int d^4y J^\mu_\text{P} (x) J^\nu_\text{P} (y)G^\text{r}_{\mu \nu}(x,y) \Big],
\label{unitaryP}
\end{align}
%%%%
where in the second line we used the Magnus expansion \cite{Magnus1954} 
\begin{equation}
\text{T}\exp\Big[-i\int_{0}^{T} dt \hat{V}_\text{I}(t)\Big]
=
\exp\Big[\sum_{k=1}^{\infty}\Omega_{k}(T,0)\Big]
\end{equation}
with
\begin{align}
\Omega_{1}(T,0)=-i\int_{0}^{T} dt \hat{V}_\text{I}(t),
\quad
\Omega_{2}(T,0)=\frac{(-i)^2}{2}\int_{0}^{T} dt_{1}\int_{0}^{t_{1}} dt_{2}[\hat{V}_\text{I}(t_{1}), \hat{V}_\text{I}(t_{2})],
\end{align}
and 
$\Omega_{k\geq3} (T,0) $ given by higher commutators, for example, 
$[[\hat{V}_\text{I}(t_1), \hat{V}_\text{I}(t_2)],\hat{V}_\text{I}(t_3)]$.
We note that the commutator $[\hat{V}_\text{I}(t_{1}), \hat{V}_\text{I}(t_{2})]$ is proportional to the identity operator and commutes with $\hat{V}_\text{I}(t)$ for any given time $t$. 
Hence, the terms $\Omega_{k\geq3}(T,0)$ involving higher commutators vanish in Eq. \eqref{unitaryP}.
$G^\text{r}_{\mu \nu} (x,y)$ in Eq. \eqref{unitaryP} is the retarded Green's function given by
%%%%
\begin{equation}
G^\text{r}_{\mu \nu} (x,y)
=-i[\hat{A}^\text{I}_\mu (x), \hat{A}^\text{I}_\nu (y)]\theta (x^0-y^0). 
\label{ret}
\end{equation}
%%%%
We obtain the reduced density matrix of the charged particle as 
%%%%
\begin{equation}
\rho_\text{p}
=\text{Tr}_\text{ph}[|\Psi(T)\rangle\langle\Psi(T)|]
=\frac{1}{2} \sum_{\text{P}, \text{P}'=\text{R}, \text{L}} 
{}_\text{ph} \langle \alpha |\hat{U}^{\dagger}_{\text{P}'} \hat{U}_{\text{P}}|\alpha\rangle_\text{ph} \,
|\text{P}_\text{f}\rangle \langle \text{P}'_\text{f}|
=\frac{1}{2} \sum_{\text{P}, \text{P}'=\text{R}, \text{L}} e^{-\Gamma_{\text{P}'\text{P}}+i \Phi_{\text{P}'\text{P}}}
|\text{P}_\text{f}\rangle
\langle \text{P}'_\text{f}|,
\label{rhop}
\end{equation}
%%%%
where
$|{\text{P}}_\text{f}\rangle=e^{-i\hat{H}_\text{p}T}|\text{P}\rangle$ is the state of the charged particle, which moved along the trajectory P($=\text{R},\text{L}$). 
$\Gamma_{\text{P}'\text{P}}$ and 
$\Phi_{\text{P}'\text{P}}$ are
%%%%
\begin{align}
\Gamma_{\text{P}^{\prime} \text{P}}
&=\frac{1}{4} \int d^{4} x \int d^{4} y\left(J_{\text{P}^{\prime}}^{\mu}(x)-J_{\text{P}}^{\mu}(x)\right)\left(J_{\text{P}^{\prime}}^{\nu}(y)-J_{\text{P}}^{\nu}(y)\right)\left\langle\left\{\hat{A}_{\mu}^{\text{I}}(x), \hat{A}_{\nu}^{\text{I}}(y)\right\}\right\rangle,
\label{GammaPP'}
\\
\Phi_{\text{P}^{\prime} \text{P}}
&=\int d^4 x 
\left(
J_{\text{P}^{\prime}}^{\mu}(x)-J_{\text{P}}^{\mu}(x)
\right) A_\mu (x)
-\frac{1}{2} \int d^{4} x \int d^{4}y 
\left(
J_{\text{P}^{\prime}}^{\mu}(x)-J_{\text{P}}^{\mu}(x)
\right)
\left(
J_{\text{P}^{\prime}}^{\nu}(y)+J_{\text{P}}^{\nu}(y)
\right) 
G_{\mu \nu}^{\text{r}}(x, y),
\label{PhiPP'}
\end{align}
%%%%
where 
$\langle\{\hat{A}_{\mu}^{\text{I}}(x), \hat{A}_{\nu}^{\text{I}}(y)\}\rangle$ is the two-point function of the vacuum given by
%%%%
\begin{align}
\left\langle\left\{\hat{A}_{\mu}^{\text{I}}(x), \hat{A}_{\nu}^{\text{I}}(y)\right\}\right\rangle=\frac{\eta_{\mu \nu}}{4 \pi^{2}}\left(\frac{1}{-\left(x^{0}-y^{0}-i \epsilon\right)^{2}+|\boldsymbol{x}-\boldsymbol{y}|^{2}}+\frac{1}{-\left(x^{0}-y^{0}+i \epsilon\right)^{2}+|\boldsymbol{x}-\boldsymbol{y}|^{2}}\right)
\label{cor}
\end{align}
%%%%
with the UV cutoff parameter
$\epsilon$, and the field $A_\mu (x)$ is 
%%%%
\begin{equation}
 A_\mu (x)= \int \frac{d^3k}{(2\pi)^{3/2} \sqrt{2k^0}}( \alpha_\mu (\bm{k})e^{ik_\nu x^\nu}+ c.c.).
 \label{A}
\end{equation}
%%%% 
The computation of the inner product 
${}_\text{ph} \langle \alpha |\hat{U}^{\dagger}_{\text{P}'} \hat{U}_{\text{P}}|\alpha\rangle_\text{ph}$ in \eqref{rhop}
and the derivation of Eqs. \eqref{GammaPP'} and \eqref{PhiPP'} are presented in Appendix \ref{sec:Tr}.
It is obvious that $\Gamma_{\text{RR}}=\Gamma_{\text{LL}}=\Phi_{\text{RR}}=\Phi_{\text{LL}}=0$. 
However, $\Gamma_{\text{RL}}$ and $\Phi_{\text{RL}}$ are given as
%%%%
\begin{align}
\Gamma_{\text{RL}}&=\frac{1}{4}\int d^4x \int d^4y (J^\mu_\text{R}(x)-J^\mu_\text{L}(x))
(J^\nu_\text{R}(y)-J^\nu_\text{L}(y))\langle \bigl\{\hat{A}^\text{I}_\mu (x), \hat{A}^\text{I}_\nu (y) \bigr\}\rangle \nonumber\\
&=\frac{e^2}{4}\oint_\text{C} dx^\mu \oint_\text{C} dy^\mu \langle \bigl\{\hat{A}^\text{I}_\mu (x), \hat{A}^\text{I}_\nu (y) \bigr\}\rangle
\label{gammaRL}
\end{align}
and
\begin{align}
\Phi_{\text{RL}}&=
\int d^4 x 
\left(
J_{\text{R}}^{\mu}(x)-J_{\text{L}}^{\mu}(x)
\right) A_\mu (x)-\frac{1}{2}\int d^4x \int d^4y (J^\mu_\text{R}(x)-J^\mu_\text{L}(x))(J^\nu_\text{R}(y)+J^\nu_\text{L}(y))G^\text{r}_{\mu \nu} (x,y)\nonumber\\
&= e \oint_\text{C} dx_\mu A^\mu (x)
-\frac{e}{2}\oint_\text{C} dx_\mu (A^\mu_\text{R}(x)+A^\mu_\text{L}(x)),
\label{phiRL}
\end{align}
%%%%
where 
$\oint_{\text{C}}dx_{\mu}=\int_{\text{R}}dx_{\mu}-\int_{\text{L}}dx_{\mu}$
is the integral along the closed trajectory composed of trajectories R and L.
Here, 
$A_{\text{P}}^{\mu}(x)$ is the retarded potential given by
%%%%
\begin{align}
A_{\text{P}}^{\mu}(x)=\int d^{4} y G_{\mu \nu}^{r}(x, y) J_{\text{P}}^{\nu}(y).
\end{align}
%%%%
According to \eqref{gammaRL},\ $\Gamma_{\text{RL}}$ is always positive, and the interference terms of 
$\rho_\text{P}$ 
(off-diagonal components) 
decay for a large 
$\Gamma_\text{RL}$. 
The quantity 
$\Gamma_\text{RL}$ is referred to as the decoherence functional.
The quantity $\Phi_{\text{RL}}=-\Phi_{\text{LR}}$ gives the phase shift in the interference pattern of the charged particle.

In Appendix E, assuming the following trajectories of the charged particle 
\begin{equation}
X^\mu_\text{P} (t) =[t,\epsilon_{\text{P}}X(t), 0,0 \Big]^\text{T},
\quad 
\epsilon_{\text{R}}=-\epsilon_{\text{L}}=1, 
\quad 
X(t)=8L\Big(1-\frac{t}{T} \Big)^2 \Big(\frac{t}{T}\Big)^2,
\end{equation}
where 
$L$ and 
$T$ are the length and time scales of the trajectories (also see Fig. \ref{fig:stra}),
we obtain the decoherence functional as 
\begin{equation}
\Gamma_\text{RL} \approx \frac{32}{3\pi^2} \frac{e^2L^2}{T^2},
\end{equation}
when the charged particle has a non-relativistic velocity 
$L/T \ll 1$.
The physical meaning of $\Gamma_{\text{RL}}$ is interpreted in the following two ways.
First, we consider that decoherence occurs through photon emission. 
The number of emitted photons is estimated as
\begin{align}
\frac{WT}{\nu} = WT^2 \sim e^2\Big(\frac{L}{T^2}\Big)^2T^2=e^2\frac{L^2}{T^2},
\label{emidec}
\end{align}
where $\nu =1/T$ is the energy of a single photon in the unit $\hbar=1$, and 
$W\sim e^2(L/T^2)^2$ is the Larmor formula of the power of radiation emitted from a non-relativistic charged particle.
This formula shows the number of emitted photons during the time $T$.
When this number exceeds one, i.e., $WT/\nu \geq 1$, the decoherence becomes significant.
The decoherence due to bremsstrahlung was also discussed in \cite{Breuer2001}. 
Second, we can deduce that the decoherence is due to the vacuum fluctuations of the photon field \cite{Stern, Ford1997}. 
The fluctuating photon field leads to dephasing effects, 
\begin{align}
\langle e^{i\phi}\rangle=e^{-\langle\phi^2\rangle/2} \sim e^{-(e\Delta E LT)^2/2},
\end{align}
where 
$\phi$ is the phase shift due to the fluctuating photon field, and
$\langle\phi^2\rangle \sim (e\Delta E LT)^2$ is its variance. 
$\Delta E$ is the vacuum fluctuation of the electric component of the photon field, which is estimated as $\Delta E \sim 1/T^2$ in~\cite{Ozawa}. 
The variance of the phase shift is
\begin{align}
(e\Delta E LT)^2\sim \Big(e\frac{1}{T^2}LT\Big)^2=e^2\frac{L^2}{T^2}.
\label{fludec}
\end{align}
This result is equivalent to Eq. \eqref{emidec}, and the decoherence becomes significant for $(e\Delta E LT)^2 \geq 1$.
%%%%%%%%

\subsection{Model of two charged particles
}

In this subsection, we extend the previous model to the model of two charged particles (for example, see Fig. \ref{fig:NTDL}).
The total Hamiltonian in the Schr\"odinger picture is composed of the local Hamiltonians of each charged particle 
$\hat{H}_1$ and $\hat{H}_2$, the free Hamiltonian of the photon field 
$\hat{H}_\text{ph}$ and the interaction term $\hat{V}$ as
%%%%
\begin{equation}
\hat{H}=\hat{H}_1+\hat{H}_2+\hat{H}_\text{ph}+\hat{V}, 
\quad 
\hat{V}=\int d^3x \Big(\hat{J}^\mu_1(\bm{x})+\hat{J}^\mu_2(\bm{x})\Big)\hat{A}^{\mu}(\bm{x}),
\end{equation}
%%%%
where 
$\hat{J}^\mu_1 $ 
and 
$\hat{J}^\mu_2 $ are the current operators of each particle, which are coupled with the photon field operator $\hat{A}^\mu $.
We consider the following initial condition at 
$t=0$,
%%%%
\begin{align}
|\Psi(0)\rangle&=\frac{1}{{2}}\sum_{\text{P, Q=R, L}}|\text{P}\rangle_{1}|\text{Q}\rangle_{2} |\alpha\rangle_{\text{ph}},
\label{inistate2}
\end{align}
%%%%
where each particle is in superposition 
$|\text{R} \rangle_{1} + |\text{L} \rangle_{1}$ and 
$|\text{R} \rangle_{2} + |\text{L} \rangle_{2}$, and the photon field is in a coherent state $|\alpha\rangle_{\text{ph}}$.
We assume that the current operators $\hat{J}^\mu_{i\text{I}}(x)=e^{i\hat{H}_0 t} \hat{J}^\mu_{i}(0,\bm{x})e^{-i\hat{H}_0 t}$ in the interaction picture with respect to 
$\hat{H}_0 =\hat{H}_1 + \hat{H}_2 +\hat{H}_\text{ph}$ are approximated by the following classical currents as
%%%%
\begin{align}
&\hat{J}_{1\text{I}}^ \mu(x)\left|{\text{P}}\right\rangle_{1}
\approx J_{1 \text{P}}^{\mu}(x)\left|{\text{P}}\right\rangle_{1},
\quad 
\hat{J}_{2\text{I}}^\mu(x)\left|{\text{Q}}\right\rangle_{2} \approx J_{2 \text{Q}}^{\mu}(x)\left|{\text{Q}}\right\rangle_{2},
\label{approx3}
\\
&J_{1 \text{P}}^{\mu}(x)
=e \int d \tau \frac{d X_{1 \text{P}}^{\mu}}{d \tau} \delta^{(4)}\left(x-X_{1 \text{P}}(\tau)\right),
\quad 
J_{2 \text{Q}}^{\mu}(x)
=e \int d \tau \frac{d X_{2 \text{Q}}^{\mu}}{d \tau} \delta^{(4)}\left(x-X_{2 \text{Q}}(\tau)\right),
\label{approx4}
\end{align}
%%%%
where 
$X^\mu_{1\text{P}}(\tau)$ and 
$X^\mu_{2\text{Q}} (\tau)$ with $\text{P}, \text{Q}=\text{R}, \text{L}$ represent the trajectories of each particle.
The initial state evolves as follows: 
%%%%
\begin{align}
|\Psi(T)\rangle 
&= 
\exp\big[-i \hat{H} T\big]|\Psi(0)\rangle
\nonumber\\
&=
e^{-i\hat{H}_0T} 
\text{T}\exp\big[-i\int_{0}^{T} dt \hat{V}_\text{I}(t) \big]
|\Psi(0)\rangle
\nonumber\\
&\approx e^{-i\hat{H}_0T}\frac{1}{2}
\sum_\text{P,Q=R,L}|\text{P}\rangle_{1} |\text{Q}\rangle_{2} \hat{U}_\text{PQ} |\alpha\rangle_\text{ph},
\end{align}
%%%%
where we used the approximations \eqref{approx3} in the third line.
The unitary operator 
$\hat{U}_{\text{PQ}}$ is given by
%%%%
\begin{align}
\hat{U}_{\text{PQ}} &=\text{T} \exp \left[-i \int_{0}^{T} d t \int d^{3} x\left(J_{1 \text{P}}^{\mu}+J_{2 \text{Q}}^{\mu}\right) \hat{A}_{\mu}^{\text{I}}(x)\right] 
\nonumber\\
&=\exp \left[-i \int d^{4} x J_{\text{PQ}}^{\mu}(x) \hat{A}_{\mu}^{\text{I}}(x)-\frac{i}{2} \int d^{4} x \int d^{4} y J_{\text{PQ}}^{\mu}(x) J_{\text{PQ}}^{\nu}(y) G_{\mu \nu}^{\text{r}}(x, y)\right],
\label{unitaryPQ}
\end{align}
%%%%
where the Magnus expansion was used, and
$J^{\mu}_{\text{PQ}}=J^{\mu}_{1\text{P}}+J^{\mu}_{2\text{Q}}$.
Tracing out the degrees of freedom of the photon field to focus on the quantum state of the charged particles, we obtain the reduced density matrix of particles 1 and 2,
%%%%
\begin{align}
\rho_{12} &=\text{Tr}_{\text{ph}}[|\Psi(T)\rangle\langle\Psi(T)|]
\nonumber\\
\quad
&=\frac{1}{4}\sum_{\text{P}, \text{Q}=\text{R}, \text{L}}
\sum_{\text{P}', \text{Q}'=\text{R}, \text{L}}
{}_{\text{ph}}\langle \alpha |\hat{U}^{\dagger}_{\text{P}'\text{Q}'}\hat{U}_{\text{PQ}}|\alpha \rangle_\text{ph}
\,|\text{P}_\text{f}\rangle_{1}\langle {\text{P}'}_\text{f}| \otimes  |\text{Q}_\text{f}\rangle_{2}\langle {\text{Q}'}_\text{f}|
\nonumber\\
\quad
&=\frac{1}{4}\sum_{\text{P}, \text{Q}=\text{R}, \text{L}}
\sum_{\text{P}', \text{Q}'=\text{R}, \text{L}}
e^{-\Gamma_{\text{P}'\text{Q}'\text{PQ}}+i\Phi_{\text{P}'\text{Q}'\text{PQ}}}
\,|\text{P}_\text{f}\rangle_{1}\langle {\text{P}'}_\text{f}| \otimes  |\text{Q}_\text{f}\rangle_{2}\langle {\text{Q}'}_\text{f}|,
\label{density12}
\end{align}
%%%%
where 
$|\text{P}_\text{f}\rangle_{1}=e^{-i\hat{H}_1T}|\text{P}\rangle_{1}$ and  $|\text{Q}_\text{f}\rangle_{2}=e^{-i\hat{H}_2T}|\text{Q}\rangle_{2}$ are the states of the charged particles 1 and 2, which moved along the trajectories P and Q, respectively.
The quantities 
$\Gamma_{\text{P}'\text{Q}'\text{PQ}}$ and $\Phi_{\text{P}'\text{Q}'\text{PQ}}$ are 
%%%%
\begin{align}
\Gamma_{\text{P}'\text{Q}'\text{PQ}}&=\frac{1}{4}\int d^4x \int d^4y (J^\mu_{\text{P}'\text{Q}'}(x)-J^\mu_{\text{PQ}}(x))
(J^\nu_{\text{P}'\text{Q}'}(y)-J^\nu_{\text{PQ}}(y))
\langle \bigl\{\hat{A}^\text{I}_\mu (x), \hat{A}^\text{I}_\nu (y)\bigr\}\rangle,
\label{GammaP'Q'PQ}
\\
\Phi_{\text{P}'\text{Q}'\text{PQ}}
&=
\int d^4x (J^\mu_{\text{P}'\text{Q}'}(x)-J^\mu_{\text{PQ}}(x))A_\mu (x)-\frac{1}{2}\int d^4x \int d^4y (J^\mu_{\text{P}'\text{Q}'}(x)-J^\mu_{\text{PQ}}(x))(J^\nu_{\text{P}'\text{Q}'}(y)+J^\nu_{\text{PQ}}(y))G^\text{r}_{\mu \nu} (x,y),
\label{PhiP'Q'PQ}
\end{align}
%%%%
where 
$\langle \bigl\{\hat{A}^\text{I}_\mu (x), \hat{A}^\text{I}_\nu (y)\bigr\}\rangle$
and 
$G^\text{r}_{\mu \nu} (x,y)$ are the two-point function \eqref{cor} and the retarded Green's function \eqref{ret}. 
$A_\mu (x)$ is the coherent photon field \eqref{A}. 
The above formulas \eqref{GammaP'Q'PQ} and \eqref{PhiP'Q'PQ} are given by replacing 
the currents 
$J^\mu_\text{P}$ and 
$J^\mu_{\text{P}'}$ in Eqs. \eqref{GammaPP'} and 
\eqref{PhiPP'} with 
$J^\mu_\text{PQ}$ and 
$J^\mu_{\text{P}'\text{Q}'}$, respectively.
In the next section, we derive the entanglement negativity of the two charged particles. 
We also demonstrate the entanglement behavior for a couple of typical configurations of the particle's trajectories.

%%%%%%%%%%%%%%%%%%%%%
\section{Entanglement behavior of two charged particles\label{secIV}}

\subsection{Formula of the negativity of two charged particles}

We evaluate the entanglement negativity with the formula \eqref{N}.
The eigenvalues of the partial transposition
$\rho^{\text{T}_1}_{12}$ with the components 
$\langle \text{P}'| \langle \text{Q}'|\rho^{\text{T}_1}_{12} |\text{P} \rangle | \text{Q} \rangle=\langle \text{P}| \langle \text{Q}'|\rho_{12} |\text{P}' \rangle | \text{Q} \rangle$ are
\begin{align}
\lambda_{\pm}&=\frac{1}{4}\Big[1-e^{-\Gamma_1-\Gamma_2} \cosh[\Gamma_\text{c}]\pm \Big\{\big(e^{-\Gamma_{1}}-e^{-\Gamma_{2}}\big)^2+4e^{-\Gamma_{1}-\Gamma_{2}}\sin^2(\Phi/2)+e^{-2\Gamma_1-2\Gamma_2} \sinh^2[\Gamma_\text{c}]\Big\}^{\frac{1}{2}}\Big],
\\
\quad
\lambda'_{\pm}&=\frac{1}{4}\Big[1+e^{-\Gamma_1-\Gamma_2} \cosh[\Gamma_\text{c}]\pm \Big\{\big(e^{-\Gamma_{1}}-e^{-\Gamma_{2}}\big)^2+4e^{-\Gamma_{1}-\Gamma_{2}}\sin^2(\Phi/2)+e^{-2\Gamma_1-2\Gamma_2} \sinh^2[\Gamma_\text{c}]\Big\}^{\frac{1}{2}}\Big].
\end{align}
We note that $\lambda_{-}$ is the minimum eigenvalue $\lambda_{\text{min}}$, and hence the negativity of the two charged particles is
%%%%
\begin{align}
\mathscr{N}&=\max[-\lambda_\text{min},0], 
\nonumber 
\\
\lambda_\text{min}&=\frac{1}{4}\Big[1-e^{-\Gamma_1-\Gamma_2} \cosh[\Gamma_\text{c}]- \Big\{\big(e^{-\Gamma_{1}}-e^{-\Gamma_{2}}\big)^2+4e^{-\Gamma_{1}-\Gamma_{2}}\sin^2(\Phi/2)+e^{-2\Gamma_1-2\Gamma_2} \sinh^2[\Gamma_\text{c}]\Big\}^{\frac{1}{2}}\Big].
 \label{mineigen}
\end{align}
%%%%
Because the density matrix 
$\rho_{12}$ of the charged particles is regarded as that of a two-qubit system, the negativity completely determines whether the particles are entangled or not. 
The quantities 
$\Gamma_i \, (i=1, 2)$, 
$\Gamma_{c}$ and $\Phi$ are given as 
%%%%
\begin{align}
\Gamma_{i}
&=\frac{1}{4}\int d^4x\int d^4y 
\Delta J^{\mu}_i(x) \Delta J^{\nu}_i(y)
\langle \{\hat{A}^\text{I}_{\mu}(x), \hat{A}^\text{I}_{\nu}(y)\}\rangle
=\frac{e^2}{4}\oint_{\text{C}_i} dx^\mu \oint_{\text{C}_i} dy^\nu 
\langle \{\hat{A}^\text{I}_{\mu}(x), \hat{A}^\text{I}_{\nu}(y)\}\rangle,
\label{gammai}
\\
\Gamma_\text{c}
&=\frac{1}{2}\int d^4x \int d^4y \Delta J_1^{\mu}(x) \Delta J_2^{\nu}(y)
\langle \{\hat{A}^\text{I}_{\mu}(x), \hat{A}^\text{I}_{\nu}(y)\}\rangle
=\frac{e^2}{2}\oint_{\text{C}_1 } dx^\mu \oint_{\text{C}_2} dy^\nu 
\, \langle \{\hat{A}^\text{I}_{\mu}(x), \hat{A}^\text{I}_{\nu}(y)\}\rangle,
\label{gammac}
\\
\Phi
&=\frac{1}{2}\int d^4x \int d^4y \Big\{\Delta J^{\mu}_{1}(x)\Delta J^{\nu}_{2}(y)+\Delta J^{\mu}_{2}(x)\Delta J^{\nu}_{1}(y)\Big\}G^{r}_{\mu \nu}(x, y)
=\frac{e}{2} 
\Big (
\oint_{\text{C}_1} dx_\mu \Delta A^\mu_2 (x)
+\oint_{\text{C}_2} dx_\mu \Delta A^\mu_1 (x)
\Big) ,
\label{phi}
\end{align}
%%%%
where 
$\Delta J^\mu_i= J^\mu_{i\text{R}}- J^\mu_{i\text{L}}$
and 
$J^\mu_{i\text{P}}$ is the current of the particle $i\,(=1,2)$ on the trajectory $\text{P}\,(=\text{R}, \text{L})$.
The line integral along the closed trajectory
$\oint_{\text{C}_{i}}dx_\mu$ is defined by $\oint_{\text{C}_{i}}dx_\mu=\int_{i\text{R}}dx_\mu-\int_{i\text{L}}dx_\mu$, where 
$i\text{P}$ denotes the trajectory $\text{P}$ of the particle $i$.
The quantity 
$\Delta A^\mu_i = A^\mu_{i\text{R}}- A^\mu_{i\text{L}}$ is the difference between the retarded potentials defined by
%%%%
\begin{align}
A_{i \text{P}}^{\mu}(x)=\int d^{4} y G_{\mu \nu}^{r}(x, y) J_{i \text{P}}^{\nu}(y).
\label{eq:A}
\end{align}
%%%%
The quantities 
$\Gamma_1$ and 
$\Gamma_2$ depend on the trajectories of each particle and have the similar form to
$\Gamma_\text{RL}$
\eqref{gammaRL}. 
These are the decoherence functionals appearing in the interference terms of each charged particle. 
In Appendix \ref{gamma12}, $\Gamma_{1}$ and $\Gamma_{2}$ are computed explicitly.
$\Gamma_\text{c}$ is characterized by the correlation function between the photon field coupled to particle 1 and the photon field coupled to particle 2. 
$\Phi$ is computed from
the phase shifts by the retarded potentials of the photon field 
$A^\mu_{i\text{P}}$, which is analogous to the Aharanov-Bohm effect. 
$\Gamma_{\text{c}}$ and $\Phi$ depend on the relative configuration of the trajectories of particles 1 and 2.
In Appendices \ref{cphil} and \ref{cphip}, we explicitly evaluate $\Gamma_{\text{c}}$ and $\Phi$ assuming two specific configurations of particles, which we refer to as the linear configuration (Figs. \ref{fig:linetdl} and \ref{fig:linedtl}) and the parallel configuration (Figs. \ref{fig:paratdl} and \ref{fig:paradtl}) in this paper.
The quantities 
$\Gamma_i$, 
$\Gamma_\text{c}$ and 
$\Phi$ are independent of the complex function $\alpha_\mu (\bm{k})$ of the initial coherent state of the photon field, and hence the negativity 
$\mathscr{N}$ also does not depend on $\alpha_\mu (\bm{k})$.
Hence, as mentioned around Eq. \eqref{D},
the entanglement between the particles does not depend on 
$\alpha_\mu (\bm{k})$.
Using the Stokes's theorem to rewrite the line integrals in Eqs. \eqref{gammai}, \eqref{gammac}, and \eqref{phi} by the surface integrals, we can express the quantities 
$\Gamma_i$,
$\Gamma_\text{c}$ and 
$\Phi$ in terms of the field strengths as
%%%%
\begin{align}
\Gamma_{i}
&
=
\frac{e^2}{16}\int_{\text{S}_i} d\sigma^{\mu\nu} \int_{\text{S}_i} d{\sigma'}^{\, \alpha \beta} 
\langle \{\hat{F}^\text{I}_{\mu \nu}(x), \hat{F}^\text{I}_{\alpha \beta}(x')\}\rangle,
\label{gammaiF}
\\
\Gamma_\text{c}
&
=
\frac{e^2}{16}\int_{\text{S}_1} d\sigma^{\mu \nu} \int_{ \text{S}_2} d{\sigma'}^{\alpha \beta} 
\, \langle \{\hat{F}^\text{I}_{\mu \nu}(x), \hat{F}^\text{I}_{\alpha \beta}(x')\}\rangle,
\label{gammacF}
\\
\Phi
&=
\frac{e}{4} 
\Big (
\int_{\text{S}_1} d\sigma_{\mu \nu} \Delta F^{\mu\nu}_2 (x)
+\int_{\text{S}_2} d\sigma_{\mu \nu} \Delta F^{\mu\nu}_1 (x)
\Big),
\label{phiF}
\end{align}
%%%%
where 
$S_i$ is the surface surrounded by the closed trajectory $C_i$,
$\hat{F}^\text{I}_{\mu\nu}=\partial_\mu \hat{A}^\text{I}_\nu -\partial_\nu \hat{A}^\text{I}_\mu$, and 
$\Delta F^{\mu\nu}_i=F^{\mu\nu}_{i\text{R}}-F^{\mu\nu}_{i\text{L}}$ with the retarded field strengths $F^{\mu\nu}_{i\text{P}}=\partial^\mu A^\nu_{i\text{P}}-\partial^\nu A^\mu_{i\text{P}}$.

In the following subsections, computing the quantities
$\Gamma_i$, $\Gamma_c$ and $\Phi$, we present 
the minimum eigenvalue \eqref{mineigen} and
entanglement negativity $\mathscr{N}$ of the charged particles.
Hereafter, we restore the reduced Planck constant 
$\hbar$ and the light velocity 
$c$ to determine
the non-relativistic limit of our analysis.

\subsection{Linear configuration}

We consider the linear configurations shown in Fig. \ref{fig:linetdl} and Fig. \ref{fig:linedtl}.
The parameters $T$, $L$, and $D$ represent the time of maintaining a superposition state of each particle, the length of separation between the superposed trajectories of each particle, and the initial distance between the charged particles 1 and 2, respectively.

\subsubsection{$cT \gg D \sim L$ or $cT\gg D \gg L$ regimes}
\begin{figure}[H]
  \centering
  \includegraphics[width=0.6\linewidth]{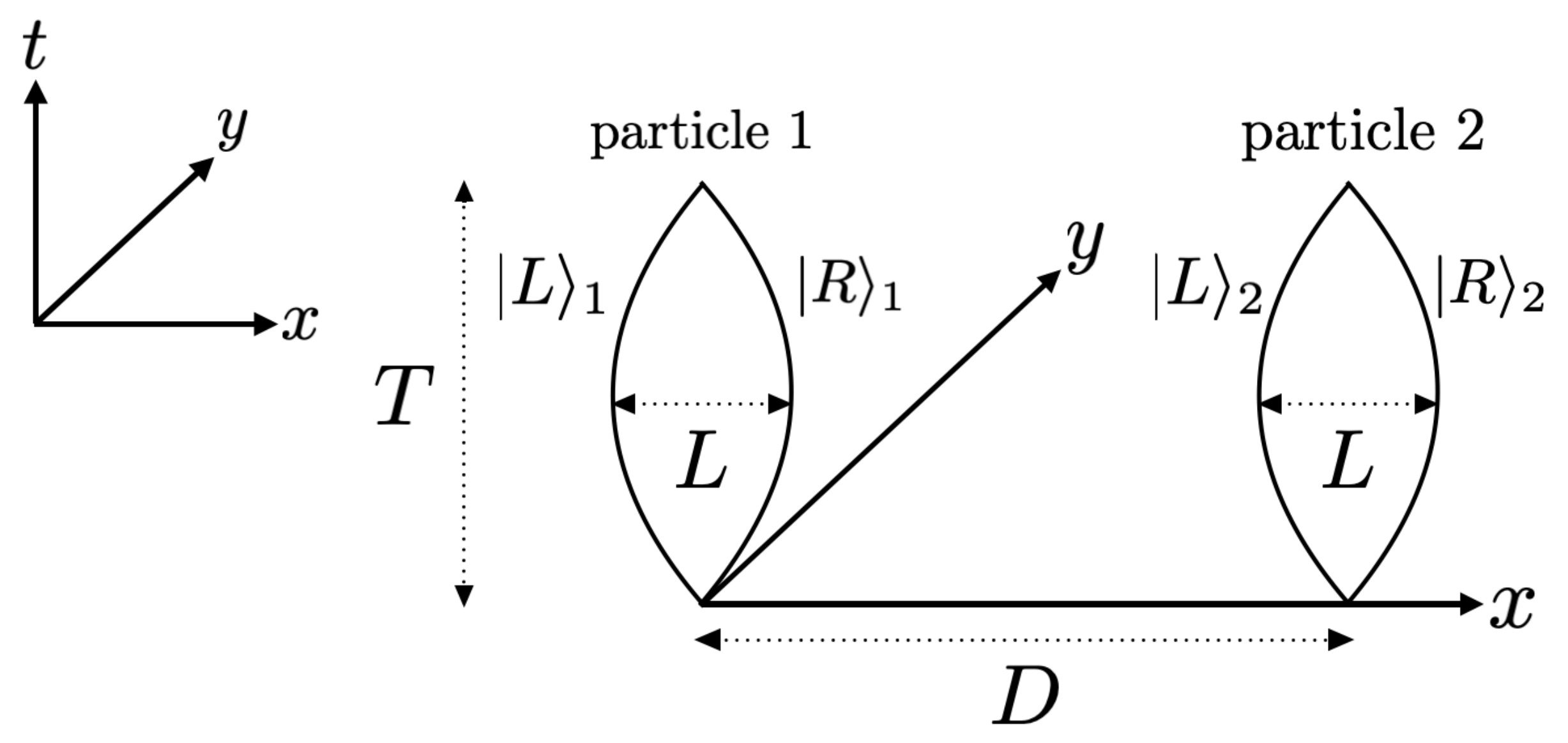}
  \caption{Linear configuration in the regimes $cT \gg D \sim L$ and 
  $cT \gg D \gg L$. 
  The left panel shows the entire view of the linear configuration.
  }
  \label{fig:linetdl}
\end{figure}

%%%%
To evaluate the minimum eigenvalue $\lambda_\text{min}$, which gives the negativity of the two charged particles, we compute the quantities 
$\Gamma_i$, $\Gamma_\text{c}$ 
and 
$\Phi$ by specifying the trajectories of the particles. 
We consider the following trajectories
\begin{equation}
X^\mu_{1\text{P}}=[t,\epsilon_{\text{P}}X(t), 0,0]^\text{T}, \quad 
X^\mu_{2\text{Q}}(t)=[t,\epsilon_{\text{Q}}X(t)+D,0,0 \Big]^\text{T},
\quad 
\epsilon_{\text{R}}=-\epsilon_{\text{L}}=1,
\quad 
X(t)=8L\Big(1-\frac{t}{T} \Big)^2 \Big(\frac{t}{T}\Big)^2,
\label{linearlTDL2}
\end{equation}
where 
$X^\mu_{1\text{P}}$ and 
$X^\mu_{2\text{Q}}$ with $\text{P},\text{Q}=\text{R},\text{L}$ describe the trajectories of particles 1 and 2, respectively. 
Fig. \ref{fig:linetdl} schematically shows the configuration of the particles. 
In the regimes
$cT\gg D \sim L$ and $cT\gg D \gg L$, the quantities 
$\Gamma_i$, 
$\Gamma_\text{c}$, and
$\Phi$ are evaluated.
As we show in Appendix \ref{gamma12}, assuming the above trajectories, we can compute 
$\Gamma_i$ for 
$cT\gg L$ as \begin{equation}
\Gamma_{1}=\Gamma_{2}\approx \frac{32 e^2}{3 \pi^2 \hbar c} \Big(\frac{L}{c T}\Big)^2.
\label{Gamma12}
\end{equation}
%%%%
%%%%
In the regime 
$cT \gg D \sim L$, the quantity  $\Gamma_\text{c}$ is analytically obtained as  
\begin{equation}
\Gamma_{\text{c}} \approx
\frac{64e^2}{3\pi^2 \hbar c} \Big(\frac{L}{cT}\Big)^2,
\label{Gammac}
\end{equation}
and the quantity $\Phi$ is numerically computed from the formula 
\begin{equation}
\Phi \approx
-\frac{e^2}{4 \pi \hbar } 
\int^T_0 dt 
\Big[
\frac{2}{D}
\Big(
 1-\frac{v^2}{c^2}
\Big)
-
\Big(
 1+\frac{v^2}{c^2}
\Big)
\Big(\frac{1}{D-2X(t)}+\frac{1}{D+2X(t)}\Big)
\Big],
\label{Phi}
\end{equation}
where 
$v=dX/dt$.
Substituting Eqs. \eqref{Gamma12}, \eqref{Gammac}, and \eqref{Phi} into Eq. \eqref{mineigen}, we evaluate the minimum eigenvalue 
$\lambda_\text{min}$ and the negativity $\mathscr{N}$. 
The behavior is shown by the red curve in Fig. \ref{fig:lTLD} (a).
The derivation of Eqs. \eqref{Gammac} and \eqref{Phi} is presented in Appendix \ref{cphilTDL}. 
In the regime 
$cT \gg D \gg L$, the quantities  
$\Gamma_\text{c}$ and 
$\Phi$ are estimated as 
%%%%
\begin{align} \Gamma_{\text{c}} \approx
\frac{64e^2}{3\pi^2 \hbar c} \Big(\frac{L}{cT}\Big)^2 \Big(1+\frac{4D^2}{(cT)^2} \ln \Big[\frac{D}{cT}\Big] \Big),
\quad \Phi \approx 
\frac{64e^2}{315\pi\hbar c}\Big(\frac{L}{cT}\Big)^2
\Big(
\Big(\frac{cT}{D}\Big)^3+\frac{6cT}{D}
\Big),
\label{GcPhitdl}
\end{align}
and we obtain the following eigenvalue $\eqref{mineigen}$
\begin{align}
\lambda_{\text{min}}
&\approx \frac{1}{4} \Big[ \Gamma_1 +\Gamma_2 -\sqrt{(\Gamma_1 -\Gamma_2)^2+\Phi^2 + \Gamma^2_\text{c}}\Big]
\nonumber 
\\
&\approx \frac{16 e^{2}}{3 \pi^{2} \hbar c} \frac{L^{2}}{(c T)^{2}}-\frac{1}{4}\sqrt{\left[\frac{64e^2}{315\pi\hbar c}
\Big(\frac{L}{cT}\Big)^2
\Big(
\Big(\frac{cT}{D}\Big)^3+
\frac{6cT}{D}
\Big)
\right]^2
+
\left[
\frac{64e^2}{3\pi^2 \hbar c} \frac{L^2}{(cT)^2} \Big(1+\frac{4D^2}{(cT)^2} \ln \Big[\frac{D}{cT}\Big] 
\Big)
\right]^2},
\label{lambda:lTDL}
\end{align}
where in the first line 
we assumed that $\Gamma_i$, 
$\Gamma_c$, and 
$\Phi$ are small, and  Eqs. \eqref{Gamma12}
and \eqref{GcPhitdl} were substituted in the second line.
Eq. \eqref{GcPhitdl} is derived in Appendix \ref{cphilTDL}.
The term $\Gamma_1+\Gamma_2$ in the first line of Eq. \eqref{lambda:lTDL} (or the first positive term in the second line) makes
$\lambda_\text{min}$ positive and reduces the negativity.
In contrast, the second term given by $\Phi$ and $\Gamma_{\text{c}}$ (or the second term in the second line) decreases 
$\lambda_\text{min}$, where $\Phi$ is much larger than  $\Gamma_{\text{c}}$ because of $\Gamma_{\text{c}}/\Phi \approx (D/cT)^3 \ll 1$.
The quantity $\Phi$ reflects the contribution of the Coulomb potential (proportional to $D^{-3}$ term) and its relativistic correction (proportional to $D^{-1}$ term).

The panels (a) and (b) in Fig. \ref{fig:lTLD} show the negativity in the regimes $cT \gg D \sim L$ and $cT \gg D \gg L$, respectively.
The blue curve in each panel presents the behavior of the negativity in Fig. \ref{fig:NlTL-D}, which is given in the non-relativistic limit and has no contributions from the dynamical degrees of freedom of the photon field.
The red curve shows the behavior of the negativity computed from our analysis.
In the panel (a) under the regime $cT \gg D \sim L$, the red curve is similar to the blue curve. This means that the Coulomb potential is dominant to determine the negativity in this regime, and the relativistic corrections are small.
However, in the panel (b) under the regime $cT \gg D \gg L$, there is the parameter region without the negativity. 
This is because the decoherence effects $\Gamma_1$ and $\Gamma_2$ are more dominant than the term $\Phi$ mainly determined by the Coulomb potential.
In this regime, the computation of the negativity in the non-relativistic limit is not valid. 
%%%%
\begin{figure}[H]
  \centering
  \begin{minipage}[b]{0.45\linewidth}
  \includegraphics[width=1\linewidth]{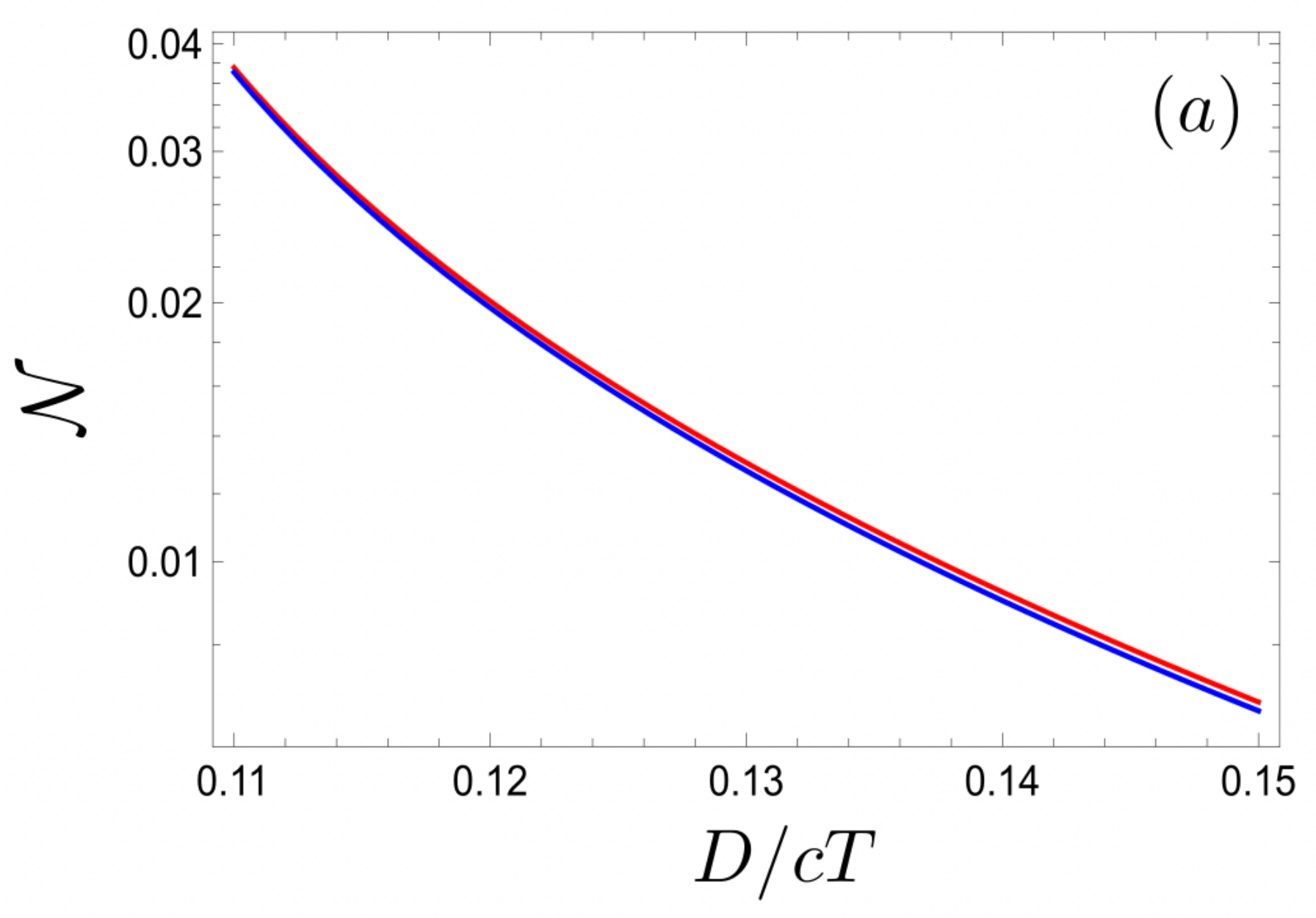}
  \end{minipage}
  \begin{minipage}[b]{0.45\linewidth}
  \includegraphics[width=1\linewidth]{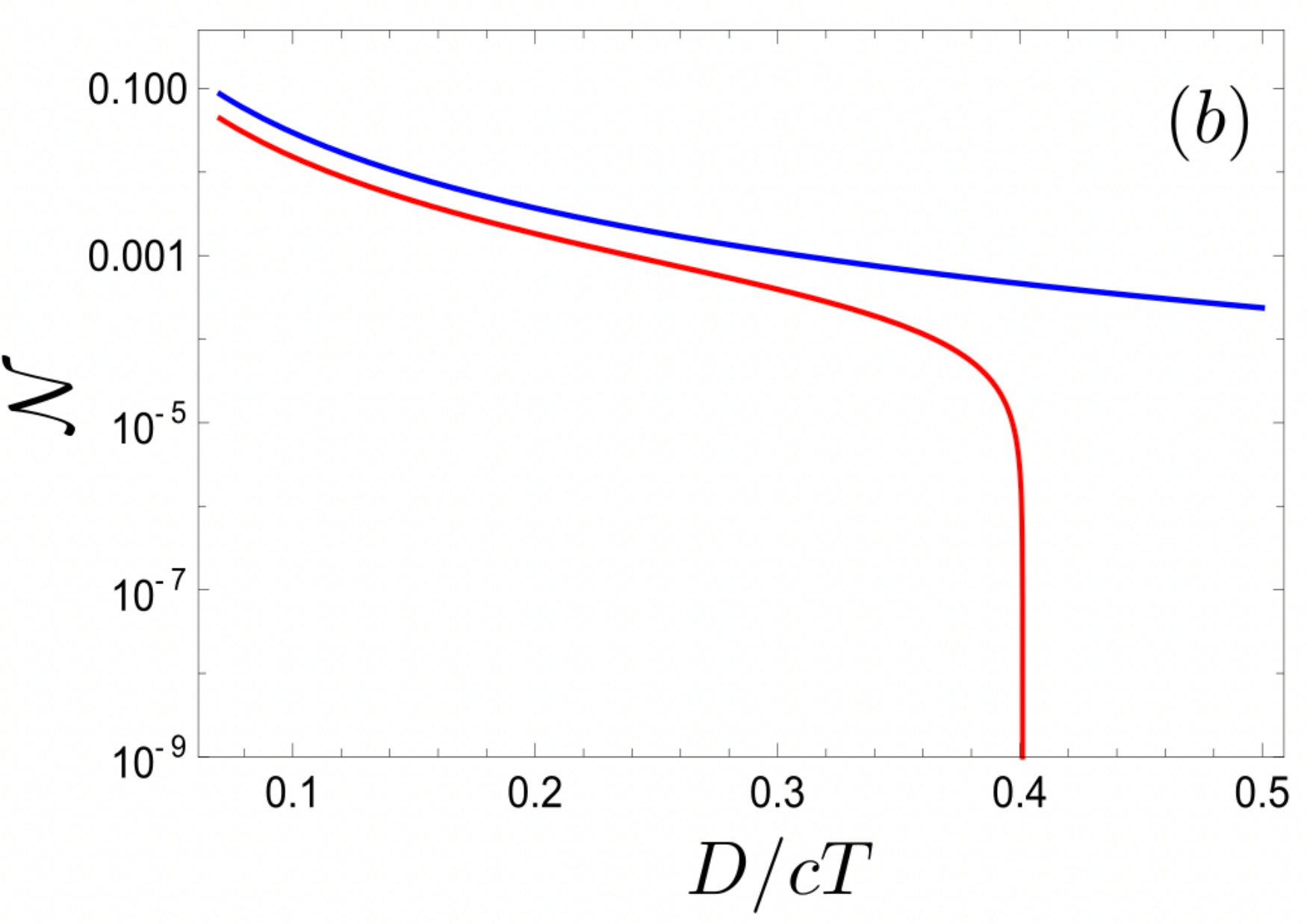}
  \end{minipage}
  \caption{Negativity $\mathscr{N}$ for the linear configuration. (a) is the case $cT \gg D \sim L$ while (b) is the case $cT \gg D \gg L$. We adopted $L/cT=0.1$.}
  \label{fig:lTLD}
\end{figure}
%%%%

\subsubsection{$D \gg cT \gg L$ regime}

\begin{figure}[H]
  \centering
  \includegraphics[width=0.5\linewidth]{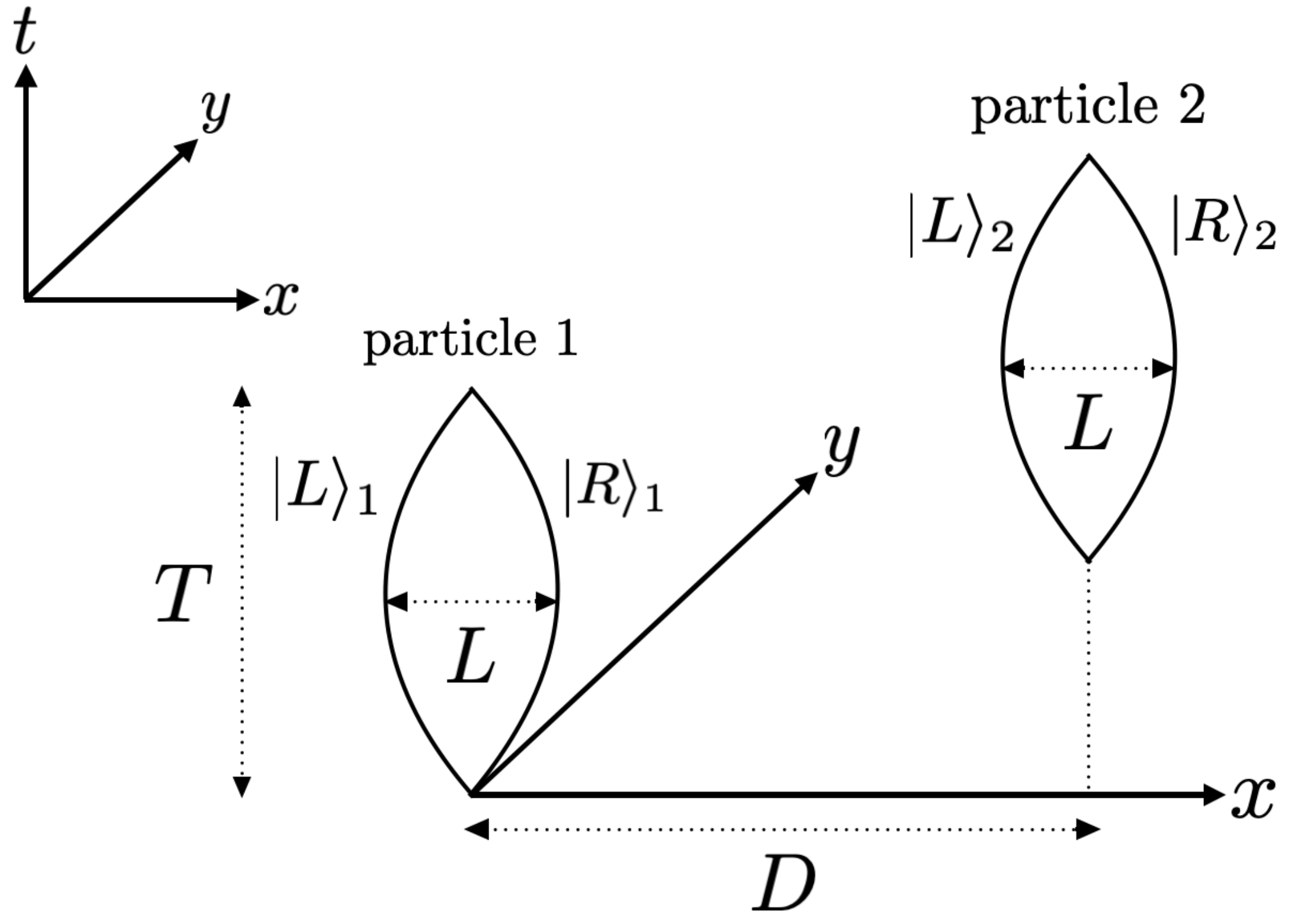}
  \caption{Linear configuration in the $D \gg cT \gg L$ regime.}
  \label{fig:linedtl}
\end{figure}

Subsequently, we present the formula of the minimum eigenvalue 
$\lambda_\text{min}$ in the regime 
$D \gg cT \gg L$. 
We assume the trajectories of the charged particles 1 and 2 given by
\begin{equation}
X^\mu_{1\text{P}}(t)=\Big[t,\epsilon_{\text{P}}X(t), 0,0 \Big]^{T}, 
\quad 
X^\mu_{2\text{Q}}(t)=\Big[t,\epsilon_{\text{Q}}X(t-D)+D, 0,0 \Big]^{T}, 
\quad 
\epsilon_{\text{R}}=-\epsilon_{\text{L}}=1, 
\quad 
X(t)=8L\Big(1-\frac{t}{T} \Big)^2 \Big(\frac{t}{T}\Big)^2,
\end{equation}
where 
$X^\mu_{2\text{Q}}$ is defined in 
$D/c \leq t \leq T+D/c$.
The whole configuration of the trajectories is shown in Fig. \ref{fig:linedtl}, in which the superposition of particle 2 is formed after particle 1 is superposed. 
The trajectories of the particles are arranged to be causally connected.
%%%%
%%%%
We obtain the following formulas for the regime $D \gg cT \gg L$,
\begin{align}
\Gamma_{1}=\Gamma_{2} \approx \frac{32 e^{2}}{3 \pi^{2}\hbar c} \frac{L^{2}}{(cT)^{2}},
\quad
\Gamma_{\text{c}} \approx-\frac{32e^{2}}{225 \pi^{2}\hbar c} \frac{L^{2} (cT)^{2}}{D^{4}},
\quad
\Phi \approx \frac{16e^2}{315\pi \hbar c} \frac{L^2 (cT)}{D^3},
\label{G12GcPhiDTL}
\end{align}
where 
$\Gamma_1$ and 
$\Gamma_2$ are the same as those given in \eqref{Gamma12} because they depend only on each particle motion, and the explicit derivation of $\Gamma_{\text{c}}$ and $\Phi$ is presented in Appendix \ref{cphilDTL}. 
We can then compute the eigenvalue \eqref{mineigen} as
%%%%
\begin{align}
\lambda_{\text{min}}
&\approx \frac{1}{4} \Big[ \Gamma_1 +\Gamma_2 -\sqrt{(\Gamma_1-\Gamma_2)^2+\Phi^2 + \Gamma^2_\text{c}}\Big]
\nonumber 
\\
&\approx \frac{16 e^{2}}{3 \pi^{2} \hbar c} \frac{L^{2}}{(c T)^{2}}-\frac{16 e^{2}}{315 \pi \hbar c} \frac{c T L^{2}}{D^3}, 
\label{lambda:lDTL}
\end{align}
%%%%
where in the first equality, the minimum eigenvalue was approximated by assuming that 
$\Gamma_i (i=1,2)$, 
$\Gamma_\text{c}$, and
$\Phi$ are small. 
In the second equality, we substituted \eqref{G12GcPhiDTL} and neglected $\Gamma_{\text{c}}$ because of $\Gamma_{\text{c}}/\Phi \approx cT/D \ll1$ for the regime $D \gg cT \gg L$. 
The positive term in the right hand side of Eq. \eqref{lambda:lDTL}, which is given by the decoherence functional 
$\Gamma_i$, comes from the vacuum fluctuations of the photon field.
The negative term in Eq. \eqref{lambda:lDTL} is given by the quantity
$\Phi$ depending on the phase shifts due to the retarded field 
(see the formula of $\Phi$ \eqref{phi} and the discussion around \eqref{eq:A}).

\begin{figure}[H]
  \centering
  \includegraphics[width=0.5\linewidth]{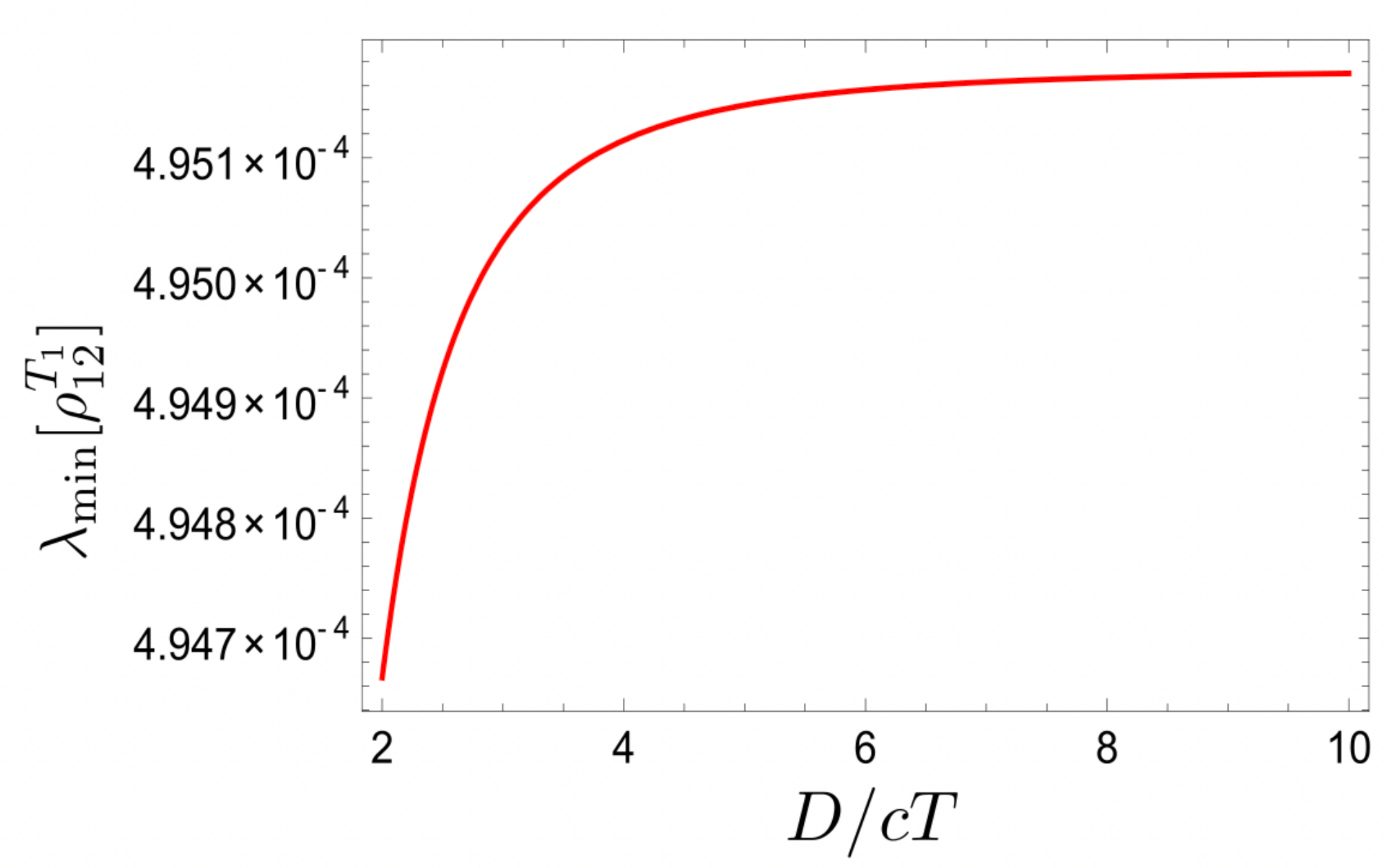}
  \caption{Minimum eigenvalue $\lambda_{\text{min}}[\rho^{\text{T}_1}_{12}]$ for the linear configuration in the regime $D \gg cT \gg L$. We adopted $L/cT=0.1.$}
  \label{fig:lDTL}
\end{figure}

Fig. \ref{fig:lDTL} shows the minimum eigenvalue \eqref{lambda:lDTL} for a fixed 
$L/cT=0.1$ as a function of 
$D/cT$ in the regime 
$D \gg cT \gg L$.
The minimum eigenvalue is always positive, and hence the charged particles 1 and 2 are not entangled.
This result shows that the decoherence due to the vacuum fluctuation of the photon field suppresses the entanglement generation due to the retarded field. 
In Sec. \ref{secV}, we will discuss that the retarded field corresponds to the longitudinal mode, that is, the non-dynamical part of the photon field. 
%%%%
%%%%

\subsection{Parallel configuration}
Here, we consider the parallel configurations shown in Fig. \ref{fig:paratdl} and Fig. \ref{fig:paradtl}. 
The parameters $T$, $L$, and $D$ play the same role as those in the linear configuration, which are the typical scales appearing in the trajectories of the particles.

\subsubsection{$cT \gg L \gg D$ or $cT\gg D \gg L$ regimes}

\begin{figure}[H]
  \centering
  \includegraphics[width=0.6\linewidth]{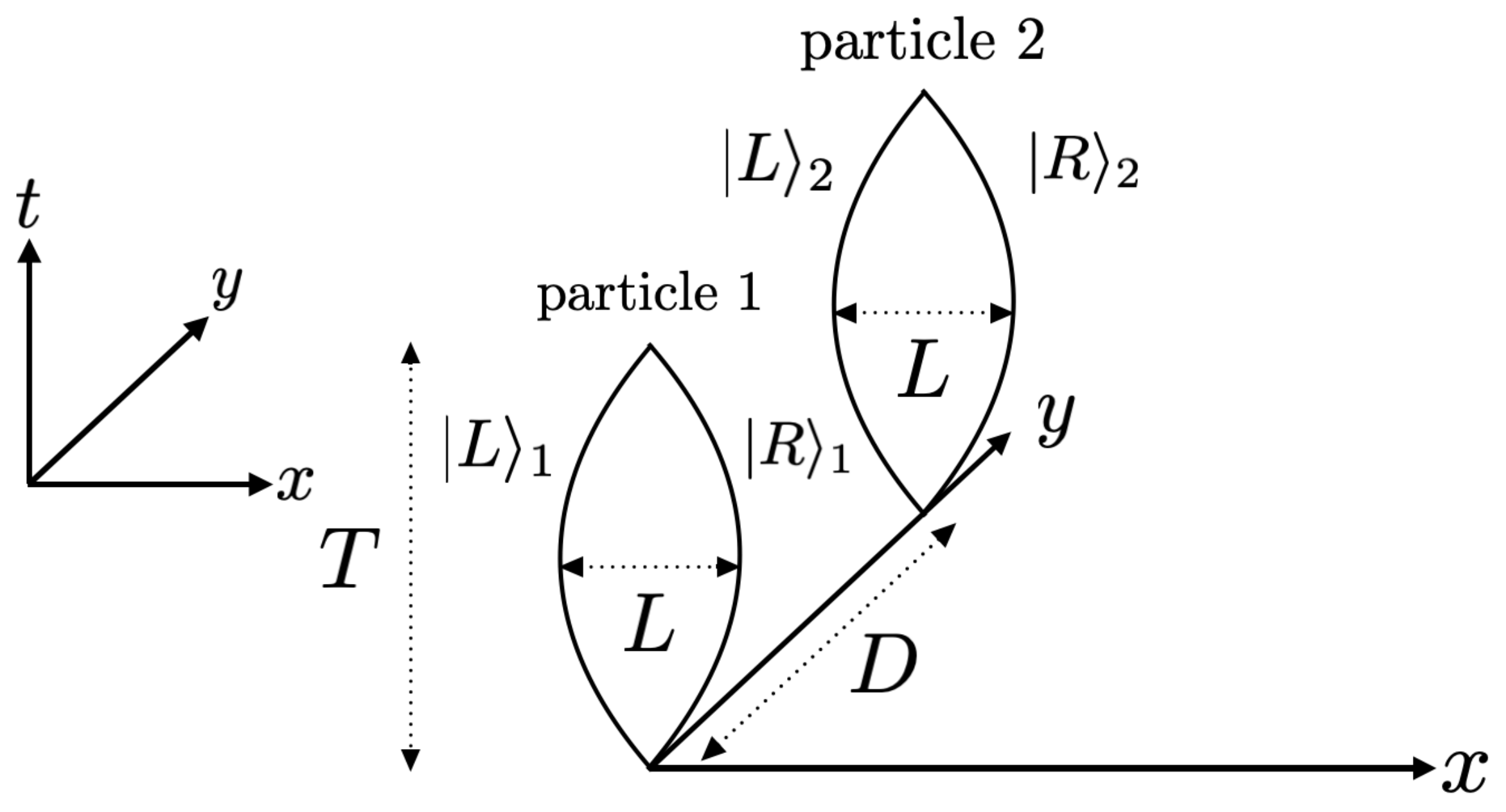}
  \caption{Parallel configuration in $cT \gg D \gg L$ regime.}
  \label{fig:paratdl}
\end{figure}

We first consider the trajectories of the two particles 1 and 2 as 
\begin{equation}
X^\mu_{1\text{P}}(t)=\Big[t,\epsilon_{\text{P}}X(t), 0,0 \Big]^\text{T}, \quad
X^\mu_{2\text{Q}}(t)=\Big[t,\epsilon_{\text{Q}}X(t), D,0 \Big]^\text{T},
\quad 
\epsilon_{\text{R}}=-\epsilon_{\text{L}}=1, 
\quad 
X(t)=8L\Big(1-\frac{t}{T} \Big)^2 \Big(\frac{t}{T}\Big)^2.
\end{equation}
The schematic configuration is shown in Fig. \ref{fig:paratdl}. 
We examine the quantities 
$\Gamma_i (i=1,2)$, $\Gamma_\text{c}$, and $\Phi$ for the regimes $cT \gg L \gg D$ and $cT\gg D \gg L$ to estimate the minimum eigenvalue $\lambda_\text{min}$. 
Even in this configuration, 
the decoherence functionals 
$\Gamma_1$ and 
$\Gamma_2$ for $cT \gg L$ are identical to those in Eq. \eqref{Gamma12}, that is, 
\begin{equation}
\Gamma_1=\Gamma_2\approx
\frac{32 e^2}{3 \pi^2 \hbar c} \frac{L^{2}}{(cT)^{2}}.
\label{G12}
\end{equation}
This is because the decoherence functionals are given by the local motions of each charged particle. 
In the following, we evaluate $\Gamma_\text{c}$ and 
$\Phi$ for each of the regimes 
$cT \gg L \gg D$ and $cT\gg D \gg L$. 
%%%%
%%%%
In the regime 
$cT \gg L \gg D$, the quantities 
$\Gamma_\text{c}$ and $\Phi$ are
%%%%
\begin{align}
\Gamma_{\text{c}}&\approx \frac{64 e^{2}}{3 \pi^{2} \hbar c} \frac{L^{2}}{(cT)^{2}},
\quad
\Phi \approx \frac{e^{2}}{4 \pi \hbar c} \frac{c T}{D}\Big(1-\frac{64 L^{2}}{105(c T)^{2}}\Big),
\end{align}
which are derived in Appendix \ref{cphipTDL}. 
The minimum eigenvalue \eqref{mineigen} for the regime $cT \gg L \gg D$ is given as 
\begin{align}
\lambda_{\text{min}}[\rho_{12}^{\mathrm{T}_{1}}] &\approx 
\frac{1}{4} 
\Big[ 
\Gamma_1 +\Gamma_2 -\sqrt{(\Gamma_1-\Gamma_2)^2+4\sin^2\Big[\frac{\Phi}{2}\Big]+\Gamma^2_\text{c}} 
\Big]
\nonumber 
\\
&\approx
\frac{16 e^{2}}{3 \pi^{2} \hbar c} \frac{L^{2}}{(cT)^{2}}
-\frac{1}{4}\sqrt{\Big(2\sin\Big[\frac{e^{2}}{4 \pi \hbar c} \frac{c T}{D}\Big(1-\frac{64 L^{2}}{105(c T)^{2}}\Big)\Big]\Big)^2+\Big(\frac{64 e^{2}}{3 \pi^{2} \hbar c} \frac{L^{2}}{(cT)^{2}}\Big)^2}.
\label{lambda:pTLD}
\end{align}
%%%% 
In the above equation, the first term coming from $\Gamma_1+\Gamma_2$ increases the minimum eigenvalue, whereas the second term given by $\Phi$ and 
$\Gamma_\text{c}$ decreases it.
It should be noted that the quantity $\Phi$ can be $\Phi \gg 1$ because of $cT/D(1-L^2/(cT)^2) \approx cT/D \gg 1$ for the regime $cT \gg L \gg D$.

In the regime $cT \gg D \gg L$, the quantities  $\Gamma_{\text{c}}$ and $\Phi$ are
%%%%
\begin{align}
\Gamma_{\text{c}}&\approx\frac{64 e^{2}}{3 \pi^{2} \hbar c} \Big(\frac{L}{c T} \Big)^2 \Big(1+\Big(\frac{2 D}{c T}\Big)^2 \ln \Big[\frac{D}{c T}\Big]\Big),
\quad
\Phi\approx\frac{32 e^{2}}{315 \pi \hbar c} \Big(\frac{L}{cT}\Big)^2\Big(\Big(\frac{cT}{D}\Big)^3-\frac{6cT}{D}\Big).
\end{align}
These formulas are derived in Appendix \ref{cphip}. 
The minimum eigenvalue \eqref{mineigen} for the regime $cT \gg D \gg L$ is approximated as
\begin{align}
\lambda_{\text{min}}[\rho_{12}^{\mathrm{T}_{1}}]
&\approx 
\frac{1}{4} 
\Big[ 
\Gamma_1 +\Gamma_2 -\sqrt{(\Gamma_1-\Gamma_2)^2+\Phi^2+\Gamma^2_\text{c}} 
\Big]
\nonumber 
\\
&\approx \frac{16 e^{2}}{3 \pi^{2} \hbar c} \frac{L^{2}}{(c T)^{2}}-\frac{1}{4}\sqrt{\left[\frac{32 e^{2}}{315 \pi \hbar c} \Big(\frac{L}{cT}\Big)^2\Big(\Big(\frac{cT}{D}\Big)^3-\frac{6cT}{D}\Big)\right]^2+\left[\frac{64 e^{2}}{3 \pi^{2} \hbar c} \Big(\frac{L}{cT} \Big)^2\Big(1+\Big(\frac{2 D}{c T}\Big)^2 \ln \Big[\frac{D}{c T}\Big]\Big)\right]^2},
\label{lambda:pTDL}
\end{align}
%%%%
This minimum eigenvalue has the very similar feature to that obtained in the case of the linear configuration. 
The first positive contribution in \eqref{lambda:pTDL} comes from the decoherence functional
$\Gamma_i$ quantifying the decoherence due to the vacuum fluctuations of the photon field. 
The second negative contribution in \eqref{lambda:pTDL} is computed from $\Gamma_\text{c}$ and $\Phi$, which is mostly from $\Phi$ because of $\Gamma_{\text{c}}/\Phi \approx (D/cT)^3 \ll 1$.
The quantities $\Gamma_\text{c}$ and $\Phi$ stem from the vacuum correlation of the photon field and the phase shifts due to the retarded field, respectively.

\begin{figure}[H]
  \centering
  \begin{minipage}[b]{0.45\linewidth}
  \includegraphics[width=1\linewidth]{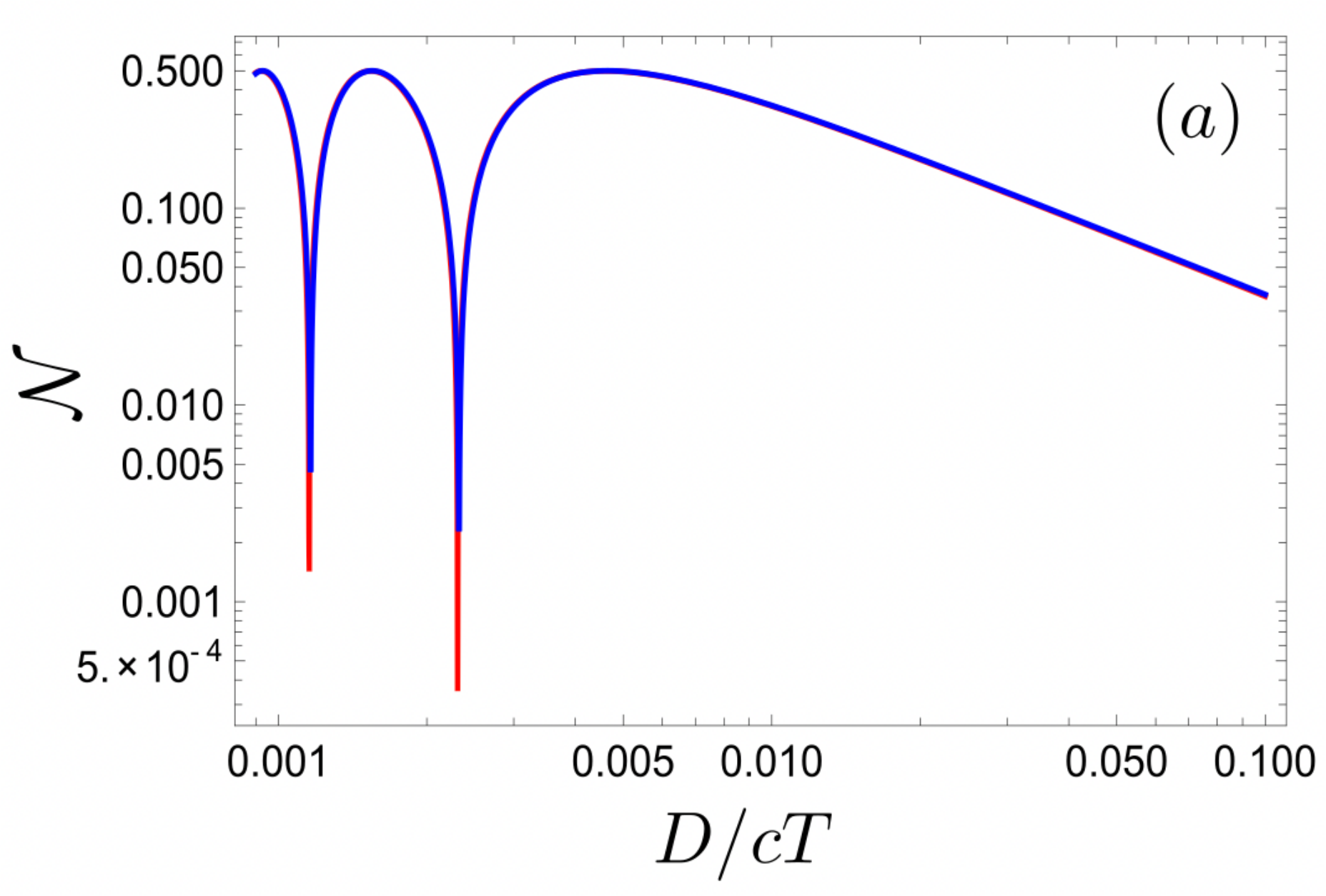}
  \end{minipage}
  \begin{minipage}[b]{0.45\linewidth}
  \includegraphics[width=1\linewidth]{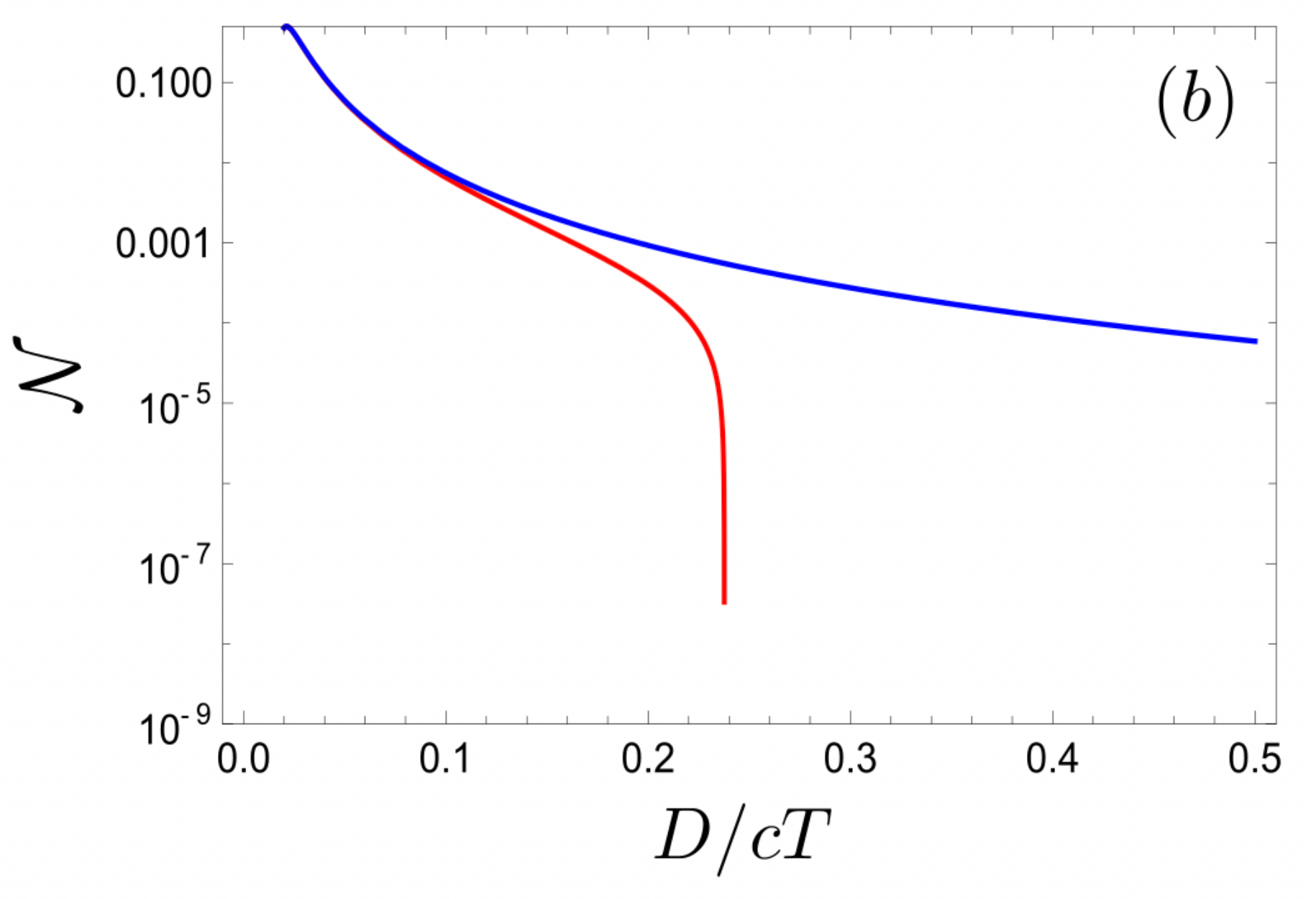}
  \end{minipage}
  \caption{Negativity $\mathscr{N}$ for the parallel configuration. (a) is the case $cT \gg L \gg D$, whereas (b) is the case $cT \gg D \gg L$. We adopted $L/cT=0.1$.}
  \label{fig:pTLD}
\end{figure}

The panels in Fig.~\ref{fig:pTLD} (a) and (b) present the behavior of the negativity in the regimes $cT \gg L \gg D$ and $cT\gg D \gg L$, respectively.
The blue curve shows the negativity in the non-relativistic limit, which corresponds to the electromagnetic version of the BMV experiment. 
The red curve is given by our analysis.
The behavior of the negativity in Fig.~\ref{fig:pTLD} (a) means that our analysis is consistent with the non-relativistic result.
However, in Fig. \ref{fig:pTLD} (b), due to the decoherence, the parameter region without the negativity appears, and hence the computation in the non-relativistic limit becomes invalid in $cT\gg D \gg L$.
%%%%

\subsubsection{$D \gg cT \gg L$ regime}

\begin{figure}[H]
  \centering
  \includegraphics[width=0.5\linewidth]{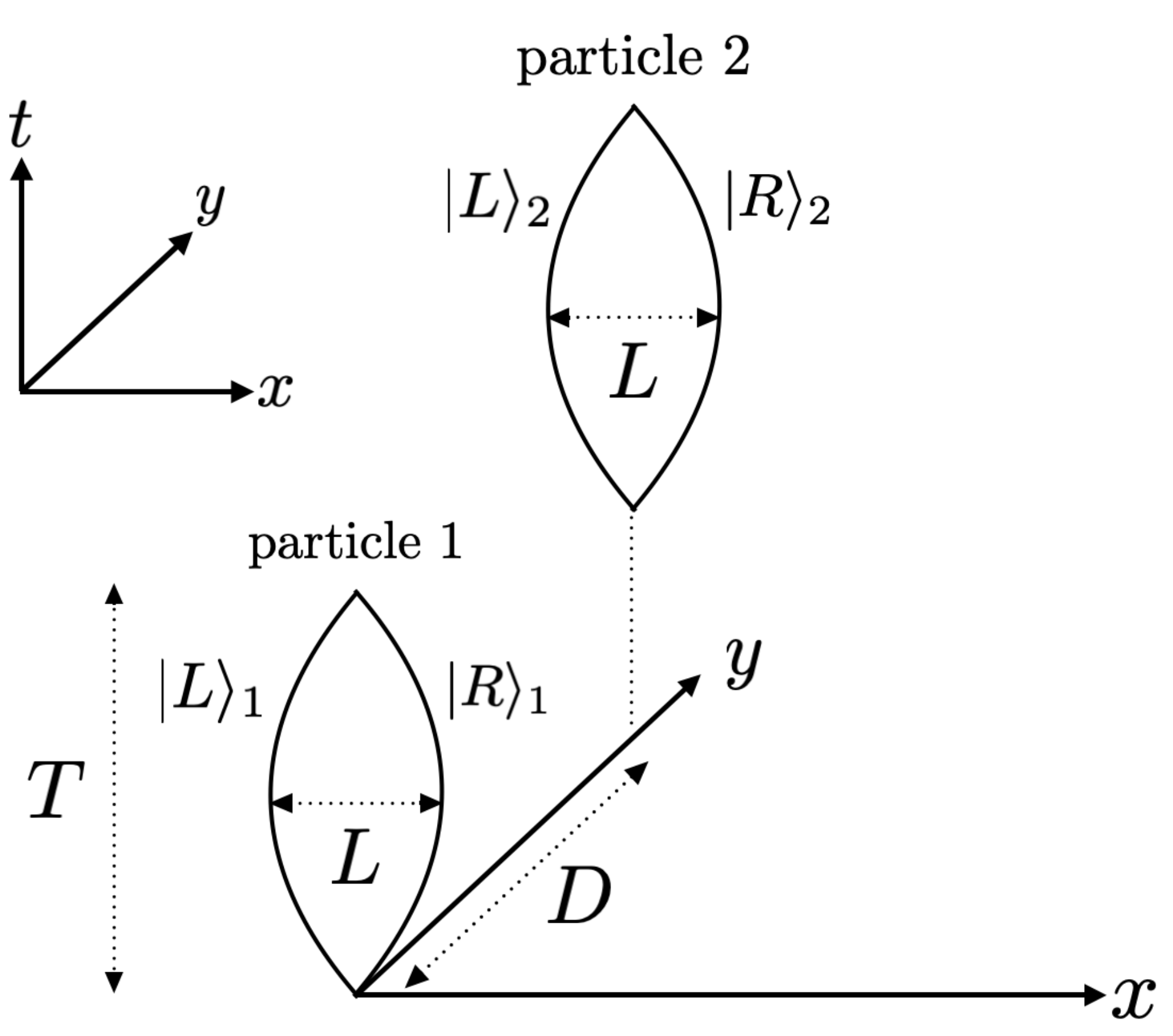}
  \caption{Parallel configuration in $D \gg cT \gg L$ regime.}
  \label{fig:paradtl}
\end{figure}

\begin{figure}[H]
  \centering
  \includegraphics[width=0.5\linewidth]{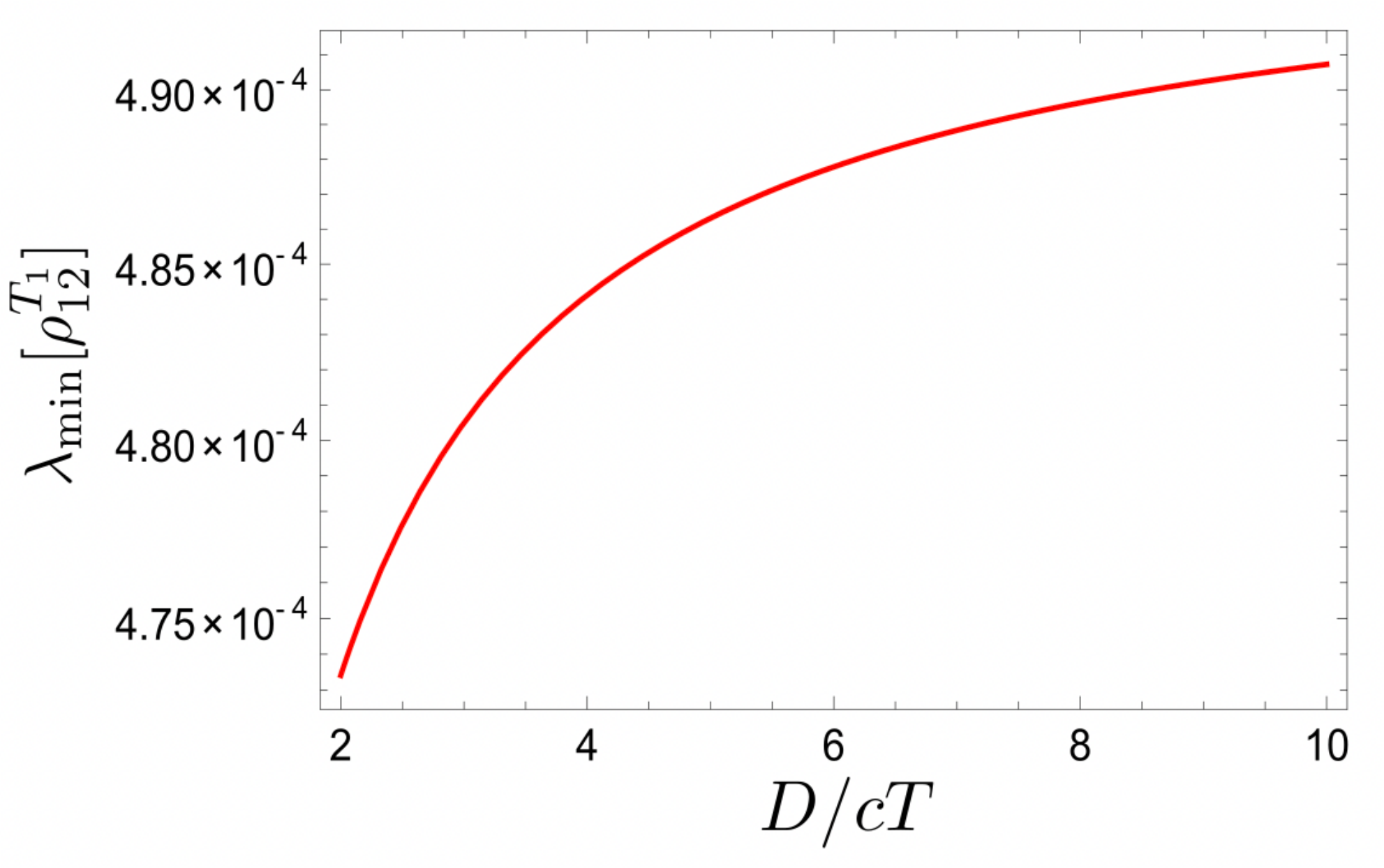}
  \caption{Minimum eigenvalue $\lambda_{\text{min}}[\rho^{\text{T}_1}_{12}]$ for the parallel configuration in the regime $D \gg cT \gg L$. We adopted $L/cT=0.1$.}
  \label{fig:pDTL}
\end{figure}

We consider the trajectories of two charged particles 1 and 2 as 
\begin{equation}
X^\mu_{1\text{P}}(t)=\Big[t,\epsilon_{\text{P}}X(t), 0,0 \Big]^\text{T},
\quad
X^\mu_{2\text{P}}(t)=\Big[t,\epsilon_{\text{P}}X(t-D/c), D,0 \Big]^\text{T}, 
\quad 
\epsilon_{\text{R}}=-\epsilon_{\text{L}}=1, 
\quad 
X(t)=8L\Big(1-\frac{t}{T} \Big)^2 \Big(\frac{t}{T}\Big)^2,
\label{parallelDTL}
\end{equation}
where 
$X^\mu_{1\text{P}}$ and 
$X^\mu_{2\text{Q}}$ with $\text{P},\text{Q}=\text{R},\text{L}$ describe the trajectory of each particle.  
Here,  
$X^\mu_{2\text{Q}}$ is defined in 
$D/c \leq  t \leq T+D/c$.
The spacetime configuration of the particles is presented in Fig. \ref{fig:paradtl}. 
We examine the minimum eigenvalue in the regime $D\gg cT \gg L$. 
%%%%
%%%%
We have the following formulas of $\Gamma_{1}$, $\Gamma_{2}$, $\Gamma_{\text{c}}$ and $\Phi$ for the regime $D\gg cT \gg L$, 
\begin{align}
\Gamma_{1}=\Gamma_{2} \approx \frac{32 e^{2}}{3 \pi^{2}\hbar c} \frac{L^{2}}{(cT)^{2}},
\quad
\Phi \approx-\frac{64 e^{2}}{105 \pi \hbar c} \frac{L^{2}}{D(cT)},
\quad
\Gamma_{\text{c}} \approx-\frac{32e^{2}}{225 \pi^{2}\hbar c} \frac{L^{2} (cT)^{2}}{D^{4}},
\end{align}
where 
$\Gamma_\text{1}$ and $\Gamma_\text{2}$ are not at all different from those given in \eqref{Gamma12} or 
\eqref{G12}, and the quantities 
$\Gamma_\text{c}$ and 
$\Phi$ are derived in Appendix \ref{cphipDTL}.
Then, we can compute the minimum eigenvalue \eqref{mineigen} as
%%%%
\begin{align}
\lambda_{\text{min}}[\rho_{12}^{\text{T}_{1}}]
&\approx \frac{1}{4} \Big[ \Gamma_1 +\Gamma_2 -\sqrt{(\Gamma_1-\Gamma_2)^2+\Phi^2 + \Gamma^2_\text{c}}\Big]
\nonumber 
\\
&\approx \frac{16 e^{2}}{3 \pi^{2} \hbar c} \frac{L^{2}}{(c T)^{2}}-\frac{16 e^{2}}{105 \pi \hbar c} \frac{L^{2}}{D(cT)},
\label{lambda:pDTL}
\end{align}
%%%%
where the first term  coming from the decoherence functional $\Gamma_i$ increases the minimum eigenvalue, and the second term given by $\Phi$ decreases it.
In the second equality, we neglected $\Gamma_{\text{c}}$ because of $\Gamma_{\text{c}}/\Phi \approx (cT/D)^3 \ll 1$.
Fig.~\ref{fig:pDTL} shows the minimum eigenvalue \eqref{mineigen} as a function of $D/cT$ in the regime $D\gg cT \gg L$, which is always positive.
Similar to the result in the case of the linear configuration (see Fig.~\ref{fig:pDTL}), the negativity remains zero, and the entanglement between the charged particles 1 and 2 does not appear in the regime $D \gg cT \gg L$. 
We come to the same conclusion that the decoherence due to the vacuum fluctuations of the photon field prevents the entanglement generation due to the retarded field. 
%%%%
%%%%
It is important to note that the parameter dependence appearing in the formulas of the minimum eigenvalue \eqref{lambda:lDTL} and \eqref{lambda:pDTL} is different.
The second terms of \eqref{lambda:lDTL} and \eqref{lambda:pDTL} are proportional to $-cTL^2/D^3$ and $-L^2/D(cT)$, respectively.
The latter is regarded as a consequence of the quantum superposition of bremsstrahlung, as we will discuss in the next section.

\section{Discussion\label{secV}}

Before the main discussion in this section, we first mention a basic property of the field strength of a charged particle.  
Generally, the field strength of a charged particle is decomposed into two terms 
$F^{\mu\nu}=F^{\mu \nu}_{\text{v}}+F^{\mu \nu}_{\text{a}}$, which are given as 
%%%%
\begin{align}
F^{\mu \nu}_{\text{v}}(x)
&=-
\frac{e}
{4\pi} \frac{(x^\mu-X^\mu(t_\text{r})) v^\nu(t_\text{r})-(\mu \leftrightarrow \nu)}{\gamma^2 [ (x-X(t_\text{r})) \cdot v(t_\text{r}) ]^3},
\\
F^{\mu \nu}_{\text{a}}(x)
&=
\frac{e}
{4\pi [ (x-X(t_\text{r})) \cdot v(t_\text{r}) ]^2 } 
\Big[ 
(x^\mu-X^\mu(t_\text{r}))
\Big(
a^\nu (t_\text{r})
-\frac{(x-X(t_\text{r})) \cdot a (t_\text{r})}{(x-X(t_\text{r})) \cdot v(t_\text{r})}
v^\nu(t_\text{r}) 
\Big)
-(\mu \leftrightarrow \nu)
 \Big] ,
\end{align}
%%%%
where 
$X^\mu$ is the spacetime position of the particle, 
$v^\mu=dX^\mu/dt$ is the velocity , 
$a^\mu=dv^\mu/dt$ is the acceleration, and 
$\gamma=1/\sqrt{-v^\mu v_\mu}$ is the Lorentz factor. 
The retarded time 
$t_\text{r}$ is given by
$-(t-t_\text{r})+|\bm{x}-\bm{X} (t_\text{r})|=0$. 
The above equations are obtained in Appendix \ref{sec:LW}.
The field strength
$F^{\mu\nu}_\text{v}$ independent of acceleration has the longitudinal mode of the retarded field. 
In fact,  the inner product of the unit vector 
$\bm{n}=(\bm{x}-\bm{X}(t_\text{r}))/|\bm{x}-\bm{X}(t_\text{r})|$ in the propagation direction and  
the electric field
$\bm{E}_\text{v}$ with 
$E^i_\text{v}=F^{0i}_\text{v}$ does not vanish, 
$\bm{n} \cdot \bm{E}_\text{v} \neq 0$. 
The field strength
$F^{\mu\nu}_\text{a}$ proportional to the acceleration only has the transverse modes of the retarded field. 
This is because the propagation direction vector 
$\bm{n}$, 
the electric field
$\bm{E}_\text{a}$ with 
$E^i_\text{a}=F^{0i}_\text{a}$ and the magnetic field 
$\bm{B}_\text{a}$ with 
$B^i_\text{a}=\varepsilon^{0i}{}_{jk} F^{jk}_\text{a}/2$ ($\varepsilon^{\mu\nu\rho\sigma}$ is the totally antisymmetric tensor) satisfy
%%%%
\begin{align}
\bm{n} \cdot \bm{E}_\text{a} =F^{0i}_\text{a} n_i = \frac{F^{0\mu}_\text{a}(x_\mu-X_\mu (t_\text{r}))}{|\bm{x}-\bm{X}(t_\text{r})|}=0, 
\quad 
\bm{n} \cdot \bm{B}_\text{a} =\frac{1}{2} \varepsilon^{0i}{}_{jk} F^{jk}_\text{a} n_i = \frac{\varepsilon^{0}{}_{\mu\nu\rho} F^{\nu\rho}_\text{a} (x^\mu-X^\mu(t_\text{r}))}{2|\bm{x}-\bm{X}(t_\text{r})|} =0,
\label{eq:EB}
\end{align}
%%%%
where the last equality of the first equation holds by the light cone condition $-(t-t_\text{r})+|\bm{x}-\bm{X} (t_\text{r})|=0$.

\begin{figure}[H]
\centering
\begin{minipage}[b]{0.4\linewidth}
  \includegraphics[width=1\linewidth]{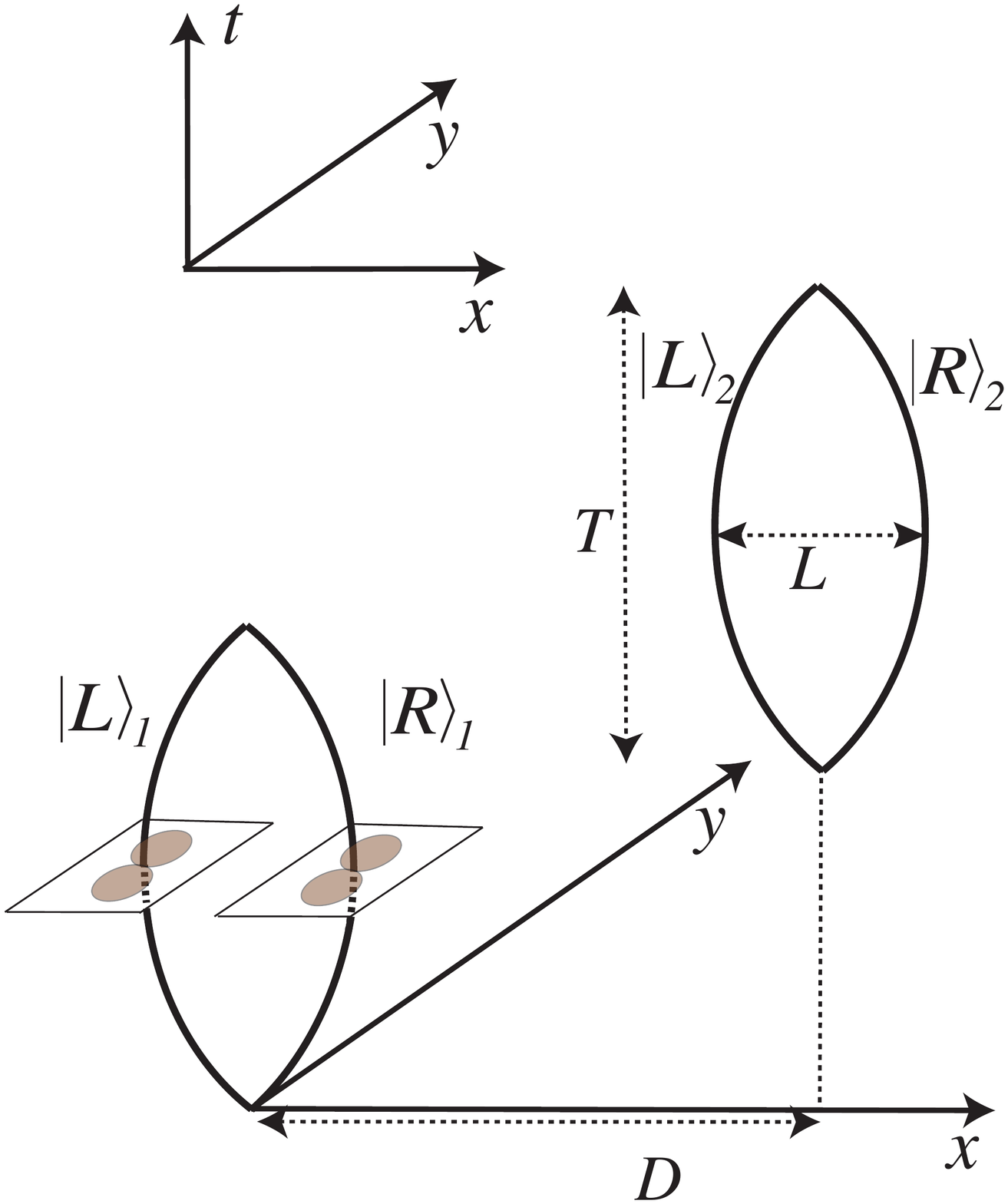}
  \subcaption{Linear configuration}
\end{minipage}
~~~~~~~~~
\begin{minipage}[b]{0.4\linewidth}
  \includegraphics[width=1\linewidth]{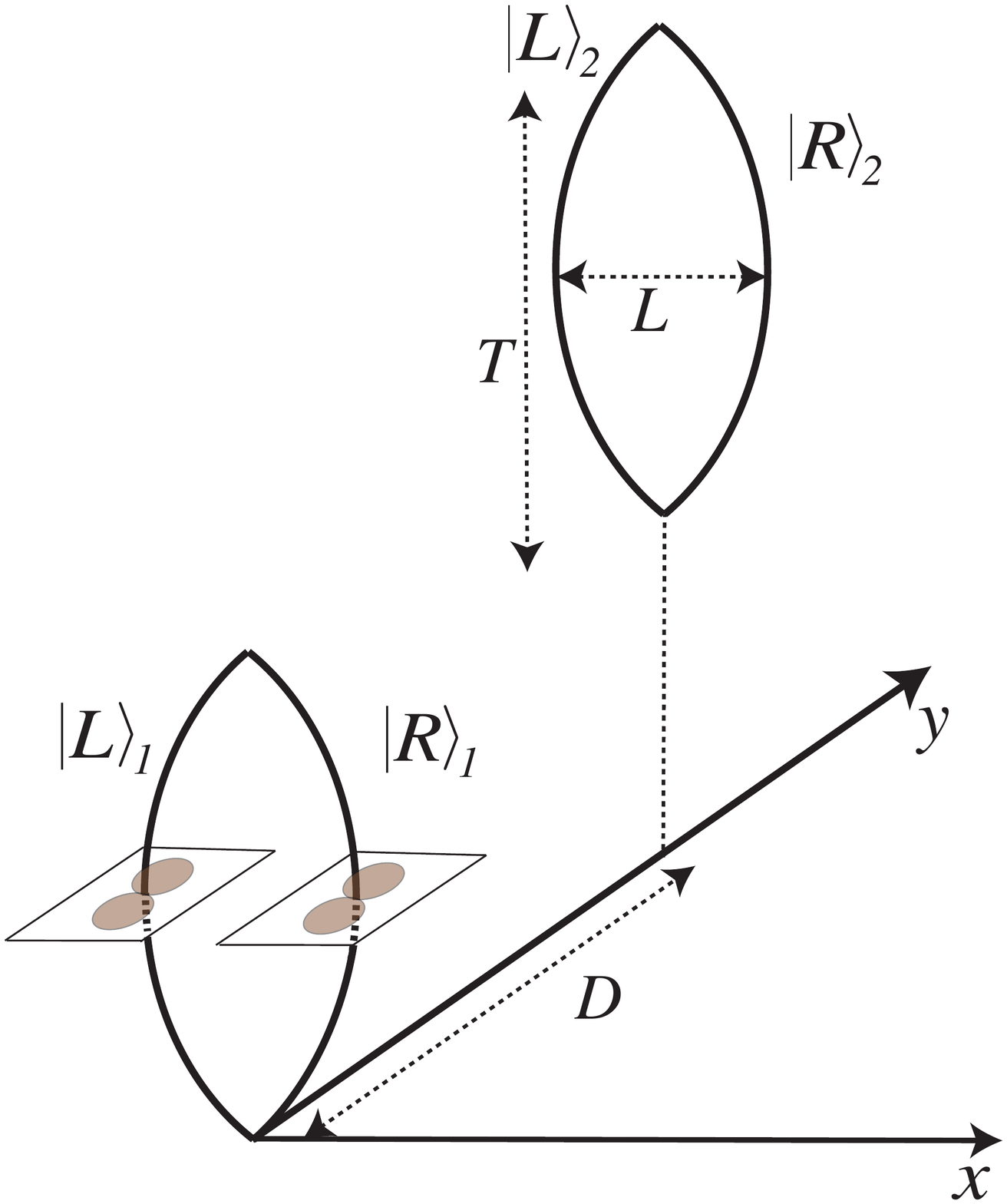}
  \subcaption{Parallel configuration}
 \end{minipage}
  \caption{Angular distribution of the photon field induced by each trajectory of the accelerating charged particle 1 for linear configuration (a) and parallel configuration (b) on the $x-y$ plane at a constant time.}
  \label{fig:elefield}
\end{figure}

With the above knowledge, we next discuss the origin of the second terms in \eqref{lambda:lDTL} and \eqref{lambda:pDTL} computed from the quantity 
$\Phi$. 
We derived those terms by assuming the regime $D\gg cT \gg L$ for each case of the linear and parallel configurations. 
The regime 
$D \gg cT $
is regarded as the wave zone in which the distance between two charged particles $D$ is much larger than the wavelength of the photon field 
$\lambda_{p}= cT$ emitted from each charged particle.
Hence it is important to understand how the radiative field affects the quantity $\Phi$.
Let us revisit the formula \eqref{phiF} of 
$\Phi$ expressed in terms of the field strengths,
%%%%
\begin{equation}
\Phi=
\frac{e}{4} 
\Big (
\int_{\text{S}_1} d\sigma_{\mu \nu} \Delta F^{\mu\nu}_2 (x)
+\int_{\text{S}_2} d\sigma_{\mu \nu} \Delta F^{\mu\nu}_1 (x)
\Big),
\end{equation}
%%%%
where 
$S_i$ is the surface surrounded by the spacetime trajectories of the particle $i (=1,2)$, and
$\Delta F^{\mu\nu}_i=F^{\mu\nu}_{i\text{R}}-F^{\mu\nu}_{i\text{L}}$.
Here, 
$F^{\mu\nu}_{i\text{P}}=\partial^\mu A^\nu_{i\text{P}}-\partial^\nu A^\mu_{i\text{P}}$ are the retarded field strengths of the charged particle 
$i$ moving along the trajectory $\text{P} (=\text{R},\text{L})$. 
As mentioned in the above paragraph, the field strengths of the particle $i$ moving the trajectory P, $F^{\mu\nu}_{i\text{P}}$, are separated into two parts 
$F^{\mu\nu}_{i\text{P}}=F^{\mu\nu}_{i\text{P},\text{v}}+F^{\mu\nu}_{i\text{P},\text{a}}$, and then the quantity 
$\Phi$ is also given as
$\Phi=\Phi_\text{v}+\Phi_\text{a}$ with 
%%%%
\begin{equation}
\Phi_\text{v}=
\frac{e}{4} 
\Big (
\int_{\text{S}_1} d\sigma_{\mu \nu} \Delta F^{\mu\nu}_{2,\text{v}} (x)
+\int_{\text{S}_2} d\sigma_{\mu \nu} \Delta F^{\mu\nu}_{1,\text{v}} (x)
\Big), 
\quad 
\Phi_\text{a}=
\frac{e}{4} 
\Big (
\int_{\text{S}_1} d\sigma_{\mu \nu} \Delta F^{\mu\nu}_{2,\text{a}} (x)
+\int_{\text{S}_2} d\sigma_{\mu \nu} \Delta F^{\mu\nu}_{1,\text{a}} (x)
\Big),
\label{eq:PhivPhia}
\end{equation}
%%%%
where 
$\Delta F^{\mu\nu}_{i,\text{v}}=F^{\mu\nu}_{i\text{R},\text{v}}-F^{\mu\nu}_{i\text{L},\text{v}}$ and 
$\Delta F^{\mu\nu}_{i,\text{a}}=F^{\mu\nu}_{i\text{R},\text{a}}-F^{\mu\nu}_{i\text{L},\text{a}}$.
The term 
$\Phi_\text{v}$ depends on the longitudinal mode (non-dynamical part) of the retarded photon field, and
$\Phi_\text{a}$ comes from the transverse modes (dynamical parts) of the retarded photon field of the accelerated charged particles. 
In the linear and parallel configurations, $\Phi_{\text{v}}$ for the regime 
$D \gg cT \gg L$ has the same formula (see \eqref{eq:PhivA1} and \eqref{eq:PhivA2}), whereas $\Phi_{\text{a}}$ for the regime 
$D \gg cT \gg L$ depends on each configuration: 
$\Phi_\text{a}$ vanishes in the linear configuration, but it does not in the parallel configuration. 
To observe this, we focus on the fact that 
$\Phi_{\text{a}}$ in the configurations shown in Fig. \ref{fig:elefield} is given as 
%%%%
\begin{equation}
\Phi_{\text{a}}= \frac{e}{4} \int_{\text{S}_2} dt dx \Delta F^{01}_{1,\text{a}}
= \frac{e}{4} \int_{\text{S}_2} dt dx (E^x_{1\text{R},a}-E^x_{1\text{L},\text{a}})
\label{phiele}
\end{equation}
%%%%
where 
$E^x_{1\text{P},\text{a}}=F^{01}_{1\text{P},\text{a}}$ is the $x$-component of the electric field induced by the accelerated motion of the charged particle 1 on the trajectory P 
$(=\text{R},\text{L})$.
Here, the first term in the formula of 
$\Phi_\text{a}$ in \eqref{eq:PhivPhia} vanished by assuming that the retarded field sourced by particle 2 is causally disconnected with particle 1. 
%%%%
%%%%
Following the Larmor radiation formula, the electromagnetic wave emitted from the charged particle 1 cannot propagate in the direction of the particle acceleration~\cite{Jackson1962}. 
The shaded region in Fig.~\ref{fig:elefield} shows the angular distribution of the photon field of the charged particle 1 on each trajectory.
In the linear configuration, because each particle moves along the $\it{x}$-axis, the electromagnetic wave from particle 1 does not propagate to particle 2. 
This leads to $E^{1\text{R},\text{a}}_{\it{x}}=E^{1\text{L},\text{a}}_{\it{x}}=0$ and hence 
$\Phi_{\text{a}}=0$.
In the parallel configuration, because the electromagnetic wave from particle 1 can reach particle 2, the electric fields $E^{1\text{R},\text{a}}_{\it{x}}$ and $E^{1\text{L},\text{a}}_{\it{x}}$ generated by the superposed particle 1 give a nontrivial 
$\Phi_{\text{a}}$.
Hence, the origin of 
$\Phi_{\text{a}}$ is regarded as the quantum superposition of bremsstrahlung from the charged particle 1 in a superposition state. 
As observed in the previous section, 
the quantity $\Phi(=\Phi_\text{v}+\Phi_\text{a})$ decreases the minimum eigenvalue $\lambda_\text{min}$. 
This suggests that the effect of the quantum superposition of bremsstrahlung appears in the formula of the entanglement. As observed in the previous section, the decoherence due to the vacuum fluctuation of the photon field suppresses the entanglement generation in the charged particles.

\section{conclusion\label{secVI}}

The BMV experiment is a proposal to detect the entanglement generation due to the Newtonian gravity, which comes from the non-dynamical component of gravity. 
To understand the entanglement generation in the context of QFT, we evaluated the entanglement generation between two charged particles coupled to a photon field on the basis of QED, motivated by a similarity of the theory between gravity and electromagnetism.
We obtained the formula of the entanglement negativity between two charged particles each in a superposition of two trajectories for the first time. 
This explicitly demonstrated the effect of a quantized photon field on the entanglement generation between two charged particles.
Our analysis automatically includes the contributions not only from the longitudinal mode (non-dynamical part) but also from the transverse mode (dynamical part) of the photon field. 
We demonstrated that the entanglement generation induced by the Coulomb potential is reproduced in the non-relativistic limit of our formula, as expected. 
We also demonstrated how the relativistic corrections to the Coulomb entanglement arise.
Particularly, the vacuum fluctuations of the photon field cause quantum decoherence, which becomes significant when the decoherence due to photon emission becomes significant simultaneously, as discussed in Sec.III. 
When the two charged particles are separated by a long distance, the decoherence effect dominates, and the entanglement generation is suppressed. However, in such a situation when the two particles are separated by a distance of a wave zone, the superposition of the electromagnetic wave from the other charged particle influences the signature of the quantum coherence.
We found that the quantum superposition of bremsstrahlung from a superposed trajectory affects the signature of the quantum coherence between the two particles; however, the entanglement is not generated because the vacuum fluctuations of the photon field dominate over the signature of the entanglement. 
This addresses the issue whether the superposition of the bremsstrahlung from a superposed trajectory could generate entanglement or not. This issue is left for future study. 

Thus, we evaluated the effect of the dynamical photon field on the entanglement generation between two charged particles each in a superposition state. 
We also demonstrated that the quantum superposition of bremsstrahlung contributes to the quantum coherence behavior between two charged particles. 
One naturally expects that similar features appear in the entanglement generation between two masses in the framework of the quantized gravitational field. 
The vacuum fluctuations of the graviton field and the quantum superposition of gravitational radiation are expected to be involved in the entanglement generation between two masses. 
It is important to extend our present work to the theory of gravity to clarify the dynamical effects of the quantized gravitational field, which remains as future work for a deeper understanding of quantum gravity.

%%%%%%%%%%%%
\acknowledgements
We thank K. Ueda, J. Soda, S. Kanno, H. Suzuki and S. Iso for useful discussions related to the topic in the present paper.
%%%%%%%%%%%%

\begin{appendix}
\section{\label{BRST}BRST formalism in QED}
\subsection{BRST formalism}

Here, we summarize the BRST formalism in QED.
The Lagrangian density in BRST formalism is written as follows
%%%%
\begin{equation}
\mathscr{L}=\mathscr{L}_{\text{QED}}+\mathscr{L}_{\text{GF}+\text{FP}},
\quad 
\mathscr{L}_{\text{QED}}=-\frac{1}{4} F_{\mu \nu} F^{\mu \nu}+\bar{\psi} (i\gamma^{\mu}D_{\mu}-m ) \psi,
\end{equation}
%%%%
where 
$F_{\mu \nu}=\partial_{\mu} A_{\nu}-\partial_{\nu}A_{\mu}$ is the field strength of the $U(1)$ gauge field $A_{\mu}$, $\psi$ is the Dirac field with mass $m$, $\bar{\psi}=\psi^{\dagger}\gamma^{0}$, $\gamma^{\mu}$ is the gamma matrix satisfying $\{\gamma^{\mu}, \gamma^{\nu}\}=2\eta^{\mu \nu}$,  $D_{\mu}=\partial_{\mu}+ieA_{\mu}$ is the covariant derivative, which
includes the electromagnetic interaction term with the coupling constant $e$, and $\mathscr{L}_{GF+FP}$ is the gauge fixing and Faddeev-Popov ghost term.
The Lagrangian density $\mathscr{L}_{\text{QED}}$ is invariant under the following transformation
 \begin{equation}
 \psi \rightarrow e^{-ie\theta(x)}\psi \simeq (1-ie\theta(x))\psi \equiv \psi +\delta \psi, 
 \quad 
 A_{\mu} \rightarrow A_{\mu}+\partial_{\mu} \theta(x) \equiv A_{\mu}+\delta A_{\mu},
 \end{equation} 
where 
$\theta(x)$ is a real function.
To give the gauge fixing and Faddeev-Popov ghost term $\mathscr{L}_{GF+FP}$
, we define $\theta(x)\equiv \lambda C(x)$, where $\lambda$ and $C(x)$ are the global and local Grassmann numbers. 
The field $C(x)$ is the scalar field but it satisfies the anti-commutation relations $\{C(x), C(y)\}=0$, which is the Faddeev-Popov ghost field.
We rewrite 
$\delta \psi$ and $\delta A_{\mu}$
as follows
\begin{align}
\delta \psi(x) =\lambda (-ieC(x)\psi(x)) \equiv \lambda \delta_{\text{B}}\psi(x),
\quad
\delta A_{\mu}=\lambda (\partial_{\mu}C(x))\equiv \lambda \delta_{\text{B}}A_{\mu},
\quad
\delta_{\text{B}}C(x)=0,
\label{BRST1}
\end{align}
where the operator $\delta_{\text{B}}$ is defined so that the nilpotency $\delta^2_{\text{B}} =0$ satisfies.
We also introduce the anti-ghost field $\bar{C}(x)$ and the Nakanishi-Lautrup field $B(x)$.
They satisfy
%%%%
\begin{align}
\delta_{\text{B}} \bar{C}(x)=iB(x),
\quad
\delta_{\text{B}} B(x)=0,
\label{BRST2}
\end{align}
%%%%
where $\alpha$ is an arbitrary parameter.
The transformation of \eqref{BRST1} and \eqref{BRST2} are referred to as the BRST transformation.
We can choose the gauge fixing and Faddeev-Popov ghost term as follows
%%%%
\begin{align}
\mathscr{L}_{\text{GF}+\text{FP}}=-i\delta_{\text{B}} (\bar{C}F),
\quad
F=\partial^{\mu} A_{\mu}+\frac{1}{2}\alpha {\text{B}}.
\end{align} 
%%%%
Consequently, the full Lagrangian density in BRST formalism is
%%%%
\begin{align}
\label{fulleom}
 \mathscr{L}&=-\frac{1}{4} F_{\mu \nu} F^{\mu \nu}+\bar{\psi} (i\gamma^{\mu}D_{\mu} \psi-m ) \psi+\frac{1}{2} \alpha B^2 -\partial^{\mu}B A_{\mu}-i\partial^{\mu}\bar{C}\partial_{\mu}C.
\end{align}
The equations of motion for fields $A_{\mu}, B, C, \bar{C}$ are given by the Euler-Lagrange equations,
\begin{align}
\label{aeom}
0&=\partial^{\nu}F_{\nu \mu}-J_{\mu}-\partial_{\mu}B,\\
\label{beom}
0&=\partial^{\mu}A_{\mu}+\alpha B,\\
0&=\Box C= \Box \bar{C},
\label{ghost}
\end{align} 
where 
$J_{\mu}=e\bar{\psi}\gamma_{\mu}\psi$.
The fields 
$C(x)$ and 
$\bar{C}(x)$ follow the free evolution and do not interact with the other fields.
Substituting \eqref{beom} into \eqref{fulleom}, we arrive at the following Lagrangian density,
\begin{align}
\mathscr{L}&=-\frac{1}{4} F_{\mu \nu} F^{\mu \nu}+\bar{\psi} (i\gamma^{\mu}D_{\mu} \psi-m ) \psi-\frac{1}{2\alpha} (\partial_{\mu} A^{\mu})^2-i\partial^{\mu}\bar{C}\partial_{\mu}C,
\end{align}
%%%
and the BRST transformations are summarized as
%%%%
\begin{align}
\delta_B A_{\mu}=\partial_{\mu} C,
\quad
\delta_{B}\psi =-ieC\psi, \delta_B C=0, \delta_B \bar{C} =\frac{i}{\alpha}(\partial_{\mu}A^{\mu}).
\end{align}
Because of the BRST transformation, the Lagrangian density has a global symmetry (BRST symmetry)
\begin{align}
\lambda \delta_{B}\mathscr{L}=0.
\end{align}
%%%%
Associated with this global symmetry, there is a conserved current referred to as the BRST current ${J}^{\mu}_\text{B}$ defined by
%%%%
\begin{align}
J^{\mu}_{\text{B}}=\sum_{\text{I}} \frac{\partial \mathscr{L}}{\partial(\partial_{\mu} \Phi_{\text{I}})} \delta_{\text{B}}\Phi_{\text{I}}
=-F^{\mu \nu}\partial_\nu C-\frac{1}{\alpha} \partial_\nu A^\nu \partial^\mu C+J^\mu C
,
\end{align}
where 
$\Phi_\text{I}=\{A_\mu, \psi, C, \bar{C}\}$.
The BRST charge 
$Q_\text{B}$ is given by
%%%%
\begin{align}
Q_\text{B} \equiv \int d^3x J^0_\text{B}(x)=\int d^3x \Big[(\partial_i C)F^{i0}+J^0 C-\frac{1}{\alpha} (\partial_{\mu}A ^{\mu})\dot{C}
\Big].
\end{align}
%%%%
We perform the canonical quantization procedure in the Feynman gauge ($\alpha =1$). 
The canonical conjugate momenta are defined as 
%%%%
\begin{align}
\pi^{\mu}_{A}\equiv \frac{\partial \mathscr{L}}{\partial \dot{A}_{\mu}}
=-F^{0\mu}-(\partial_{\nu}A^{\nu})\eta^{0\mu}
,
\quad
\pi_{\psi}\equiv \frac{\partial \mathscr{L}}{\partial \dot{\psi}}
=i\bar{\psi}\gamma^{0}
,
\quad
\pi_{c}\equiv \frac{\partial \mathscr{L}}{\partial \dot{C}}
=i\dot{\bar{C}}
,
\quad
\pi_{\bar{c}}\equiv \frac{\partial \mathscr{L}}{\partial \dot{\bar{C}}}
=i\dot{C}
,
\end{align}
%%%%
where $``\cdot"$ denotes the derivative with respect to time $x^{0}=t$.
The commutation relations are assigned as follows
%%%%
\begin{align}
\{\hat{\psi}(x), \hat{\pi}_{\psi}(y)\}|_{x^0=y^0}=i\delta^3(\mathbf{x}-\mathbf{y}),
\nonumber\\
\{\hat{C}(x), \hat{\pi}_{c}(y)\}|_{x^0=y^0}=i\delta^3(\mathbf{x}-\mathbf{y}),
\nonumber\\
\{\hat{\bar{C}}(x), \hat{\pi}_{\bar{c}}(y)\}|_{x^0=y^0}=i\delta^3(\mathbf{x}-\mathbf{y}),
\nonumber\\
[\hat{A}_{\mu}(x), \hat{\pi}^{\nu}(y)]|_{x^0=y^0}=i\delta^{\nu}_{\mu}\delta^3(\mathbf{x}-\mathbf{y})
.\nonumber
\end{align}
%%%%
The quantized BRST charge is given by 
%%%%
\begin{equation}
\hat{Q}_\text{B}=\int d^3x[(\partial_i \hat{C})\hat{F}^{i0}+\hat{J}^0\hat{C}-(\partial_{\mu}\hat{A}^{\mu})\dot{\hat{C}}]=\int d^3x[-(\partial_{i}\hat{\pi}^{i})\hat{C}+\hat{J}^{0}\hat{C}+i\hat{\pi}^{0}\hat{\pi}_{\bar{c}}].
\label{intqb}
\end{equation}

As is well known, when we quantize a gauge theory while maintaining the Lorentz covariance, a state space $\mathcal{V}$ with an indefinite metric is required.
For the standard probabilistic interpretation of quantum mechanics , a physical state 
$|\Psi_\text{phys} \rangle$ has no negative norm. 
Such a state with the non-negative norm is identified by imposing the following condition (the BRST condition)
%%%%
\begin{align}
\hat{Q}_\text{B}|\Psi_{\text{phys}}\rangle=0,
\label{eq:QPsi=0}
\end{align}
%%%%
where the physical state $|\Psi_{\text{phys}}\rangle$ satisfies $\langle \Psi_{\text{phys}}|\Psi_{\text{phys}}\rangle \geq0$.

\subsection{BRST charge in the interaction picture and in the Schr\"odinger picture}
%%%%
We derive a useful form of the BRST charge for our computation. 
Using \eqref{intqb}, we obtain the BRST charge in the interaction picture,
%%%%
\begin{align}
\hat{Q}^{\text{I}}_\text{B}(t)=e^{i\hat{H}_0t}\hat{Q}_\text{B} e^{-i\hat{H}_0t}=\int d^3x[-(\partial_{i}\hat{\pi}^{i\text{I}})\hat{C}+\hat{J}^{0}_{\text{I}}\hat{C}^{\text{I}}+i\hat{\pi}^{0\text{I}}\hat{\pi}^{\text{I}}_{\bar{c}}],
\end{align}
where $\hat{\phi}^{\text{I}}=e^{i\hat{H}_0t}\hat{\phi} e^{-i\hat{H}_0t}, \hat{\phi}=\{\hat{A}_{\mu}, \hat{\pi}^{\mu}, \hat{C}, \hat{\bar{C}}, \hat{\pi}_{c}, \hat{\pi}_{\bar{c}}, \hat{J}^{0}\}$, and they satisfy the Heisenberg equation
%%%%
\begin{align}
i\dot{\hat{\phi}}^{\text{I}}=[\hat{\phi}^{\text{I}}, \hat{H}_{0}].
\end{align}
%%%%
The gauge field $\hat{A}^{\text{I}}_{\mu}(x)$ and the ghost field $\hat{C}^{\text{I}}(x)$ satisfy the Klein-Gordon equation.
The solutions are
%%%%
\begin{align}
\label{solutiona}
\hat{A}^{\text{I}}_{\mu}(x)&=\int \frac{d^3k}{\sqrt{(2\pi)^32k^{0}}}(\hat{a}_{\mu}(\bm{k})e^{ik\cdot x}+h.c.),\\
\label{solutionc}
\hat{C}^{\text{I}}(x)&=\int \frac{d^3k}{\sqrt{(2\pi)^32k^{0}}}(\hat{c}(\bm{k})e^{ik\cdot x}+h.c.),
\end{align}
%%%%
where 
$k^0=|\bm{k}|$, $\hat{a}_{\mu}(\bm{k})$ and $\hat{c}(\bm{k})$ are the annihilation operators of the gauge field $\hat{A}^{\text{I}}_{\mu}(x)$, and the ghost field $\hat{C}^{\text{I}}(x)$, respectively. 
The annihilation operators  $\hat{a}_{\mu}(\bm{k})$, $\hat{c}(\bm{k})$, and the creation operators satisfy
%%%%
\begin{equation}
\bigl[\hat{a}_{\mu}(\bm{k}),\hat{a}^{\dagger}_{\nu}(\bm{k}^{\prime})\bigr]=\eta_{\mu \nu}\delta (\bm{k}-\bm{k}^{\prime}), 
\quad
\big\{\hat{c}(\bm{k}),\hat{c}^{\dagger}(\bm{k}^{\prime})\bigr\}=\delta (\bm{k}-\bm{k}^{\prime}).
\end{equation}
Substituting \eqref{solutiona} and \eqref{solutionc} into \eqref{intqb}, we obtain the BRST charge in the interaction picture
\begin{align}
\hat{Q}^{\text{I}}_\text{B}(t)=\int \frac{d^3k}{\sqrt{(2\pi)^3}} \left[\left(k^{\mu}\hat{a}_{\mu}(\bm{k})+\frac{\hat{\tilde{J}}^{0}_{\text{I}}(t,\bm{k})}{\sqrt{2k^0}}e^{ik^0t}\right)c^{\dagger}(\bm{k})+\text{h.c.}\right],
\end{align}
where $\hat{\tilde{J}}^{0}_{\text{I}}(t,\bm{k})$ is the Fourier transformation of $\hat{J}^{0}_{\text{I}}(t,\bm{x})$
\begin{align}
\hat{J}^{0}_{\text{I}}(t,\bm{x})=\int \frac{d^3k}{\sqrt{(2\pi)^3}}\hat{\tilde{J}}^{0}_{\text{I}}(t,\bm{k})e^{i\bm{k}\cdot \bm{x}}.
\end{align}
%%%%
Using the BRST charge in the interaction picture and (\eqref{intqb}, the BRST charge in the Schr\"{o}dinger picture is obtained as
%%%%
\begin{align}
\hat{Q}_\text{B}=e^{-i\hat{H}_0t}\hat{Q}^{\text{I}}_\text{B}(t)e^{i\hat{H}_0t}=\int \frac{d^3k}{\sqrt{(2\pi)^3}} \left[\left(k^{\mu}\hat{a}_{\mu}(\bm{k}+\frac{\hat{\tilde{J}}^{0}(\bm{k})}{\sqrt{2k^0}}\right)c^{\dagger}(\bm{k})+\text{h.c.}\right],
\label{eq:QBSch}
\end{align}
%%%%
where we used
\begin{equation}
e^{-i\hat{H}_0t}\hat{a}_{\mu}(\bm{k})e^{i\hat{H}_0t}=\hat{a}_{\mu}(\bm{k})e^{ik^0t}, 
\quad 
e^{-i\hat{H}_0t}\hat{c}^{\dagger}(\bm{k})e^{i\hat{H}_0t}=\hat{c}^{\dagger}(\bm{k})e^{-ik^0t},
\quad 
e^{-i\hat{H}_0t}\hat{\tilde{J}}^0_{\text{I}}(t, \bm{k})e^{i\hat{H}_0t}=\hat{\tilde{J}}^0(\bm{k}),
\end{equation}
%%%
Here, 
$\hat{\tilde{J}}^0$ is the Fourier transform of the matter current in the Schr\"{o}dinger picture.

\subsection{BRST condition for our models with charged particles}

We use the explicit form of the BRST charge in the Schr\"{o}dinger picture \eqref{eq:QBSch} to derive the BRST condition for our models. 
Assuming a physical state $|\Psi_{\text{phys}}\rangle=|\Psi^{\prime}_{\text{phys}}\rangle \otimes |0\rangle_{c}$, where $|0\rangle_{c}$ is the ground state of the ghost field, and using \eqref{eq:QBSch}, we can reduce the BRST condition \eqref{eq:QPsi=0} as 
%%%%
\begin{equation}
\label{coh}
\Big(k^{\mu}\hat{a}_{\mu}(\bm{k})+\frac{\hat{\tilde{J}}^{0}(\bm{k})}{\sqrt{2k^{0}}}
\Big)|\Psi^{\prime}_{\text{phys}} \rangle 
= 0.
\end{equation}
%%%%
When 
$|\Psi^{\prime}_{\text{phys}}\rangle$ is the initial state given in \eqref{inistate1}, \eqref{coh} gives the equation,
%%%%
\begin{align}
0&=\Big(k^{\mu}\hat{a}_{\mu}(\bm{k})+\frac{\hat{\tilde{J}}^{0}(\bm{k})}{\sqrt{2k^{0}}}
\Big)|\Psi^{\prime}_{\text{phys}} \rangle 
\nonumber 
\\
&= \Big(k^{\mu}\hat{a}_{\mu}(\bm{k})+\frac{\hat{\tilde{J}}^{0}(\bm{k})}{\sqrt{2k^{0}}}
\Big) \frac{1}{\sqrt{2}}\big(|\text{R}\rangle + |\text{L} \rangle\big)\otimes |\alpha \rangle_\text{ph}
\nonumber 
\\
&
\approx \frac{1}{\sqrt{2}}\big(|\text{R}\rangle + |\text{L} \rangle\big) \otimes  \Big(k^{\mu}\hat{a}_{\mu}(\bm{k})+\frac{\tilde{J}^{0}(\bm{k})}{\sqrt{2k^{0}}}
\Big) |\alpha \rangle_\text{ph},
\end{align}
%%%%
where the approximation \eqref{approx1} was used in the second line, and note that 
$\tilde{J}^0_\text{R}(\bm{k})=\tilde{J}^0_\text{L}(\bm{k})=\tilde{J}^0(\bm{k})$ at the initial time. 
Hence the initial coherent state of the photon field must satisfy
%%%%
\begin{align}
\label{coh2}
\Big(k^{\mu}\hat{a}_{\mu}(\bm{k})+\frac{\tilde{J}^{0}(\bm{k})}{\sqrt{2k^{0}}}
\Big) |\alpha\rangle_\text{ph}=0. 
\end{align}
%%%%
Because the displacement operator $\hat{D}(\alpha)$ given in \eqref{D} has the following relation
\begin{align}
\hat{D}^{\dagger}(\alpha)\hat{a}_{\mu}(\bm{k})\hat{D}(\alpha)=\hat{a}_{\mu}(\bm{k})+\alpha_{\mu}(\bm{k}),
\label{DdgaD}
\end{align}
we obtain the constraint for the complex function 
$\alpha^\mu(\bm{k})$ as 
%%%%
\begin{equation}
k^{\mu} \alpha_{\mu}(\bm{k})=-\frac{\tilde{J}^{0}(\bm{k})}{\sqrt{2k^{0}}}. 
\end{equation}
%%%%
This is the BRST condition for the model of a single charged particle. 
The BRST condition for the model of two charged particles is obtained using the same procedure. 

\section{Computation of the inner product in Eq. \eqref{rhop} and derivation of Eqs. \eqref{GammaPP'} and \eqref{PhiPP'}}
\label{sec:Tr}

Here, we compute the inner product 
${}_\text{ph} \langle\alpha|\hat{U}^{\dagger}_{\text{P}'}\hat{U}_{\text{P}}|\alpha\rangle_\text{ph}$ in Eq. \eqref{rhop}.
The inner product is rewritten as 
\begin{align}
{}_\text{ph}\langle\alpha|\hat{U}^{\dagger}_{\text{P}'}\hat{U}_{\text{P}}|\alpha\rangle_\text{ph}
&={}_{\text{ph}}\langle0|\hat{D}^{\dagger}(\alpha)\hat{U}^{\dagger}_{\text{P}'}\hat{D}(\alpha)\hat{D}^{\dagger}(\alpha)\hat{U}_{\text{P}}\hat{D}(\alpha)|0\rangle_{\text{ph}}
\nonumber\\
\quad
&={}_{\text{ph}}\langle0|(\hat{D}^{\dagger}(\alpha)\hat{U}_{\text{P}'}\hat{D}(\alpha))^{\dagger}(\hat{D}^{\dagger}(\alpha)\hat{U}_{\text{P}}\hat{D}(\alpha))|0\rangle_{\text{ph}},
\label{inn}
\end{align}
where we used $|\alpha\rangle=\hat{D}(\alpha)|0\rangle_{\text{ph}}$, and the identity operator 
$\hat{I}=\hat{D}(\alpha)\hat{D}^{\dagger}(\alpha)$ was inserted between the unitary operators 
$\hat{U}^{\dagger}_{P'}$ and 
$\hat{U}_{\text{P}}$ in the first equality.
Because the displacement operator 
$\hat{D}(\alpha)$ satisfies Eq. \eqref{DdgaD}, we obtain  
\begin{align}
\hat{D}^{\dagger}(\alpha)\hat{A}^{\text{I}}_{\mu}(x)\hat{D}(\alpha)=\hat{A}^{\text{I}}_{\mu}(x)+A_{\mu}(x),
\label{dicop}
\end{align}
where 
$A_\mu (x)$ is defined in Eq. \eqref{A}. 
Subsequently, we obtain
\begin{align}
\hat{D}^{\dagger}(\alpha)\hat{U}_{\text{P}}(x)\hat{D}(\alpha)
&=\exp \left[-\frac{i}{2} \int d^{4} x \int d^{4} y J_{\text{P}}^{\mu}(x) J_{\text{P}}^{\nu}(y) G_{\mu \nu}^{\text{r}}(x, y)\right]\hat{D}^{\dagger}(\alpha)\exp \left[-i \int d^{4} x J_{\text{P}}^{\mu}(x) \hat{A}_{\mu}^{\text{I}}(x)\right]\hat{D}(\alpha)
\nonumber\\
\quad
&=\exp \left[-\frac{i}{2} \int d^{4} x \int d^{4} y J_{\text{P}}^{\mu}(x) J_{\text{P}}^{\nu}(y) G_{\mu \nu}^{\text{r}}(x, y)\right]\exp \left[-i \int d^{4} x J_{\text{P}}^{\mu}(x) \hat{D}^{\dagger}(\alpha)\hat{A}_{\mu}^{\text{I}}(x)\hat{D}(\alpha)\right]
\nonumber\\
\quad
&=\exp \left[-\frac{i}{2} \int d^{4} x \int d^{4} y J_{\text{P}}^{\mu}(x) J_{\text{P}}^{\nu}(y) G_{\mu \nu}^{\text{r}}(x, y)-i \int d^{4} x J_{\text{P}}^{\mu}(x) {A}_{\mu}(x)\right]
\nonumber\\
\quad
&\times\exp \left[-i \int d^{4} x J_{\text{P}}^{\mu}(x) \hat{A}_{\mu}^{\text{I}}(x)\right],
\end{align}
where the formula of the unitary operator 
$\hat{U}_\text{P}$ 
\eqref{unitaryP} was substituted and $G^{\text{r}}_{\mu \nu}(x, y)$ denotes the retarded Green's function given in Eq. \eqref{ret}. 
In the third equality we used Eq. \eqref{dicop}.
We further obtain
\begin{align}
&(\hat{D}^{\dagger}(\alpha)\hat{U}_{\text{P}'}\hat{D}(\alpha))^{\dagger}(\hat{D}^{\dagger}(\alpha)\hat{U}_{\text{P}}\hat{D}(\alpha))
\nonumber 
\\
&
=\exp \left[\frac{i}{2} \int d^{4} x \int d^{4} y \big(J_{\text{P}'}^{\mu}(x) J_{\text{P}'}^{\nu}(y)-J_{\text{P}}^{\mu}(x) J_{\text{P}}^{\nu}(y)\big) G_{\mu \nu}^{\text{r}}(x, y)+i \int d^{4} x \big(J_{\text{P}'}^{\mu}(x)-J_{\text{P}}^{\mu}(x)\big) {A}_{\mu}(x)\right]
\nonumber
\\
&
\quad \times 
\exp \left[i \int d^{4} x J_{\text{P}'}^{\mu}(x) \hat{A}_{\mu}^{\text{I}}(x)\right]\exp \left[-i \int d^{4} x J_{\text{P}}^{\mu}(x) \hat{A}_{\mu}^{\text{I}}(x)\right]
\nonumber
\\
&
=\exp \left[\frac{i}{2} \int d^{4} x \int d^{4} y \big(J_{\text{P}'}^{\mu}(x) J_{\text{P}'}^{\nu}(y)-J_{\text{P}}^{\mu}(x) J_{\text{P}}^{\nu}(y)\big) G_{\mu \nu}^{\text{r}}(x, y)+i \int d^{4} x \big(J_{\text{P}'}^{\mu}(x)-J_{\text{P}}^{\mu}(x)\big) {A}_{\mu}(x)\right]
\nonumber
\\
&
\quad \times 
\exp \left[i \int d^{4} x \big(J_{\text{P}'}^{\mu}(x)-J_{\text{P}}^{\mu}(x)\big) \hat{A}_{\mu}^{\text{I}}(x)+\frac{1}{2}\int d^4x d^4yJ_{\text{P}'}^{\mu}(x)J_{\text{P}}^{\nu}(y)[\hat{A}^{\text{I}}_{\mu}(x), \hat{A}^{\text{I}}_{\nu}(y)]\right]
\nonumber
\\
&
=\exp \left[\frac{i}{2} \int d^{4} x \int d^{4} y \big(J_{\text{P}'}^{\mu}(x) J_{\text{P}'}^{\nu}(y)-J_{\text{P}}^{\mu}(x) J_{\text{P}}^{\nu}(y)\big) G_{\mu \nu}^{\text{r}}(x, y)+i \int d^{4} x \big(J_{\text{P}'}^{\mu}(x)-J_{\text{P}}^{\mu}(x)\big) {A}_{\mu}(x)\right]
\nonumber
\\
&
\quad \times 
\exp \left[i \int d^{4} x \big(J_{\text{P}'}^{\mu}(x)-J_{\text{P}}^{\mu}(x)\big) \hat{A}_{\mu}^{\text{I}}(x)+\frac{i}{2}\int d^4x d^4y\big(J_{\text{P}'}^{\mu}(x)J_{\text{P}}^{\nu}(y)-J_{\text{P}'}^{\nu}(y)J_{\text{P}}^{\mu}(x)\big)G^{\text{r}}_{\mu \nu}(x, y)\right]
\nonumber 
\\
&=
\exp\left[i\int d^{4} x \big(J_{\text{P}'}^{\mu}(x)-J_{\text{P}}^{\mu}(x)\big) {A}_{\mu}(x)+\frac{i}{2} \int d^{4} x \int d^{4} y \big(J_{\text{P}'}^{\mu}(x) -J_{\text{P}}^{\mu}(x)\big)\big(J_{\text{P}'}^{\nu}(y)+J_{\text{P}}^{\nu}(y)\big)G_{\mu \nu}^{\text{r}}(x, y)\right]
\nonumber
\\
&
\quad \times 
\exp \left[i \int d^{4} x \big(J_{\text{P}'}^{\mu}(x)-J_{\text{P}}^{\mu}(x)\big) \hat{A}_{\mu}^{\text{I}}(x)\right]
\nonumber
\\
&=\exp\left[i\Phi_{\text{P}'\text{P}}+i \hat{\Theta}_{\text{PP}'}\right],
\label{duud}
\end{align}
where the Baker–Campbell–Hausdorff formula 
$e^{\hat{A}}e^{\hat{B}}=e^{\hat{A}+\hat{B}+[\hat{A},\hat{B}]/2 + \cdots}$ was used in the second equality, and the relation $[\hat{A}^{\text{I}}_{\mu}(x), \hat{A}^{\text{I}}_{\nu}(y)]=iG^{\text{r}}_{\mu \nu}(x, y)-iG^{\text{r}}_{\nu }(y, x)$ was substituted in the third equality. 
$``\cdots"$ in the Baker–Campbell–Hausdorff formula 
%It should be noted that $``\cdots"$
indicates the terms involving the higher commutators of $\hat{A}$ and $\hat{B}$. 
In our case, the commutator $[\hat{A}^{\text{I}}_{\mu}(x), \hat{A}^{\text{I}}_{\nu}(y)]$ is proportional to the identity operator, so the higher commutators vanish.
In the last equality, we defined 
$\hat{\Theta}_{\text{PP}'}$ and 
$\Phi_{\text{P}'\text{P}}$ as
\begin{align}
\hat{\Theta}_{\text{PP}'}&=\int d^{4} x \big(J_{\text{P}'}^{\mu}(x)-J_{\text{P}}^{\mu}(x)\big) \hat{A}_{\mu}^{\text{I}}(x), 
\label{ThetaPP'}
\\
\Phi_{\text{P}'\text{P}}&=\int d^{4} x \big(J_{\text{P}'}^{\mu}(x)-J_{\text{P}}^{\mu}(x)\big) {A}_{\mu}(x)+\frac{1}{2} \int d^{4} x \int d^{4} y \big(J_{\text{P}'}^{\mu}(x) -J_{\text{P}}^{\mu}(x)\big)\big(J_{\text{P}'}^{\nu}(y)+J_{\text{P}}^{\nu}(y)\big)G_{\mu \nu}^{\text{r}}(x, y).
\end{align}
Using the cumulant expansion for a given
%the 
density matrix 
$\rho$,
\begin{align}
\langle e^{i\lambda \hat{A}}\rangle_\rho
&=\text{Tr}[\rho e^{i\lambda \hat{A}}]=\exp\left[i\lambda \langle\hat{A}\rangle_\rho-\frac{1}{2}\lambda^2 \langle\big(\hat{A}-\langle\hat{A}\rangle_\rho \big)^2\rangle_\rho + \cdots \right],
\end{align}
where $\lambda$ is a c-number parameter, $\hat{A}$ is an operator, and 
$``\cdots"$ is the term with the third or higher cumulant, we can compute the inner product \eqref{inn} as  
\begin{align}
\ _{\text{ph}}\langle0|\hat{D}^{\dagger}(\alpha)\hat{U}^{\dagger}_{\text{P}'}\hat{U}_{\text{P}}\hat{D}(\alpha)|0\rangle_{\text{ph}}
&=e^{i\Phi_{\text{P}'\text{P}}}\ _{\text{ph}}\langle0|e^{i \hat{\Theta}_{\text{PP}'}}|0\rangle_{\text{ph}}
\nonumber 
\\
&
=e^{i\Phi_{\text{P}'\text{P}}}
\exp\left[i \langle \hat{\Theta}_{\text{PP}'} \rangle 
-\frac{1}{2}\langle (\hat{\Theta}_{\text{PP}'}-\langle \hat{\Theta}_{\text{PP}'} \rangle)^2\rangle + \cdots   \right]
\nonumber 
\\
&
=e^{i\Phi_{\text{P}'\text{P}}}\exp\left[-\frac{1}{2}\int d^4xd^4y\big(J_{\text{P}'}^{\mu}(x)-J_{\text{P}}^{\mu}(x)\big)\big(J_{\text{P}'}^{\nu}(y)-J_{\text{P}}^{\nu}(y)\big)\ _{\text{ph}}\langle0|\hat{A}_{\mu}^{\text{I}}(x)\hat{A}_{\nu}^{\text{I}}(y)|0\rangle_{\text{ph}}\right]
\nonumber\\
\quad
&=
e^{-\Gamma_{\text{P}'\text{P}}+i\Phi_{\text{P}'\text{P}}}.
\end{align}
We used Eq. \eqref{duud} and the cumulant expansion with 
$\rho=|0\rangle_\text{ph} \langle 0|$,
$\lambda=1$ and $\hat{A}=\hat{\Theta}_{\text{PP}'}$ in the first and second lines, respectively. 
$\langle \cdot \rangle$ denotes the vacuum expectation value. 
In the third equality, we substituted Eq. \eqref{ThetaPP'}, and the term
$``\cdots"$ with the $n$-th cumulant for $n\geq 3$ vanishes because the free vacuum state 
$|0\rangle_\text{ph}$ is Gaussian. 
In the last equality, we defined 
$\Gamma_{\text{P}'\text{P}}$ as
\begin{align}
\Gamma_{\text{P}'\text{P}}
&=\frac{1}{2}\int d^4xd^4y\big(J_{\text{P}'}^{\mu}(x)-J_{\text{P}}^{\mu}(x)\big)\big(J_{\text{P}'}^{\nu}(y)-J_{\text{P}}^{\nu}(y)\big)\ _{\text{ph}}\langle0|\hat{A}_{\mu}^{\text{I}}(x)\hat{A}_{\nu}^{\text{I}}(y)|0\rangle_{\text{ph}}
\nonumber\\
\quad
&=\frac{1}{4}\int d^4xd^4y\big(J_{\text{P}'}^{\mu}(x)-J_{\text{P}}^{\mu}(x)\big)\big(J_{\text{P}'}^{\nu}(y)-J_{\text{P}}^{\nu}(y)\big)\langle \bigl\{\hat{A}^\text{I}_{\mu} (x), \hat{A}^\text{I}_{\nu} (y)\bigr\}\rangle.
\nonumber
\end{align}
Replacing the currents 
$J^\mu_\text{P}$ and $J^\mu_{\text{P}'}$ with 
$J^\mu_\text{PQ}$ and $J^\mu_{\text{P}'\text{Q}'}$ in the above procedure, we can also derive \eqref{density12} .

\section{Li\'{e}nard-Wiechert potentials and field strength}
\label{sec:LW}
In this section, we derive the field strength induced by a charged particle \cite{Jackson1962}. 
The current of a charged particle is given as a four-vector current in a covariant form with 
%%%%
\begin{equation}
J^\mu (x)=e \int d\tau \frac{dX^\mu}{d\tau} \delta^{(4)}(x-X(\tau)), 
\label{eq:pJ}
\end{equation}
%%%%
where 
$X^{\mu}(\tau)$ is the trajectory of the charged particle parameterized by a proper time 
$\tau$. 
Using this current and the retarded Green's function, 
%%%%
\begin{equation}
G^r_{\mu\nu} (x,y)=-\frac{\eta_{\mu\nu}}{4\pi |\bm{x}-\bm{y}|}
\delta \big(|\bm{x}-\bm{y}|-(x^0-y^0) \big), 
\label{eq:expGr}
\end{equation}
%%%%
we obtain the retarded potential as
%%%%
\begin{equation}
A^\mu (x)
= \int d^y G^\text{r} {}^\mu_\nu (x,y) J^\nu (y)
=
\frac{e}{4\pi} 
\frac{u^\mu (\tau_\text{r})}
{(x-X(\tau_\text{r})) \cdot u(\tau_\text{r})},
\end{equation}
%%%%
where $u^\mu= {dX^\mu}/{d\tau}$ is the four-velocity of the charge, and 
$\tau_\text{r}$ is determined by the light-cone condition
%%%%
\begin{equation}
-(t-X^0(\tau_\text{r}))+|\bm{x}-\bm{X} (\tau_\text{r})|=0. 
\label{eq:taur}
\end{equation}
%%%%
From the definition of the field strength $F^{\mu \nu}=\partial^{\mu}A^{\nu}-\partial^{\nu}A^{\mu}$, we obtain 
%%%%
\begin{align}
F^{\mu \nu}
&= F^{\mu \nu}_{\text{v}}+F^{\mu \nu}_{\text{a}},
\\
F^{\mu \nu}_{\text{v}}
&=-
\frac{e}
{4\pi } 
\frac{(x^\mu-X^\mu(\tau_\text{r})) u^\nu(\tau_\text{r})- (x^\nu-X^\nu(\tau_\text{r})) u^\mu(\tau_\text{r})}
{ [ (x-X(\tau_\text{r})) \cdot u(\tau_\text{r}) ]^3 },
\label{eq:Fv}
\\
F^{\mu \nu}_{\text{a}}
&=
\frac{e}
{4\pi [ (x-X(\tau_\text{r})) \cdot u(\tau_\text{r}) ]^2 } 
\Big( 
(x^\mu-X^\mu(\tau_\text{r})) 
\big(
 \dot{u}^\nu(\tau_\text{r})
 -\frac{(x-X(\tau_\text{r})) \cdot \dot{u}(\tau_\text{r})}
{ (x-X(\tau_\text{r})) \cdot u(\tau_\text{r})  } u^\nu(\tau_\text{r}) 
\big)
\nonumber 
\\
&
\quad
-
(x^\nu-X^\nu(\tau_\text{r})) 
\big(
 \dot{u}^\mu(\tau_\text{r})
 -\frac{(x-X(\tau_\text{r})) \cdot \dot{u}(\tau_\text{r})}
{ (x-X(\tau_\text{r})) \cdot u(\tau_\text{r})  } u^\mu(\tau_\text{r}) 
\Big),
\label{eq:Fa}
\end{align}
%%%%
where $\dot{u}^\mu=d u^\mu /d\tau$ is the four-acceleration.
We use the coordinate time 
$t$ instead of the proper time 
$\tau$ to rewrite the above field strengths.
The four-vector and four-acceleration as a function of $t$ are
%%%%
\begin{equation}
u^\mu=\frac{dX^\mu}{d\tau}=\gamma \frac{dX^\mu}{dt}= \gamma v^\mu,
\quad
\dot{u}^\mu=\frac{du^\mu}{d\tau}=\gamma\frac{d{\gamma}}{dt}v^\mu + \gamma^2 a^\mu,
\label{eq:utov}
\end{equation}
%%%%
where 
$v^{\mu}$ and 
$a^{\mu}$ are the velocity and acceleration measured in the coordinate time $t$, and $\gamma$ is the Lorentz factor. 
These are defined by
%%%%
\begin{equation}
v^\mu
=\frac{dX^\mu}{dt}
=\Big[1, \frac{d\bm{X}}{dt} \Big]^\text{T},
\quad
a^\mu
=\frac{dv^\mu}{dt}
=\Big[0, \frac{d^2 \bm{X}}{dt^2} \Big]^\text{T},
\quad
\gamma=\frac{1}{\sqrt{-v^2}}=\frac{1}{\sqrt{1-\bm{v}^2}}.
\label{eq:vgamma}
\end{equation}
%%%%
We then determine the following retarded potential and its field strength as
%%%%
\begin{align}
A^\mu (x)
&=
\frac{e}{4\pi} 
\frac{v^\mu (t_\text{r})}
{(x-X(t_\text{r})) \cdot v(t_\text{r})},
\label{eq:A2}
\\
F^{\mu \nu}_{\text{v}}
&=-
\frac{e}
{4\pi} \frac{(x^\mu-X^\mu(t_\text{r})) v^\nu(t_\text{r})- (x^\nu-X^\nu(t_\text{r})) v^\mu(t_\text{r})}{\gamma^2 [ (x-X(t_\text{r})) \cdot v(t_\text{r}) ]^3},
\label{eq:Fv2}
\\
F^{\mu \nu}_{\text{a}}
&=
\frac{e}
{4\pi [ (x-X(t_\text{r})) \cdot v(t_\text{r}) ]^2 } 
\Big[ 
(x^\mu-X^\mu(t_\text{r}))
\Big(
a^\nu (t_\text{r})
-\frac{(x-X(t_\text{r})) \cdot a (t_\text{r})}{(x-X(t_\text{r})) \cdot v(t_\text{r})}
v^\nu(t_\text{r}) 
\Big)
\nonumber 
\\
&
\quad 
-(x^\nu-X^\nu(t_\text{r})) 
\Big(
a^\mu (t_\text{r})
-\frac{(x-X(t_\text{r})) \cdot a (t_\text{r})}{(x-X(t_\text{r})) \cdot v(t_\text{r})}
v^\mu(t_\text{r}) 
\Big)
 \Big] ,
\label{eq:Fa2}
\end{align}
%%%%
where the retarded time 
$t_\text{r}$ is given by
%%%%
\begin{equation}
-(t-t_\text{r})+|\bm{x}-\bm{X} (t_\text{r})|=0. 
\label{eq:tr}
\end{equation}

\section{\label{expansion}1/c expansion of $\Phi$}
We present the 
$1/c$ expansion of the quantity
%%%%
\begin{equation}
\Phi
=\frac{e}{2\hbar c} 
\Big (
\oint_{\text{C}_1} dx_\mu \Delta A^\mu_2 (x)
+\oint_{\text{C}_2} dx_\mu \Delta A^\mu_1 (x)
\Big),
\label{eq:PhiPhi}
\end{equation}
%%%%
where 
%%%%
\begin{equation}
\Delta A_i^\mu (x)
=\sum_{\text{P}=\text{R},\text{L}}
\epsilon_\text{P}
\frac{e}{4\pi} 
\Big
[\frac{v^\mu_{i\text{P}} (t_{i\text{P}})}
{(x-X_{i\text{P}}(t_{i\text{P}})) \cdot v_{i\text{P}}(t_{i\text{P}})}
\Big], 
\label{eq:DeltaAi}
\end{equation} 
%%%%
and  
$v^\mu = [c, \bm{v}]^\text{T}$, 
$\epsilon_\text{R}=1$, 
$\epsilon_\text{L}=-1$ and 
$t_{i\text{P}}$ satisfies the light cone condition $-c(t-t_{i\text{P}})+|\bm{x}-\bm{X}_{i\text{P}}(t_{i\text{P}})|=0$.
We restored the reduced Planck constant 
$\hbar$ and 
the light velocity 
$c$. 
Substituting \eqref{eq:DeltaAi} into \eqref{eq:PhiPhi}, 
we obtain 
%%%%
\begin{align}
\Phi
&=\frac{e^2}{8 \pi \hbar c} 
\Big (
\oint_{\text{C}_1} dx_\mu 
\sum_{\text{Q}=\text{R},\text{L}}
\epsilon_\text{Q} 
\Big[
\frac{v^\mu_{2\text{Q}} (t_{2\text{Q}})}
{(x-X_{2\text{Q}}(t_{2\text{Q}})) \cdot v_{2\text{Q}}(t_{2\text{Q}})}
\Big]
+(1 \leftrightarrow 2)
\Big) 
\nonumber 
\\
&=\frac{e^2}{8 \pi \hbar } 
\int dt 
\sum_{\text{P},\text{Q}=\text{R},\text{L}}
\epsilon_\text{P} \epsilon_\text{Q} 
\Big[
\frac{v_{1\text{P}}(t) \cdot v_{2\text{Q}} (t_{2\text{Q}})}
{c(X_{1\text{P}}(t)-X(t_{2\text{Q}})) \cdot v_{2\text{Q}}(t_{2\text{Q}})}
\Big]
+(1 \leftrightarrow 2),
\end{align}
%%%%
where we changed the integral as  
$\oint_{\text{C}_i} dx^{\mu}=\sum_{\text{P}=\text{R}, \text{L}}\epsilon_{\text{P}}\int (dX^\mu_{i\text{P}}/dt)dt=\sum_{\text{P}=\text{R}, \text{L}}\epsilon_{\text{P}}\int v^{\mu}_{i\text{P}}(t)dt\ (i=1,2)$ in the second line.
The integrands have the form
%%%%
\begin{align}
\frac{v_1 (t) \cdot v_2 (t_\text{r})}
{c(X_1(t)-X_2(t_\text{r})) \cdot v_2 (t_\text{r})}
&
=\frac{c^2-\bm{v}_1(t) \cdot \bm{v}_2 (t_\text{r})}
{c(-c(t-t_\text{r})+(\bm{X}_1 (t)-\bm{X}_2 (t_\text{r})) \cdot \bm{v}_2 (t_\text{r}))} 
\nonumber 
\\
&
=\frac{-1 }
{|\bm{X}_1(t)-\bm{X}_2 (t_\text{r})|- (\bm{X}_1 (t)-\bm{X}_2 (t_\text{r})) \cdot \bm{v}_2 (t_\text{r})/c} 
\Big(
1-\frac{\bm{v}_1(t) \cdot \bm{v}_2 (t_\text{r})}{c^2}
\Big),
\label{eq:integrand}
\end{align}
%%%% 
where the light cone condition 
$-c(t-t_\text{r})+|\bm{X}_1 (t)-\bm{X}_2 (t_\text{r})|=0$ was used in the second line. 
The 
$1/c$ expansion of the retarded time 
$t_\text{r}$ is 
%%%%
\begin{align}
t_\text{r}
&=t-\frac{1}{c} |\bm{X}_1 (t)-\bm{X}_2(t_\text{r})|
\nonumber 
\\
&=t-\frac{1}{c} \sqrt{(\bm{X}_1 (t)-\bm{X}_2(t_\text{r}))^2}
\nonumber 
\\
&=t-\frac{1}{c} \sqrt{
\Big(
\bm{X}_1 (t)-\bm{X}_2(t)+\frac{\bm{v}_2 (t)}{c} |\bm{X}_1-\bm{X}_2 (t)|
\Big)^2} + O\Big(\frac{1}{c^3}\Big)
\nonumber 
\\
&=t-\frac{1}{c} \sqrt{
(\bm{X}_1 (t)-\bm{X}_2(t))^2+(\bm{X}_1 (t)-\bm{X}_2 (t))\cdot \frac{2\bm{v}_2 (t)}{c} |\bm{X}_1 (t)-\bm{X}_2 (t)|
} + O\Big(\frac{1}{c^3}\Big)
\nonumber 
\\
&=t-\frac{|\bm{X}_1 (t)-\bm{X}_2 (t)|}{c} 
\Big(
1+\frac{\bm{X}_1 (t)-\bm{X}_2(t)}{|\bm{X}_1 (t)-\bm{X}_2(t)|} \cdot 
 \frac{\bm{v}_2 (t)}{c} 
\Big)
 + O\Big(\frac{1}{c^3}\Big)
\nonumber 
\\
&=t-\frac{r(t)}{c} 
-\bm{r}(t) \cdot \frac{\bm{v} (t)}{c^2} 
+ \mathcal{O}\Big(\frac{1}{c^3}\Big),
\label{eq;tr1/c}
\end{align}
%%%%
where 
$\bm{r}(t)=\bm{X}_1 (t)-\bm{X}_2(t)$ and
$r(t)=|\bm{r}(t)|$. 
The denominator of the integrand \eqref{eq:integrand} is
%%%%
\begin{align}
&
|\bm{X}_1(t)-\bm{X}_2 (t_\text{r})|- (\bm{X}_1 (t)-\bm{X}_2 (t_\text{r})) \cdot \frac{\bm{v}_2 (t_\text{r})}{c} 
\nonumber
\\
\quad
&=
\sqrt{(\bm{X}_1(t)-\bm{X}_2 (t_\text{r}))^2}- (\bm{X}_1 (t)-\bm{X}_2 (t_\text{r})) \cdot \frac{\bm{v}_2 (t_\text{r})}{c} 
\nonumber
\\
\quad
&=
\sqrt{
\Big(
\bm{r}
+\bm{v}_2 
\Big(
\frac{r}{c}+\frac{\bm{r} \cdot \bm{v}_2}{c^2} 
\Big)
- \frac{r^2 \bm{a}_2}{2c^2}
\Big)^2}
-\Big(
\bm{r}+\bm{v}_2 
\frac{r}{c}
\Big) \cdot \frac{1}{c} 
\Big(
\bm{v}_2 -\frac{r}{c} \bm{a}_2
\Big) 
+O\Big(\frac{1}{c^3} \Big)
\nonumber
\\
\quad
&=
\sqrt{
r^2
+2 \bm{r} \cdot \bm{v}_2 
\Big(
\frac{r}{c}+\frac{\bm{r} \cdot \bm{v}_2}{c^2} 
\Big)
- 2 \bm{r} \cdot \frac{r^2 \bm{a}_2}{2c^2}
+ \frac{r^2 v^2_2}{c^2}
}
-\Big(
\frac{\bm{r}\cdot \bm{v}_2}{c}+\frac{r v^2_2}{c^2} -
\frac{r}{c^2} \bm{r} \cdot \bm{a}_2 
\Big) 
+O\Big(\frac{1}{c^3} \Big)
\nonumber
\\
\quad
&=
r
\Big(
1
+ \frac{\bm{r} \cdot \bm{v}_2}{r^2} 
\Big(
\frac{r}{c}+\frac{\bm{r} \cdot \bm{v}_2}{c^2} 
\Big)
-  \bm{r} \cdot \frac{ \bm{a}_2}{2c^2}
+ \frac{v^2_2}{2c^2}
-\frac{(\bm{r} \cdot \bm{v}_2)^2}{2r^2c^2}
\Big)
-\Big(
\frac{\bm{r}\cdot \bm{v}_2}{c}+\frac{r v^2_2}{c^2} -
\frac{r}{c^2} \bm{r} \cdot \bm{a}_2 
\Big) 
+O\Big(\frac{1}{c^3} \Big)
\nonumber
\\
\quad
&=
r \Big[
1
+ \frac{\bm{r} \cdot \bm{v}_2}{r^2} 
\Big(
\frac{r}{c}+\frac{\bm{r} \cdot \bm{v}_2}{c^2} 
\Big)
-  \bm{r} \cdot \frac{ \bm{a}_2}{2c^2}
+ \frac{v^2_2}{2c^2}
-\frac{(\bm{r} \cdot \bm{v}_2)^2}{2r^2c^2}
-
\frac{\bm{r}\cdot \bm{v}_2}{rc}
-\frac{v^2_2}{c^2} 
+\frac{\bm{r} \cdot \bm{a}_2 }{c^2} 
\Big]
\nonumber
\\
\quad
&=
r \Big[
1
+\frac{(\bm{r} \cdot \bm{v}_2)^2}{2r^2c^2}
-\frac{v^2_2}{2c^2} 
+\frac{\bm{r} \cdot \bm{a}_2 }{2c^2} 
\Big]
+\mathcal{O}\Big(\frac{1}{c^3} \Big),
\label{eq:den}
\end{align}
%%%%
and the numerator of \eqref{eq:integrand} is
%%%%
\begin{equation}
1-\frac{\bm{v}_1(t) \cdot \bm{v}_2 (t_\text{r})}{c^2}=1-\frac{\bm{v}_1 \cdot \bm{v}_2 }{c^2}+\mathcal{O}\Big(\frac{1}{c^3} \Big),
\label{eq:num}
\end{equation}
%%%%
where the light cone condition and the Taylor expansion were used and the argument $t$ was omitted.
Then, \eqref{eq:integrand} reduces to
%%%%
\begin{align}
&\frac{v^\mu_{1}(t)v_2 {}_\mu (t_\text{r})}
{c(X_1(t)-X_2(t_\text{r})) \cdot v_2 (t_\text{r})}
\nonumber 
\\
\quad
&
=\frac{-1 }
{|\bm{X}_1(t)-\bm{X}_2 (t_\text{r})|- (\bm{X}_1 (t)-\bm{X}_2 (t_\text{r})) \cdot \bm{v}_2 (t_\text{r})/c} 
\Big(
1-\frac{\bm{v}_1(t) \cdot \bm{v}_2 (t_\text{r})}{c^2}
\Big)
\nonumber 
\\
\quad
&
=\frac{-1 }
{r \Big[
1
+\frac{(\bm{r} \cdot \bm{v}_2)^2}{2r^2c^2}
-\frac{v^2_2}{2c^2} 
+\frac{\bm{r} \cdot \bm{a}_2 }{2c^2} 
\Big]} 
\Big(
1-\frac{\bm{v}_1 \cdot \bm{v}_2 }{c^2}
\Big)+O\Big(\frac{1}{c^3} \Big)
\nonumber 
\\
\quad
&
=-\frac{1}{r} 
\Big[
1
-\frac{(\bm{r} \cdot \bm{v}_2)^2}{2r^2c^2}
+\frac{v^2_2}{2c^2} 
-\frac{\bm{r} \cdot \bm{a}_2 }{2c^2} 
-\frac{\bm{v}_1 \cdot \bm{v}_2 }{c^2}
\Big] 
+ O\Big(\frac{1}{c^3} \Big)
\nonumber 
\\
\quad
&
\approx -\frac{1}{|\bm{X}_1 -\bm{X}_2 |} 
\Big[
1-\frac{\bm{v}_1 \cdot \bm{v}_2 }{c^2}
+\frac{1}{2c^2} 
\Big \{
v^2_2-\Big(\frac{\bm{X}_1 -\bm{X}_2}{|\bm{X}_1 -\bm{X}_2 |} \cdot \bm{v}_2 \Big)^2
\Big \}
-\frac{(\bm{X}_1 -\bm{X}_2) \cdot \bm{a}_2 }{2c^2} 
\Big].
\label{eq:integrand2}
\end{align}
%%%% 
We find that the 
$1/c$ expansion of 
$\Phi$ is 
%%%%
\begin{align}
\Phi
&=\frac{e^2}{8 \pi \hbar } 
\int dt 
\sum_{\text{P},\text{Q}=\text{R},\text{L}}
\epsilon_\text{P} \epsilon_\text{Q} 
\Big[
\frac{v_{1\text{P}}(t) \cdot v_{2\text{Q}} (t_{2\text{Q}})}
{c(X_{1\text{P}}(t)-X(t_{2\text{Q}})) \cdot v_{2\text{Q}}(t_{2\text{Q}})}
\Big]
+(1 \leftrightarrow 2)
\nonumber 
\\
&\approx 
-\frac{e^2}{8 \pi \hbar } 
\int dt 
\sum_{\text{P},\text{Q}=\text{R},\text{L}}
\frac{\epsilon_\text{P} \epsilon_\text{Q} }{|\bm{X}_{1\text{P}} -\bm{X}_{2\text{Q}} |} 
\Big[
1-\frac{\bm{v}_{1\text{P}} \cdot \bm{v}_{2\text{Q}} }{c^2}
\nonumber 
\\
&
\quad
+\frac{1}{2c^2} 
\Big \{
v^2_{2\text{Q}}-\Big(\frac{\bm{X}_{1\text{P}} -\bm{X}_{2\text{Q}}}{|\bm{X}_{1\text{P}} -\bm{X}_{2\text{Q}} |} \cdot \bm{v}_{2\text{Q}} \Big)^2
\Big \}
-\frac{(\bm{X}_{1\text{P}} -\bm{X}_{2\text{Q}}) \cdot \bm{a}_{2\text{Q}} }{2c^2} 
\Big]
+(1 \leftrightarrow 2).
\label{phiex}
\end{align}
%%%%
For the non-relativistic limit $c\rightarrow \infty$, the quantity $\Phi$ is
\begin{align}
\Phi 
&\rightarrow 
-\frac{e^2}{8 \pi \hbar } 
\int dt 
\sum_{\text{P},\text{Q}=\text{R},\text{L}}
\frac{\epsilon_\text{P} \epsilon_\text{Q} }{|\bm{X}_{1\text{P}} -\bm{X}_{2\text{Q}} |} 
+(1 \leftrightarrow 2)
\nonumber 
\\
&=
-\frac{e^2}{4 \pi \hbar } 
\int dt 
\Big(
\frac{1}{|\bm{X}_{1\text{R}} -\bm{X}_{2\text{R}} |}- \frac{1}{|\bm{X}_{1\text{R}} -\bm{X}_{2\text{L}}|}-
\frac{1}{|\bm{X}_{1\text{L}} -\bm{X}_{2\text{R}}|}
+\frac{1}{|\bm{X}_{1\text{L}} -\bm{X}_{2\text{L}}|}
\Big)
.
\label{nphiex}
\end{align}
This result is equivalent to the quantity \eqref{phic} (in the unit $\hbar=1$) computed in the non-relativistic regime.

\section{\label{detailc}Detail derivation of $\Gamma_{\text{RL}}$,\ $\Gamma_{1}$,\ $\Gamma_{2}$,\ $\Gamma_{\text{c}}$ and $\Phi$}

We present the detailed calculation of $\Gamma_{\text{RL}}, \Gamma_{\text{1}}, \Gamma_{2}, \Gamma_{\text{c}}$, and $\Phi$.
In this calculation, we assume that the charged particle has the non-relativistic velocity.
We recover the constants $c$ and $\hbar$ when we show the result of the calculation or use the formula of the $1/c$ expansion of $\Phi$ derived as \eqref{phiex}.

\subsection{\label{gamma12}Computations of $\Gamma_{\text{RL}}$,\ $\Gamma_{1}$ and $\Gamma_{{2}}$}

We first calculate the quantity $\Gamma_{\text{RL}}$.
We assume the following trajectories
\begin{equation}
X^\mu_\text{P} (t)  =[t,\epsilon_{\text{P}}X(t), 0,0 \Big]^\text{T},
\quad 
\epsilon_{\text{R}}=-\epsilon_{\text{L}}=1, 
\quad 
X(t)=8L\Big(1-\frac{t}{T} \Big)^2 \Big(\frac{t}{T}\Big)^2.
\label{xt}
\end{equation}
Using Eq. \eqref{gammaRL}, we obtain
\begin{align}
\Gamma_{\text{RL}}
&=\frac{e^2}{4}\oint_\text{C} dx^\mu \oint_\text{C} dy^\mu \langle \bigl\{\hat{A}^\text{I}_\mu (x), \hat{A}^\text{I}_\nu (y) \bigr\}\rangle \nonumber \\
&\approx
\frac{e^2}{4}\oint_\text{C} dx^\mu \oint_\text{C} dy^\mu \langle \bigl\{\hat{A}^\text{I}_\mu (x^0, \bm{0}), \hat{A}^\text{I}_\nu (y^0, \bm{0}) \bigr\}\rangle \nonumber \\
&=
\frac{e^2}{4}\oint_\text{C} dx^\mu \oint_\text{C} dy^\mu \frac{\eta_{\mu \nu} }{4\pi^2} \Big(\frac{1}{-(t-t'-i\epsilon)^2}+\frac{1}{-(t-t'+i\epsilon)^2} \Big) \nonumber \\
&=
\frac{e^2}{16\pi^2}\int^T_0 dt \Big(\frac{d X^\mu_\text{R}}{dt}-\frac{d X^\mu_\text{L}}{dt}\Big) \int^T_0 dt'\Big(\frac{d X_{\text{R}\,\mu}}{dt'}-\frac{d X_{\text{L}\,\mu}}{dt'}\Big) \Big(\frac{1}{-(t-t'-i\epsilon)^2}+\frac{1}{-(t-t'+i\epsilon)^2} \Big) \nonumber \\
&=\frac{e^2}{16 \pi^2}\int^T_0 dt \int^T_0 dt'\Big(\frac{d \bm{X}_\text{R}}{dt}-\frac{d \bm{X}_\text{L}}{dt}\Big) \cdot\Big(\frac{d \bm{X}_\text{R}}{dt'}-\frac{d \bm{X}_\text{L}}{dt'}\Big) \Big(\frac{1}{-(t-t'-i\epsilon)^2}+\frac{1}{-(t-t'+i\epsilon)^2} \Big)\nonumber\\
&=
\frac{32e^2}{3\pi^2} \frac{L^2}{T^2},
\end{align}
where we took the limit $\epsilon \rightarrow 0$ after the integration, and in the second line we used the dipole approximation \cite{Mazzitelli, Hsiang} which ignores the spatial dependence of the photon field.
The dipole approximation is valid when the wave length of the photon field $\lambda_{\text{p}}=T$ is considerably larger than the typical size ($\sim L$) of the region  where the charge exists.
This condition is always satisfied if we assume the non-relativistic velocity $L/T \ll 1$.

We next consider the quantity 
$\Gamma_i$ \eqref{gammai} given in the model of two charged particles. 
Because of the time and spatial translation invariance of the vacuum state, $\Gamma_i$ is independent of the choice of the origin.
Assuming that each of the charged particles 1 and 2 follows the trajectories defined by \eqref{xt} up to the choice of the origin of the time or spatial axis, we can evaluate 
$\Gamma_1$ and 
$\Gamma_2$ as 
%%%%
\begin{align}
\Gamma_1=\Gamma_2=\Gamma_{\text{RL}}\approx \frac{32e^2}{3\pi^2\hbar c} \frac{L^2}{(cT)^2},
\end{align}
%%%%
where we recovered the constants 
$c$ and $\hbar$.

\subsection{\label{cphil}Computations of $\Gamma_{\text{c}}$ and $\Phi$ for the linear configuration}
\subsubsection{\label{cphilTDL}$T \gg D \sim L$ or $T \gg D \gg L$ regimes}
Here, we focus on the regime $T \gg D \sim L$ or $T \gg D \gg L$ for the linear configuration. 
We assume the trajectories of two charged particles 1 and 2 as follows
\begin{equation}
X^\mu_{1\text{P}}=[t,\epsilon_{\text{P}}X(t), 0,0]^\text{T}, \quad 
X^\mu_{2\text{Q}}(t)=[t,\epsilon_{\text{Q}}X(t)+D,0,0 \Big]^\text{T},
\quad 
\epsilon_{\text{R}}=-\epsilon_{\text{L}}=1, 
\quad 
X(t)=8L\Big(1-\frac{t}{T} \Big)^2 \Big(\frac{t}{T}\Big)^2.
\label{linearlTDL}
\end{equation}
The parameters $L$ and $D$ should be $D>L\geq2X(t)$ to avoid overlapping each trajectory of particles 1 and 2.
First, we focus on the regime $T \gg D \sim L$.
The quantity $\Gamma_{\text{c}}$ is computed by Eq. \eqref{gammac} as
\begin{align}
\Gamma_{c}
&=\frac{e^2}{2}\oint_{\text{C}_1} dx^\mu \oint_{\text{C}_2} dy^\nu  
\, 
\langle \{\hat{A}^\text{I}_{\mu}(x), \hat{A}^\text{I}_{\nu}(y)\}\rangle
\nonumber 
\\
&\approx 
\frac{e^2}{2}\oint_{\text{C}_1} dx^\mu \oint_{\text{C}_2} dy^\nu  
\, 
\langle \{\hat{A}^\text{I}_{\mu}(x^0,\bm{0}), \hat{A}^\text{I}_{\nu}(y^0, \bm{0})\}\rangle
\nonumber 
\\
&=\frac{e^2}{2}\oint_{\text{C}_1} dx^\mu \oint_{\text{C}_2} dy^\nu  
\, 
\frac{\eta_{\mu \nu} }{4\pi^2} 
\Big(
\frac{1}{-(x^0-y^0-i\epsilon)^2}
+\frac{1}{-(x^0-y^0+i\epsilon)^2} 
\Big)
\nonumber 
\\
&=
\frac{e^2}{8\pi^2}\int^T_0 dt 
\Big( 
\frac{d X^\mu_{1\text{R}}}{dt}
-\frac{d X^\mu_{1\text{L}}}{dt} 
\Big) 
\int^{T}_0 dt' 
\Big( 
\frac{d X_{2\text{R}\,\mu}}{dt'}
-\frac{d X_{2\text{L}\,\mu}}{dt'} 
\Big)
\Big(
\frac{1}{-(t-t'-i\epsilon)^2}
+\frac{1}{-(t-t'+i\epsilon)^2} 
\Big)
\nonumber 
\\
&=
\frac{e^2}{8\pi^2}\int^T_0 dt \int^{T}_0 dt' 
\Big( 
\frac{d \bm{X}_{1\text{R}}}{dt}
-\frac{d \bm{X}_{1\text{L}}}{dt} 
\Big) 
\cdot 
\Big( 
\frac{d \bm{X}_{2\text{R}}}{dt'}
-\frac{d \bm{X}_{2\text{L}}}{dt'} 
\Big) 
\Big(
\frac{1}{-(t-t'-i\epsilon)^2}
+\frac{1}{-(t-t'+i\epsilon)^2} 
\Big)
\nonumber 
\\
&
= \frac{64e^2}{3\pi^2} \frac{L^2}{T^2},
\label{eq:GammacA3}
\end{align}
where the dipole approximation was used in the second line because of the condition $T \gg L$.
The quantity $\Phi$ is evaluated using the result of \eqref{phiex} as
\begin{align}
\Phi
&=
-\frac{e^2}{8 \pi \hbar } 
\int dt 
\sum_{\text{P},\text{Q}=\text{R},\text{L}}
\frac{\epsilon_\text{P} \epsilon_\text{Q} }
{|\bm{X}_{1\text{P}} -\bm{X}_{2\text{Q}} |} 
\Big[1-\frac{\bm{v}_{1\text{P}} \cdot \bm{v}_{2\text{Q}} }{c^2}
\nonumber 
\\
&
\quad
+\frac{1}{2c^2} 
\Big \{v^2_{2\text{Q}}-\Big(\frac{\bm{X}_{1\text{P}} -\bm{X}_{2\text{Q}}}{|\bm{X}_{1\text{P}} -\bm{X}_{2\text{Q}} |} \cdot \bm{v}_{2\text{Q}} \Big)^2\Big\}
-\frac{(\bm{X}_{1\text{P}}-\bm{X}_{2\text{Q}}) \cdot \bm{a}_{2\text{Q}} }{2c^2} \Big]+(1 \leftrightarrow 2)\nonumber\\
&=
-\frac{e^2}{8 \pi \hbar } 
\int dt 
\sum_{\text{P},\text{Q}=\text{R},\text{L}}
\frac{\epsilon_\text{P} \epsilon_\text{Q} }{|D-(\epsilon_{\text{P}}-\epsilon_{\text{Q}})X(t)| }
\Big[
1-\epsilon_{\text{P}}\epsilon_{\text{Q}}\frac{v^2(t)}{c^2}
-\epsilon_{\text{Q}}\frac{\{-D+(\epsilon_{\text{P}}-\epsilon_{\text{Q}})X(t)\}  a(t) }{2c^2} 
\Big]
+(1 \leftrightarrow 2)
\nonumber 
\\
&=
-\frac{e^2}{4 \pi \hbar } 
\int dt 
\Big[
\frac{2}{D}
\Big(
 1-\frac{v^2}{c^2}
\Big)
-
\Big(
 1+\frac{v^2}{c^2}
\Big)
\Big(\frac{1}{|D-2X(t)|}+\frac{1}{|D+2X(t)|}\Big)
+\frac{a(t)}{2c^2}\Big(\frac{D-2X(t)}{|D-2X(t)|}-\frac{D+2X(t)}{|D+2X(t)|}\Big)\Big]
\nonumber 
\\
&=
-\frac{e^2}{4 \pi \hbar } 
\int dt 
\Big[
\frac{2}{D}
\Big(
 1-\frac{v^2}{c^2}
\Big)
-
\Big(
 1+\frac{v^2}{c^2}
\Big)
\Big(\frac{1}{D-2X(t)}+\frac{1}{D+2X(t)}\Big)
\Big],
\label{phicex}
\end{align}
where we have recovered the natural units $c$ and $\hbar$ to show the result of the $1/c$ expansion.
Next, we consider the regime $T \gg D \gg L$.
In this regime, we obtain the $\Gamma_{\text{c}}$ and $\Phi$ using \eqref{gammac} and \eqref{phicex} as follows,
\begin{align}
\Gamma_{c}
&=\frac{e^2}{2}\oint_{\text{C}_1} dx^\mu \oint_{\text{C}_2} dy^\nu  
\, 
\langle \{\hat{A}^\text{I}_{\mu}(x), \hat{A}^\text{I}_{\nu}(y)\}\rangle
\nonumber 
\\
&=\frac{e^2}{2}\oint_{\text{C}_1} dx^\mu \oint_{\text{C}_2} dy^\nu  
\, 
\frac{\eta_{\mu \nu} }{4\pi^2} 
\Big(
\frac{1}{-(x^0-y^0-i\epsilon)^2+|\bm{x}-\bm{y}|^2}
+\frac{1}{-(x^0-y^0+i\epsilon)^2+|\bm{x}-\bm{y}|^2} 
\Big)
\nonumber 
\\
&\approx 
\frac{e^2}{2}\oint_{\text{C}_1} dx^\mu \oint_{\text{C}_2} dy^\nu  
\, 
\frac{\eta_{\mu \nu} }{4\pi^2} 
\Big(
\frac{1}{-(x^0-y^0-i\epsilon)^2+D^2}
+\frac{1}{-(x^0-y^0+i\epsilon)^2+D^2} 
\Big)
\nonumber 
\\
&=
\frac{e^2}{8\pi^2}\int^T_0 dt 
\Big( 
\frac{d X^\mu_{1\text{R}}}{dt}
-\frac{d X^\mu_{1\text{L}}}{dt} 
\Big) 
\int^{T}_0 dt' 
\Big( 
\frac{d X_{2\text{R}\,\mu}}{dt'}
-\frac{d X_{2\text{L}\,\mu}}{dt'} 
\Big) 
\Big(
\frac{1}{-(t-t'-i\epsilon)^2+D^2}
+\frac{1}{-(t-t'+i\epsilon)^2+D^2} 
\Big)
\nonumber 
\\
&=
\frac{e^2}{8\pi^2}\int^T_0 dt \int^{T}_0 dt' 
\Big( 
\frac{d \bm{X}_{1\text{R}}}{dt}
-\frac{d \bm{X}_{1\text{L}}}{dt} 
\Big) 
\cdot 
\Big( 
\frac{d \bm{X}_{2\text{R}}}{dt'}
-\frac{d \bm{X}_{2\text{L}}}{dt'} 
\Big) 
\Big(
\frac{1}{-(t-t'-i\epsilon)^2+D^2}
+\frac{1}{-(t-t'+i\epsilon)^2+D^2} 
\Big)
\nonumber 
\\
&
\approx 
\frac{64e^2}{3\pi^2} \frac{L^2}{T^2} \Big(1+\frac{4D^2}{T^2} \ln \Big[\frac{D}{T}\Big] \Big),
\label{gammactdl}
\end{align}
where the distance between the particles $|\bm{x}-\bm{y}|$ was approximated as $D$ because of $D \gg L$ in the third line, and in the final line we took the limit $\epsilon \rightarrow 0$ and the leading order of $T/D \ll 1$ after the integration, and 
\begin{align}
\Phi
&=
-\frac{e^2}{4 \pi \hbar } 
\int dt 
\Big[
\frac{2}{D}
\Big(
 1-\frac{v^2}{c^2}
\Big)
-
\Big(
 1+\frac{v^2}{c^2}
\Big)
\Big(\frac{1}{D-2X(t)}+\frac{1}{D+2X(t)}\Big)
\Big]
\nonumber
\\
&\approx
-\frac{e^2}{4 \pi \hbar } 
\int dt 
\Big[
\frac{2}{D}
\Big(
 1-\frac{v^2}{c^2}
\Big)
-
\frac{2}{D}
\Big(
 1+\frac{v^2}{c^2}
\Big)
\Big(1+\frac{4X^2(t)}{D^2}\Big)
\Big]
\nonumber
\\
&\approx
-\frac{e^2}{4 \pi \hbar } 
\int dt 
\Big[
-\frac{4}{D}\frac{v^2}{c^2}
-\frac{8X^3(t)}{D^3}
\Big]
\nonumber 
\\
&
=\frac{64e^2}{315\pi\hbar c}\Big(\frac{L}{cT}\Big)^2\Big(\frac{6cT}{D}+\Big(\frac{cT}{D}\Big)^3\Big),
\end{align}
where we took the leading order of $4X^2(t)/D^2 \sim O(L^2/D^2) \ll 1$ in the second line, and neglected $\mathcal{O}(L^4/D^4)$ in the last line.
Therefore, we obtain the result in the linear configuration in $cT \gg D \gg L$ regime as
\begin{align}
\Gamma_1=\Gamma_2 \approx \frac{32 e^2}{3\pi^2 \hbar c} \frac{L^2}{(cT)^2}, 
\quad 
\Gamma_\text{c} \approx \frac{64e^2}{3\pi^2 \hbar c} \frac{L^2}{(cT)^2} \Big(1+\frac{4D^2}{(cT)^2} \ln \Big[\frac{D}{cT}\Big] \Big),
\quad
\Phi \approx \frac{64e^2}{315\pi\hbar c}\Big(\frac{L}{cT}\Big)^2\Big(\frac{6cT}{D}+\Big(\frac{cT}{D}\Big)^3\Big).
\label{lTDL}
\end{align}

\subsubsection{\label{cphilDTL}$D \gg T \gg L$ regime}
Here, we focus on the regime $D \gg T \gg L$ and calculate the quantities $\Gamma_{\text{c}}$ and $\Phi$.
We assume the following trajectories of the two charged particles 1 and 2 as
\begin{equation}
X^\mu_{1\text{P}}(t)=\Big[t,\epsilon_{\text{P}}X(t), 0,0 \Big]^{T}, 
\quad 
X^\mu_{2\text{Q}}(t)=\Big[t,\epsilon_{\text{Q}}X(t-D)+D, 0,0 \Big]^{T}, 
\quad 
\epsilon_{\text{R}}=-\epsilon_{\text{L}}=1, 
\quad 
X(t)=8L\Big(1-\frac{t}{T} \Big)^2 \Big(\frac{t}{T}\Big)^2,
\label{linearlDTL}
\end{equation}
where 
$X^\mu_{2\text{Q}}$ is defined in 
$D \leq t \leq T+D$.
First, we calculate the quantity $\Gamma_{\text{c}}$ in this regime by
using $\eqref{gammac}$ as
\begin{align}
\Gamma_{c}
&=\frac{e^2}{2}\oint_{\text{C}_1} dx^\mu \oint_{\text{C}_2} dy^\nu  
\, 
\langle \{\hat{A}^\text{I}_{\mu}(x), \hat{A}^\text{I}_{\nu}(y)\}\rangle
\nonumber 
\\
&=\frac{e^2}{2}\oint_{\text{C}_1} dx^\mu \oint_{\text{C}_2} dy^\nu  
\, 
\frac{\eta_{\mu \nu} }{4\pi^2} 
\Big(
\frac{1}{-(x^0-y^0-i\epsilon)^2+|\bm{x}-\bm{y}|^2}
+\frac{1}{-(x^0-y^0+i\epsilon)^2+|\bm{x}-\bm{y}|^2} 
\Big)
\nonumber 
\\
&\approx 
\frac{e^2}{2}\oint_{\text{C}_1} dx^\mu \oint_{\text{C}_2} dy^\nu  
\, 
\frac{\eta_{\mu \nu} }{4\pi^2} 
\Big(
\frac{1}{-(x^0-y^0-i\epsilon)^2+D^2}
+\frac{1}{-(x^0-y^0+i\epsilon)^2+D^2} 
\Big)
\nonumber 
\\
&=
\frac{e^2}{8\pi^2}\int^T_0 dt 
\Big( 
\frac{d X^\mu_{1\text{R}}}{dt}
-\frac{d X^\mu_{1\text{L}}}{dt} 
\Big) 
\int^{T+D}_D dt' 
\Big( 
\frac{d X_{2\text{R}\,\mu}}{dt'}
-\frac{d X_{2\text{L}\,\mu}}{dt'} 
\Big) 
\Big(
\frac{1}{-(t-t'-i\epsilon)^2+D^2}
+\frac{1}{-(t-t'+i\epsilon)^2+D^2} 
\Big)
\nonumber 
\\
&=
\frac{e^2}{8\pi^2}\int^T_0 dt \int^{T+D}_D dt' 
\Big( 
\frac{d \bm{X}_{1\text{R}}}{dt}
-\frac{d \bm{X}_{1\text{L}}}{dt} 
\Big) 
\cdot 
\Big( 
\frac{d \bm{X}_{2\text{R}}}{dt'}
-\frac{d \bm{X}_{2\text{L}}}{dt'} 
\Big) 
\Big(
\frac{1}{-(t-t'-i\epsilon)^2+D^2}
+\frac{1}{-(t-t'+i\epsilon)^2+D^2} 
\Big)
\nonumber 
\\
&
\approx
\frac{e^2}{8\pi^2}\frac{4}{D^2}\int^T_0 dt \int^{T+D}_D dt' 
\frac{d {X}(t)}{dt}
\cdot 
\frac{d {X}(t'-D)}{dt'}\Big\{1+\frac{(t-t'-i\epsilon)^2}{D^2}+1+\frac{(t-t'+i\epsilon)^2}{D^2}\Big\}
\nonumber 
\\
&
=
\frac{e^2}{2\pi^2D^4}\int^T_0 dt \int^{T+D}_D dt' 
\frac{d {X}(t)}{dt}
\cdot 
\frac{d {X}(t'-D)}{dt'}\Big\{{(t-t'-i\epsilon)^2}+{(t-t'+i\epsilon)^2}\Big\}
\nonumber 
\\
&
=
-\frac{32e^2}{225\pi^2} \frac{L^2 T^2}{D^4}.
\label{eq:GammacA1}
\end{align}
where the distance between the particles $|\bm{x}-\bm{y}|$ was approximated as $D$ because of $D \gg L$ in the third line. We used the geometric series expansion because of $|(t-t^{'}\pm i\epsilon)| < T \ll D$ in the third to last line, and in the final line, we took the limit $\epsilon \rightarrow 0$ after the integration.
We next calculate
the quantity $\Phi$ using Eq. \eqref{phiF} in this regime.
The quantity $\Phi$ is
\begin{align}
\Phi
&=
\frac{e}{4} 
\Big (
\int_{\text{S}_1} d\sigma_{\mu \nu} \Delta F^{\mu\nu}_2 (x)
+\int_{\text{S}_2} d\sigma_{\mu \nu} \Delta F^{\mu\nu}_1 (x)
\Big)
\nonumber 
\\
&=
\frac{e}{4} 
\int_{\text{S}_2} d\sigma_{\mu \nu} \Delta F^{\mu\nu}_1 (x)
\nonumber 
\\
&=
\frac{e}{2} 
\int^{T+D}_{D} dt \int^{X_{2\text{R}}(t)+D}_{X_{2\text{L}}(t)+D}dx \, \Delta F^{01}_1 (t,x,0,0)
\nonumber
\\
&=
\frac{e}{2} 
\int^{T+D}_{D} dt \int^{X_{2\text{R}}(t)}_{X_{2\text{L}}(t)}dx \, \Delta F^{01}_1 (t,x+D,0,0),
\end{align}
where the region $S_2= \{D \leq t \leq T+D,\ X_{2\text{L}}(t)+D\leq x \leq X_{2\text{R}}(t)+D,\ y=0,\ z=0\}$, and the first term in the first line vanishes because, in this configuration, particle 1 does not experience the retarded field of particle 2.
We changed the variable $x\rightarrow x+D$ in the final line.
The quantity $\Phi$ is decomposed into two terms $\Phi = \Phi_{\text{v}}+\Phi_{\text{a}}$, which are given as follows (see Eqs. \eqref{eq:Fv2} and \eqref{eq:Fa2} in  Appendix \ref{sec:LW})
\begin{align}
\Phi_\text{v}
&=
\frac{e}{2} 
\int^{T+D}_{D} dt \int^{X_{2\text{R}}(t)}_{X_{2\text{L}}(t)}dx \, \Delta F^{01}_{1,\text{v}} (t,x+D,0,0)
\nonumber 
\\
&=
\frac{e}{2} 
\int^{T+D}_{D} dt \int^{X_{2\text{R}}(t)}_{X_{2\text{L}}(t)}dx \, \sum_{\text{P}=\text{R},\text{L}}
\epsilon_\text{P}
\Big 
[
\frac{e}
{4\pi} \frac{(t-t_{1\text{P}}) v_{1\text{P}} (t_{1\text{P}})- (x+D-X_{1\text{P}}(t_{1\text{P}}))}{\gamma^2_{1\text{P}} [ t-t_{1\text{P}}-(x+D-X_{1\text{P}}(t_{1\text{P}})) v_{1\text{P}}(t_{1\text{P}}) ]^3}
\Big],
\\
\Phi_\text{a}
&=
\frac{e}{2} 
\int^{T+D}_{D} dt \int^{X_{2\text{R}}(t)}_{X_{2\text{L}}(t)}dx \, \Delta F^{01}_{1,\text{a}} (t,x+D,0,0)
\nonumber 
\\
&=
\frac{e}{2} 
\int^{T+D}_{D} dt \int^{X_{2\text{R}}(t)}_{X_{2\text{L}}(t)}dx \sum_{\text{P}=\text{R},\text{L}}
\epsilon_\text{P}
\frac{e}
{4\pi [t-t_{1\text{P}}-(x+D-X_{1\text{P}}(t_{1\text{P}})) v_{1\text{P}}(t_{1\text{P}}) ]^2 } 
\nonumber 
\\
&
\quad 
\times
\Big [
(t-t_{1\text{P}})
\Big(
a_{1\text{P}} (t_{1\text{P}})
+\frac{(x+D-X_{1\text{P}}(t_{1\text{P}}))a_{1\text{P}}(t_{1\text{P}})}
{t-t_{1\text{P}}-(x+D-X_{1\text{P}}(t_{1\text{P}})) v_{1\text{P}}(t_{1\text{P}})}
v_{1\text{P}}(t_{1\text{P}}) 
\Big)
\nonumber 
\\
&
\quad
-\frac{(x+D-X_{1\text{P}}(t_{1\text{P}}))^2a_{1\text{P}}(t_{1\text{P}})}
{(t-t_{1\text{P}})-(x+D-X_{1\text{P}}(t_{1\text{P}})) v_{1\text{P}}(t_{1\text{P}})}
\Big)
\Big],
\nonumber
\\
&=
\frac{e^2}{8\pi} 
\int^{T+D}_{D} dt \int^{X_{2\text{R}}(t)}_{X_{2\text{L}}(t)}dx \sum_{\text{P}=\text{R},\text{L}}
\epsilon_\text{P}
\Big[
\frac{(t-t_{1\text{P}})^2-(x+D-X_{1\text{P}}(t_{1\text{P}}))^2}
{ [t-t_{1\text{P}}-(x+D-X_{1\text{P}}(t_{1\text{P}})) v_{1\text{P}}(t_{1\text{P}}) ]^3 } 
\Big]a_{1\text{P}}(t_{1\text{P}}),
\end{align}
where the retarded time $t_{1\text{p}}$ is approximated by neglecting $\mathcal{O}(L^2/D^2)$ as 
\begin{align}
t_{1\text{P}}
&
=t-|\bm{x}-\bm{X}_{1\text{P}}(t_{1\text{P}})|
=t-\sqrt{(x+D-X_{1\text{P}}(t_{1\text{P}}))^2}
\approx 
t-{D}, 
\label{retlin}
\end{align}
where $(x-X_{1\text{P}}(t_{1\text{P}})) \sim \mathcal{O}(L)$. 
For $D \gg cT \gg L$, we can approximate $\Phi_{\text{v}}$ as
\begin{align}
\Phi_\text{v}
&=
\frac{e}{2} 
\int^{T+D}_{D} dt \int^{X_{2\text{R}}(t)}_{X_{2\text{L}}(t)}dx \, \sum_{\text{P}=\text{R},\text{L}}
\epsilon_\text{P}
\Big 
[
\frac{e}
{4\pi} \frac{(t-t_{1\text{P}}) v_{1\text{P}} (t_{1\text{P}})- (x+D-X_{1\text{P}}(t_{1\text{P}}))}{\gamma^2_{1\text{P}} [ t-t_{1\text{P}}-(x+D-X_{1\text{P}}(t_{1\text{P}})) v_{1\text{P}}(t_{1\text{P}}) ]^3}
\nonumber 
\\
&\approx
\frac{e^2}{8\pi} 
\int^{T+D}_{D} dt \int^{X_{2\text{R}}(t)}_{X_{2\text{L}}(t)}dx 
\sum_{\text{P}=\text{R},\text{L}}
\epsilon_\text{P}
\Big 
[
\frac{v_{1\text{P}}(t-D)}{D^2}
+\frac{X_{1\text{P}}(t-D)}{D^3}
-\frac{x+D}{D^3}
\Big]
\nonumber 
\\
&=
\frac{e^2}{8\pi} 
\int^{T+D}_{D} dt 
(X_{2\text{R}}(t)-X_{2\text{L}}(t)) 
\Big 
[
\frac{v_{1\text{R}}(t-D)-v_{1\text{L}}(t-D)}{D^2}
+\frac{X_{1\text{R}}(t-D)-X_{1\text{L}}(t-D)}{D^3}
\Big]
\nonumber 
\\
&
= \frac{16e^2}{315\pi} \frac{L^2 T}{D^3}.
\label{eq:PhivA1}
\end{align}
Moreover, in the second line of the above equation, we substituted the retarded condition \eqref{retlin} into Eq. \eqref{eq:PhivA1} and approximated the denominator as
\begin{align}
\gamma^2_{1\text{P}} [ t-t_{1\text{P}}-(x+D-X_{1\text{P}}(t_{1\text{P}})) v_{1\text{P}}(t_{1\text{P}})]
&\approx \big(1-v^2_{1\text{P}}(t-D)\big)^{-1}[D-(x+D-X_{1\text{P}}(t-D)) v_{1\text{P}}(t-D)]
\nonumber\\
\quad
&=
D\big(1-v^2_{1\text{P}}(t-D)\big)^{-1}[1-(1+(x-X_{1\text{P}}(t-D))/D) v_{1\text{P}}(t-D)]
\nonumber\\
\quad
&
\approx
D,
\label{denapp}
\end{align}
where $v_{1\text{P}}\sim \mathcal{O}(L/T)$, $v^2_{1\text{P}} \sim \mathcal{O}(L^2/T^2)$, and $(x-X_{1\text{P}})/D \sim \mathcal{O}(L/D)$ were neglected in the last line.
However, the quantity $\Phi_{\text{a}}$ is exactly equal to zero because of the retarded time condition \eqref{retlin}.
This result indicates that in the context of equation \eqref{phiele}, the electric field $E^{1\text{R}, \text{a}}_{x}\ (E^{1\text{L}, \text{a}}_{x})$ is equal to zero because the electromagnetic wave cannot propagate the direction of the acceleration of the charged particle 1.
Therefore, we summarize the result in the linear configuration in $D \gg cT \gg L$ regime as follows
\begin{align}
\Gamma_{1}=\Gamma_{2} \approx \frac{32 e^{2}}{3 \pi^{2}\hbar c} \frac{L^{2}}{(cT)^{2}},
\quad
\Gamma_{\text{c}} \approx-\frac{32e^{2}}{225 \pi^{2}\hbar c} \frac{L^{2} (cT)^{2}}{D^{4}},
\quad
\Phi \approx \frac{16e^2}{315\pi \hbar c} \frac{L^2 (cT)}{D^3}.
\label{lDTL}
\end{align}
\subsection{\label{cphip}Computation of $\Gamma_{\text{c}}$ and $\Phi$ for parallel configuration}
\subsubsection{\label{cphipTDL}$T \gg L \gg D$ or $T\gg D \gg L$ regimes}
Here, we focus on the regimes $T \gg L \gg D$ or $T\gg D \gg L$ and calculate the quantities $\Gamma_{\text{c}}$ and $\Phi$.
We assume the following trajectories of the two charged particles 1 and 2 as
\begin{equation}
X^\mu_{1\text{P}}(t)=\Big[t,\epsilon_{\text{P}}X(t), 0,0 \Big]^\text{T}, 
\quad
X^\mu_{2\text{Q}}(t)=\Big[t,\epsilon_{\text{Q}}X(t), D,0 \Big]^\text{T}, 
\quad 
\epsilon_{\text{R}}=-\epsilon_{\text{L}}=1, 
\quad 
X(t)=8L\Big(1-\frac{t}{T} \Big)^2 \Big(\frac{t}{T}\Big)^2,
\label{parallelTDL}
\end{equation}
In these regimes, the approximate form of $\Gamma_{\text{c}}$ is equal to $\eqref{gammactdl}$.
Neglecting $\mathcal{O}(D^2/T^2)$ in $T \gg L \gg D$, we obtain the quantity $\Gamma_{\text{c}}$ as
\begin{align}
\Gamma_{\text{c}}
\approx 
\frac{64e^2}{3\pi^2} \frac{L^2}{T^2},
\end{align}
The quantity $\Phi$ up to $\mathcal{O}(1/c^2)$ obtained from \eqref{phiex} is
\begin{align}
\Phi
&=
-\frac{e^2}{8 \pi \hbar } 
\int dt 
\sum_{\text{P},\text{Q}=\text{R},\text{L}}
\frac{\epsilon_\text{P} \epsilon_\text{Q} }
{|\bm{X}_{1\text{P}} -\bm{X}_{2\text{Q}} |} 
\Big[1-\frac{\bm{v}_{1\text{P}} \cdot \bm{v}_{2\text{Q}} }{c^2}
\nonumber 
\\
&
\quad
+\frac{1}{2c^2} 
\Big \{v^2_{2\text{Q}}-\Big(\frac{\bm{X}_{1\text{P}} -\bm{X}_{2\text{Q}}}{|\bm{X}_{1\text{P}} -\bm{X}_{2\text{Q}} |} \cdot \bm{v}_{2\text{Q}} \Big)^2\Big\}
-\frac{(\bm{X}_{1\text{P}}-\bm{X}_{2\text{Q}}) \cdot \bm{a}_{2\text{Q}} }{2c^2} \Big]+(1 \leftrightarrow 2)\nonumber\\
&=
-\frac{e^2}{8 \pi \hbar } 
\int dt 
\sum_{\text{P},\text{Q}=\text{R},\text{L}}
\frac{\epsilon_\text{P} \epsilon_\text{Q} }{\sqrt{(X_{1\text{P}} -X_{2\text{Q}})^2 +D^2}} 
\Big[
1-\frac{v_{1\text{P}} v_{2\text{Q}} }{c^2}
\nonumber 
\\
&
\quad
+\frac{1}{2c^2} 
\Big \{
v^2_{2\text{Q}}-\Big(\frac{X_{1\text{P}} -X_{2\text{Q}}}{\sqrt{(X_{1\text{P}} -X_{2\text{Q}})^2+D^2}} v_{2\text{Q}} \Big)^2
\Big \}
-\frac{(X_{1\text{P}} -X_{2\text{Q}})  a_{2\text{Q}} }{2c^2} 
\Big]
+(1 \leftrightarrow 2)
\nonumber 
\\
&=
-\frac{e^2}{4 \pi \hbar } 
\int dt 
\Big(
\frac{2}{D}
\Big[
 1-\frac{v^2}{2c^2}
\Big]
-
\frac{2 }{\sqrt{4X^2 +D^2}} 
\Big[
1+\Big(1+\frac{D^2}{2(4X^2+D^2)} \Big)\frac{v^2}{c^2}
+\frac{X a }{c^2} 
\Big]
\Big).
\end{align}
For $cT \gg L \gg D$, the quantity $\Phi$ is approximated as
\begin{align}
\Phi
&=
-\frac{e^2}{4 \pi \hbar } 
\int dt 
\Big(
\frac{2}{D}
\Big[
 1-\frac{v^2}{2c^2}
\Big]
-
\frac{2 }{\sqrt{4X^2 +D^2}} 
\Big[
1+\Big(1+\frac{D^2}{2(4X^2+D^2)} \Big)\frac{v^2}{c^2}
+\frac{X a }{c^2} 
\Big]
\Big)
\nonumber 
\\
&\approx
-\frac{e^2}{4 \pi \hbar } 
\int dt 
\frac{2}{D}
\Big[
 1-\frac{v^2}{2c^2}
\Big]
\nonumber 
\\
&=
-\frac{e^2}{2 \pi \hbar c} \frac{cT}{D}
\Big(1
-\frac{64 L^2}{105(cT)^2}
\Big),
\end{align}
where we neglected $\mathcal{O}(D/L)$ in the second line. 
In the regime $cT \gg D \gg L$, we obtain
\begin{align}
\Phi
&=
-\frac{e^2}{4 \pi \hbar } 
\int dt 
\Big(
\frac{2}{D}
\Big[
 1-\frac{v^2}{2c^2}
\Big]
-
\frac{2 }{\sqrt{4X^2 +D^2}} 
\Big[
1+\Big(1+\frac{D^2}{2(4X^2+D^2)} \Big)\frac{v^2}{c^2}
+\frac{X a }{c^2} 
\Big]
\Big)
\nonumber 
\\
&\approx
-\frac{e^2}{4 \pi \hbar } 
\int dt 
\Big[ 
\frac{4X^2}{D^3}-\frac{4v^2+2X a}{c^2 D}
\Big]
\nonumber 
\\
&=
-\frac{32e^2}{315 \pi \hbar c} \frac{cT L^2}{D^3}
\Big(1
-\frac{6 D^2}{(cT)^2}
\Big),
\label{eq:PhiA2}
\end{align}
where we used the Taylor expansion $(4X^2+D^2)^{\alpha} \approx D^{2\alpha}(1+4\alpha X^2/D^2)$ in the first line and neglected $\mathcal{O}(L^3/T^3)$ in the second line.
Consequently, $\Gamma_{1}, \Gamma_{2}, \Gamma_{\text{c}}$, and $\Phi$ in the parallel configuration are obtained as
\begin{align}
\Gamma_{1}=\Gamma_{2} \approx \frac{32 e^{2}}{3 \pi^{2} \hbar c} \frac{L^{2}}{(c T)^{2}},
\quad
\Gamma_{\text{c}} \approx \frac{64 e^{2}}{3 \pi^{2} \hbar c} \frac{L^{2}}{(c T)^{2}},
\quad
\Phi \approx-\frac{e^{2}}{2 \pi \hbar c} \frac{c T}{D}\left(1-\frac{64 L^{2}}{105(c T)^{2}}\right),
\label{pTLD}
\end{align}
for $cT \gg L \gg D$, and 
\begin{align}
\Gamma_{1}=\Gamma_{2} \approx \frac{32 e^{2}}{3 \pi^{2} \hbar c} \frac{L^{2}}{(c T)^{2}},
\quad
\Gamma_{\text{c}} \approx \frac{64 e^{2}}{3 \pi^{2} \hbar c} \frac{L^{2}}{(c T)^{2}}\left(1+\frac{4 D^{2}}{(c T)^{2}} \ln \left[\frac{D}{c T}\right]\right),
\quad
\Phi \approx-\frac{32 e^{2}}{315 \pi \hbar c} \frac{c T L^{2}}{D^{3}}\left(1-\frac{6 D^{2}}{(c T)^{2}}\right),
\label{pTDL}
\end{align}
for $cT \gg D \gg L$, respectively.
\subsubsection{\label{cphipDTL}$D \gg T \gg L$ regime}
Here, we consider the $D \gg T \gg L$ regime and calculate the quantities $\Gamma_{\text{c}}$ and $\Phi$.
In this regime, the trajectories of the two charged particles 1 and 2 are assumed as follows
\begin{equation}
X^\mu_{1\text{P}}(t)=\Big[t,\epsilon_{\text{P}}X(t), 0,0 \Big]^\text{T}, 
\quad
X^\mu_{2\text{P}}(t)=\Big[t,\epsilon_{\text{P}}X(t-D), D,0 \Big]^\text{T}, 
\quad 
\epsilon_{\text{R}}=-\epsilon_{\text{L}}=1, 
\quad 
X(t)=8L\Big(1-\frac{t}{T} \Big)^2 \Big(\frac{t}{T}\Big)^2,
\end{equation}
where 
$X^\mu_{2\text{Q}}$ is defined in 
$D \leq t \leq T+D$.
The quantity $\Gamma_{\text{c}}$ is equal to the Eq. \eqref{eq:GammacA1} because we can approximate the difference of the distance of the two charged particles $|\bm{x}-\bm{y}| \approx D$ and use the geometric series expansion because of $|(t-t'\pm i\epsilon)|/D < T/D \ll 1$ in this regime (detailed derivation, see the Eq. \eqref{eq:GammacA1}). 
The quantity $\Phi$ is obtained as
\begin{align}
\Phi
&=
\frac{e}{4} 
\Big (
\int_{\text{S}_1} d\sigma_{\mu \nu} \Delta F^{\mu\nu}_2 (x)
+\int_{\text{S}_2} d\sigma_{\mu \nu} \Delta F^{\mu\nu}_1 (x)
\Big)
\nonumber 
\\
&=
\frac{e}{4} 
\int_{\text{S}_2} d\sigma_{\mu \nu} \Delta F^{\mu\nu}_1 (x)
\nonumber 
\\
&=
\frac{e}{2} 
\int^{T+D}_{D} dt \int^{X_{2\text{R}}(t)}_{X_{2\text{L}}(t)}dx \, \Delta F^{01}_1 (t,x,D,0),
\end{align}
where we note that the region $S_2=\{D\leq t \leq T+D, X_{2\text{L}}(t)\leq x \leq X_{2\text{R}}, y=D, z=0\}$; in this configuration of interest, the first term in the first line vanishes because the retarded field from particle 2 is causally disconnected with particle 1.
The retarded time $t_{1\text{p}}$ is approximated as
\begin{align}
t_{1\text{P}}
=t-|\bm{x}-\bm{X}_{1\text{P}}(t_{1\text{P}})|
=t-\sqrt{(x-X_{1\text{P}}(t_{1\text{P}}))^2 +D^2}
\approx 
t-{D} - \frac{(x-X_{1\text{P}}(t-D))^2}{2D},
\label{retpar}
\end{align}
where $(x-X_{1\text{P}}(t_{1\text{P}})) \sim \mathcal{O}(L)$ and $\mathcal{O}(L^2/D^2)$ was neglected.
We therefore obtain the quantity $\Phi_{\text{v}}$ and $\Phi_{\text{a}}$ as
\begin{align}
\Phi_\text{v}
&=
\frac{e}{2} 
\int^{T+D}_{D} dt \int^{X_{2\text{R}}(t)}_{X_{2\text{L}}(t)}dx \, \sum_{\text{P}=\text{R},\text{L}}
\epsilon_\text{P}
\Big 
[
\frac{e}
{4\pi} \frac{(t-t_{1\text{P}}) v_{1\text{P}} (t_{1\text{P}})- (x-X_{1\text{P}}(t_{1\text{P}}))}{\gamma^2_{1\text{P}} [ t-t_{1\text{P}}-(x-X_{1\text{P}}(t_{1\text{P}})) v_{1\text{P}}(t_{1\text{P}}) ]^3}
\Big]
\nonumber 
\\
&\approx
\frac{e^2}{8\pi} 
\int^{T+D}_{D} dt \int^{X_{2\text{R}}(t)}_{X_{2\text{L}}(t)}dx 
\sum_{\text{P}=\text{R},\text{L}}
\epsilon_\text{P}
\Big 
[
\frac{v_{1\text{P}}(t-D)}{D^2}
-\frac{x-X_{1\text{P}}(t-D)}{D^3}
\Big]
\nonumber 
\\
&=
\frac{e^2}{8\pi} 
\int^{T+D}_{D} dt 
(X_{2\text{R}}(t)-X_{2\text{L}}(t)) 
\Big 
[
\frac{v_{1\text{R}}(t-D)-v_{1\text{L}}(t-D)}{D^2}
+\frac{X_{1\text{R}}(t-D)-X_{1\text{L}}(t-D)}{D^3}
\Big]
\nonumber 
\\
&
= \frac{16e^2}{315\pi} \frac{L^2 T}{D^3},
\label{eq:PhivA2}
\end{align}
where in the second line of the above equation, the denominator was approximated in the same manner performed in \eqref{denapp},
and
\begin{align}
\Phi_\text{a}
&=
\frac{e}{2} 
\int^{T+D}_{D} dt \int^{X_{2\text{R}}(t)}_{X_{2\text{L}}(t)}dx \sum_{\text{P}=\text{R},\text{L}}
\epsilon_\text{P}
\frac{e}
{4\pi [t-t_{1\text{P}}-(x-X_{1\text{P}}(t_{1\text{P}})) v_{1\text{P}}(t_{1\text{P}}) ]^2 } 
\nonumber 
\\
&
\quad 
\times
\Big [
(t-t_{1\text{P}})
\Big(
a_{1\text{P}} (t_{1\text{P}})
+\frac{(x-X_{1\text{P}}(t_{1\text{P}}))a_{1\text{P}}(t_{1\text{P}})}
{t-t_{1\text{P}}-(x-X_{1\text{P}}(t_{1\text{P}})) v_{1\text{P}}(t_{1\text{P}})}
v_{1\text{P}}(t_{1\text{P}}) 
\Big)
-\frac{(x-X_{1\text{P}}(t_{1\text{P}}))^2a_{1\text{P}}(t_{1\text{P}})}
{(t-t_{1\text{P}})-(x-X_{1\text{P}}(t_{1\text{P}})) v_{1\text{P}}(t_{1\text{P}})}
\Big)
\Big]
\nonumber
\\
&=
\frac{e^2}{8\pi} 
\int^{T+D}_{D} dt \int^{X_{2\text{R}}(t)}_{X_{2\text{L}}(t)}dx \sum_{\text{P}=\text{R},\text{L}}
\epsilon_\text{P}
\Big[
\frac{(t-t_{1\text{P}})^2-(x-X_{1\text{P}}(t_{1\text{P}}))^2}
{ [t-t_{1\text{P}}-(x-X_{1\text{P}}(t_{1\text{P}})) v_{1\text{P}}(t_{1\text{P}}) ]^3 } 
\Big]a_{1\text{P}}(t_{1\text{P}})
\nonumber 
\\
&=
\frac{e^2}{8\pi} 
\int^{T+D}_{D} dt \int^{X_{2\text{R}}(t)}_{X_{2\text{L}}(t)}dx \sum_{\text{P}=\text{R},\text{L}}
\epsilon_\text{P}
\Big[
\frac{D^2}
{ [t-t_{1\text{P}}-(x-X_{1\text{P}}(t_{1\text{P}})) v_{1\text{P}}(t_{1\text{P}}) ]^3 } 
\Big]a_{1\text{P}}(t_{1\text{P}})
\nonumber 
\\
&\approx 
\frac{e^2}{8\pi} 
\int^{T+D}_{D} dt \int^{X_{2\text{R}}(t)}_{X_{2\text{L}}(t)}dx
\sum_{\text{P}=\text{R},\text{L}}
\epsilon_\text{P}
\frac{a_{1\text{P}}(t-D)}
{D} 
\nonumber 
\\
&=
\frac{e^2}{8\pi} 
\int^{T+D}_{D} dt (X_{2\text{R}}(t)-X_{2\text{L}}(t))
\Big [
\frac{a_{1\text{R}}(t-D)-a_{1\text{L}}(t-D)}
{D} 
\Big]
\nonumber 
\\
&=
-\frac{64e^2}{105\pi} \frac{L^2}{DT},
\end{align}
where we substituted the retarded time condition $\eqref{retpar}$ into the second line of the above equation and neglected the $\mathcal{O}(L^2/D^2)$ and $v \sim \mathcal{O}(L/T)$ in the third line of the denominator.
Consequently, the quantity $\Phi$ is
\begin{equation}
\Phi
\approx-\frac{64 e^{2}}{105 \pi} \frac{L^{2}}{DT}\left(1-\frac{ T^{2}}{12D^{2}}\right)
\approx -\frac{64 e^{2}}{105 \pi} \frac{L^{2}}{DT},
\end{equation}
where we neglected the second term because of $D \gg T$ in the last equality.
Thus, $\Gamma_{1}, \Gamma_{2}, \Gamma_{\text{c}}$, and $\Phi$ in the parallel configuration in the regime $D \gg cT \gg L$ are 
\begin{align}
\Gamma_{1}=\Gamma_{2} \approx \frac{32 e^{2}}{3 \pi^{2}\hbar c} \frac{L^{2}}{(cT)^{2}},
\quad
\Gamma_{\text{c}} \approx-\frac{32e^{2}}{225 \pi^{2}\hbar c} \frac{L^{2} (cT)^{2}}{D^{4}},
\quad
\Phi \approx-\frac{64 e^{2}}{105 \pi \hbar c} \frac{L^{2}}{D(cT)}.
\label{pDTL}
\end{align}

\end{appendix}

\end{document}